\documentclass[a4paper,11pt]{article}

\usepackage{amsmath}
\usepackage{amssymb}
\usepackage[page]{appendix}
\usepackage{authblk}
\usepackage{bbm}
\usepackage{bm}
\usepackage{cancel}
\usepackage{cite}
\usepackage{comment}
\usepackage{enumitem}
\usepackage{geometry}
\usepackage{graphicx}
\usepackage[scr=boondox]{mathalpha}
\usepackage{mathtools}
\usepackage{physics}
\usepackage{slashed}
\usepackage{subcaption}
\usepackage{subfiles}
\usepackage[normalem]{ulem}
\usepackage{upgreek}
\usepackage{xcolor}
\usepackage{tikz}
\usepackage{tikz}
\usetikzlibrary{arrows,shapes}
\usetikzlibrary{trees,patterns}
\usetikzlibrary{matrix,arrows} 				
\usetikzlibrary{positioning}				  
\usetikzlibrary{calc,through}				  
\usetikzlibrary{decorations.pathreplacing}  
\usepackage{pgffor}							

\usetikzlibrary{decorations.pathmorphing}	
\usetikzlibrary{decorations.markings}
\tikzset{
>=stealth', 
vector/.style={decorate, decoration={snake}, draw},
provector/.style={decorate, decoration={snake,amplitude=2.5pt}, draw},
antivector/.style={decorate, decoration={snake,amplitude=-2.5pt}, draw},
bigvector/.style={decorate, decoration={snake,amplitude=4pt}, draw},
fermion/.style={draw=black, postaction={decorate}, 
	decoration={markings,mark=at position .55 with {\arrow[draw=black]{>}}}},
fermionbar/.style={draw=black, postaction={decorate},
    decoration={markings,mark=at position .55 with {\arrow[draw=black]{<}}}},
fermionnoarrow/.style={draw=black},
doublefermion/.style={draw=black,double, postaction={decorate},
	decoration={markings,mark=at position .57 with {\arrow[draw=black]{>}}}},
doublefermionbar/.style={draw=black,double, postaction={decorate},
	decoration={markings,mark=at position .57 with {\arrow[draw=black]{<}}}},
doublefermionnoarrow/.style={draw=black,double},
gluon/.style={decorate, draw=black,
    decoration={coil,amplitude=4pt, segment length=5pt}},
scalar/.style={dashed,draw=black, postaction={decorate},
	decoration={markings,mark=at position .55 with {\arrow[draw=black]{>}}}},
scalarbar/.style={dashed,draw=black, postaction={decorate},
    decoration={markings,mark=at position .55 with {\arrow[draw=black]{<}}}},
scalarnoarrow/.style={dashed,draw=black},
momentum/.style={draw=black, postaction={decorate},
    decoration={markings,mark=at position 1 with {\arrow[draw=black]{>}}}},
antimomentum/.style={draw=black, postaction={decorate},
    decoration={markings,mark=at position 0.1 with {\arrow[draw=black]{<}}}}
}

\tikzstyle{block} = [draw, rectangle, minimum height=3em, minimum width=6em]

\usepackage[colorlinks=true,breaklinks=true]{hyperref}
\hypersetup{
allcolors=[rgb]{0.0 0.0 0.6},
linkcolor=[rgb]{0.0 0.0 1}
}

\newcommand{\ann}{{\rm Ann}}
\newcommand{\arho}{\alpha_{\varrho}}
\newcommand{\aV}{\alpha_{\scriptscriptstyle{V}}}
\newcommand{\average}[1]{\left\langle #1 \right\rangle}
\newcommand{\BSD}{{\scriptscriptstyle{\rm BSD}}}

\newcommand{\DM}{{\scriptscriptstyle{\rm DM}}}
\newcommand{\elas}{{\rm elas}}
\newcommand{\fin}{{\rm fin}}
\newcommand{\im}{\mathbbm{i}}
\newcommand{\inel}{{\rm inel}}

\newcommand{\med}{{\rm med}}
\newcommand{\mrho}{m_{\varrho}}
\newcommand{\mT}{m_{\scriptscriptstyle{T}}}
\newcommand{\mV}{m_{\scriptscriptstyle{V}}}
\newcommand{\mX}{m_{\scriptscriptstyle{X}}}
\newcommand{\nDM}{n_{\scriptscriptstyle{\rm DM}}}
\newcommand{\SIDM}{{\scriptscriptstyle{\rm SIDM}}}
\newcommand{\sym}{\mathscr{C}}
\newcommand{\ThreejSymbol}[6]{
\begin{pmatrix}
#1 & #3 & #5\\
#2 & #4 & #6
\end{pmatrix}}

\newcommand{\twoPI}{{\scriptscriptstyle{\rm 2PI}}}
\newcommand{\UNreg}{{\rm (un)reg}}
\newcommand{\unreg}{{\rm unreg}}
\newcommand{\reg}{{\rm reg}}
\newcommand{\ren}{{\rm ren}}

\newcommand{\vesc}{v_{\rm esc}}
\newcommand{\vrel}{v_{\rm rel}}
\newcommand{\vrelren}{v_{{\rm rel},\diamond}}
\newcommand{\y}{\mathrm{y}}

\newcommand{\zetaRenorm}{\zeta_{\diamond}}
\newcommand{\xiH}{\xi^{\scriptscriptstyle{H}}}

\newcommand{\T}{{\scriptscriptstyle{\rm T}}}
\newcommand{\U}{{\scriptscriptstyle{\rm U}}}
\newcommand{\UT}{{\scriptscriptstyle{\rm U,T}}}
\newcommand{\UV}{{\scriptscriptstyle{\rm U,V}}}
\newcommand{\V}{{\scriptscriptstyle{\rm V}}}
\newcommand{\CC}{\mathbbm{C}}
\newcommand{\ID}{\mathbbm{1}}

\newcommand{\NN}{\mathbbm{N}}
\newcommand{\RR}{\mathbbm{R}}
\newcommand{\WW}{\mathbbm{W}}

\newcommand{\ZZ}{\mathbbm{Z}}

\colorlet{lightblue}{blue!30}
\colorlet{lightred}{red!30}

\AtBeginEnvironment{appendices}{\crefalias{section}{appendix}}
\numberwithin{equation}{section}

\title{\bf 
Annihilation feedback on elastic scattering in self-interacting dark matter
}
\author{\Large 
Kalliopi Petraki\footnote{kalliopi.petraki@phys.ens.psl.eu} }
\author{Thomas Pignard\footnote{thomas.pignard@phys.ens.psl.eu}}
\affil{\large\it
Laboratoire de Physique de l'École Normale Supérieure, 
ENS, 
Université PSL, 
CNRS, 
Sorbonne Université, 
Université Paris Cité,
Paris, F-75005, France}

\usepackage[capitalise]{cleveref}
\crefformat{pluralequation}{#2{\color{black} Eqs.~(}#1{\color{black} )}#3}
\Crefformat{pluralequation}
{#2{\color{black} Equations~(}#1{\color{black} )}#3}

\begin{document}

\maketitle
\begin{abstract}
Sizable dark-matter self-interactions, motivated by observations of galactic structure, often arise from the exchange of light force mediators. Light mediators can render scattering nonperturbative and make inelastic processes important. Unitarity, via the optical theorem, implies that elastic and inelastic interactions cannot be treated independently: inelasticity induces absorptive contributions to the two-particle self-energy kernel that determines the elastic amplitude, while the resulting incoming-state dynamics also modifies the inelastic amplitudes. We study the feedback of dark-matter annihilation on elastic scattering in the presence of a long-range force. Although annihilation is a short-distance process, it can be strongly enhanced by such a force. We find that its feedback becomes significant near Sommerfeld resonances, where it suppresses the low-velocity growth of the baseline elastic cross section. Near baseline elastic-scattering zeros, it may instead generate a nonzero elastic floor required by unitarity. A companion paper studies the feedback of radiative bound-state formation on elastic scattering, which can generate sizable self-interactions.
\end{abstract}

\clearpage
\noindent\rule{\textwidth}{0.4pt}
\tableofcontents
\noindent\rule{\textwidth}{0.4pt}

\clearpage
\section{Introduction}
\label{Sec:Intro}

The collisionless cold dark matter paradigm successfully describes structure formation on cosmological and large astrophysical scales. On galactic and subgalactic scales, however, several challenges have motivated scenarios in which dark matter (DM) possesses sizable self-interactions~\cite{Spergel:1999mh}. Such interactions can redistribute energy and momentum within halos, modifying their density profiles and potentially affecting star formation~\cite{Bullock:2017xww,Tulin:2017ara}. The self-interaction strength relevant for halo dynamics varies considerably across astrophysical scales. At cluster velocities, $\vrel \sim 10^3~{\rm km/s}$, observations constrain the self-scattering cross section per unit mass to $\sigma/m_{\DM} \lesssim 1~{\rm barn/GeV}$~\cite{Markevitch:2003at,Randall:2008ppe,Eckert:2022qia}. At dwarf-galaxy velocities, $\vrel \sim 10~{\rm km/s}$, cross sections $\sigma/m_{\DM} \gtrsim 1~{\rm barn/GeV}$ are typically required to appreciably modify halo structure~\cite{Tulin:2017ara}. Strong-lensing analyses have also begun to reveal exceptionally compact substructures with inferred inner density profiles substantially steeper than typical collisionless-DM predictions. Such structures may arise from gravothermal core collapse in self-interacting dark matter (SIDM) halos~\cite{Dutra:2024qac,Enzi:2024ygw,Vegetti:2026mmx,Gilman:2026wul}. Under this interpretation, the cross sections at the characteristic internal velocities of the relevant subhalos may need to be larger by two to three orders of magnitude~\cite{Zeng:2023fnj}. Reconciling these different scales motivates self-interaction cross sections that decrease with increasing velocity~\cite{Firmani:2000ce,Yoshida:2000uw}.

A natural particle-physics realization of this behavior is provided by interactions mediated by light particles. When the mediator mass is comparable to or smaller than the characteristic momentum transfer, the interaction becomes effectively long-ranged and scattering requires a nonperturbative treatment.  This amounts to resumming the repeated exchange of light mediators and is achieved in the nonrelativistic regime by solving the Schrödinger equation with a Coulomb- or Yukawa-type potential derived from the one-boson-exchange kernel. The resulting cross sections can exhibit strong velocity dependence and resonant behavior~\cite{Feng:2009mn,Feng:2009hw,Loeb:2010gj,Kaplinghat:2015aga,Tulin:2012wi,Tulin:2013teo}.

Light force mediators can also render inelastic processes important. In addition to annihilation, DM particles may radiate force mediators in bremsstrahlung~\cite{Foot:2004wz,Rosenberg:2017qia,Chang:2018bgx} or radiative bound-state formation (BSF) processes~\cite{Pospelov:2008jd,March-Russell:2008klu,Pearce:2013ola,vonHarling:2014kha,Petraki:2015hla,Petraki:2016cnz}. These inelastic processes, occurring in the early Universe or inside halos, can have a profound impact on DM phenomenology. Less appreciated, however, is that they may also affect the computation of elastic scattering amplitudes through their contribution to the two-particle self-energy. This is the focus of the present work.

Indeed, elastic and inelastic scattering cannot be treated independently. Unitarity, via the optical theorem, imposes coupled constraints on the elastic and inelastic amplitudes in each partial wave, which encompass global upper bounds. A calculation that resums only the purely elastic interaction kernel respects the global elastic unitarity bound, but not in general the coupled constraints. A consistent treatment must also include the absorptive contributions due to inelastic processes in the self-energy kernel of the incoming two-particle state.  Equivalently, the nonrelativistic Schrödinger equation must be solved with a generally non-Hermitian effective potential, whose anti-Hermitian part describes probability flux out of the elastic channel. This feedback alters both elastic and inelastic cross sections. Its effect on elastic scattering is not purely suppressive: inelasticity may decrease elastic scattering when the latter is already substantial, owing to probability conservation, yet enhance it when elastic scattering is weak or absent via transitions of the incoming particles into intermediate states. The effect is most important when the baseline partial-wave inelastic cross sections approach their global unitarity limits~\cite{Flores:2024sfy,Flores:2025uoh}.

The unitarity limits on partial-wave cross sections scale as $\sigma_\ell^\U \propto k^{-2}\propto \vrel^{-2}$, where $k$ is the momentum in the center-of-momentum (CM) frame and $\vrel$ is the relative velocity. Away from threshold resonances, contact interactions typically remain well below these limits at low velocities. In contrast, the nonperturbative Sommerfeld enhancement from long-range forces~\cite{Sommerfeld:1931,Sakharov:1948yq} can make elastic and inelastic cross sections grow rapidly as $\vrel$ decreases and, in the Coulomb regime, scale with momentum as the unitarity bounds~\cite{vonHarling:2014kha,Baldes:2017gzw,Petraki:2015hla,Petraki:2016cnz,Flores:2024sfy}. For sufficiently large couplings, the corresponding partial-wave cross sections can then approach or saturate their unitarity limits. Moreover, Yukawa-type interactions can produce parametric resonances, known also as Sommerfeld resonances, at bound-state thresholds, near which elastic and inelastic scattering cross sections may approach the unitarity limits even for small couplings~\cite{Hisano:2003ec}.

In this paper, we focus on the feedback of annihilation on elastic scattering. Although annihilation is a contact-like interaction in the nonrelativistic effective theory, the long-range potential modifies the annihilation amplitude through the Sommerfeld effect. Near Sommerfeld resonances, the baseline calculation that neglects the feedback of annihilation on the incoming state can become highly inaccurate, leading to inelastic cross sections that approach or even exceed their unitarity bounds. In this regime, the annihilation kernel must be resummed~\cite{Blum:2016nrz}. Because annihilation is contact-like in the leading nonrelativistic approximation, resumming its contribution to the two-particle kernel requires renormalization~\cite{Flores:2025uoh,Blum:2016nrz}. We review and apply the unitarization framework of Refs.~\cite{Flores:2024sfy,Flores:2025uoh}, which relates the baseline (unregulated) annihilation amplitude to the full (regulated) inelastic and elastic partial-wave cross sections.

We illustrate these effects in a dark sector where particle-antiparticle scattering is governed by a Yukawa potential. We find that annihilation feedback is important primarily near Sommerfeld resonances, where it curtails the low-velocity growth of the baseline elastic cross section, and near zeros of the baseline elastic amplitude, where it generates the nonzero elastic contribution required by unitarity. Away from these regions, the baseline Hermitian calculation remains a good approximation. A companion paper studies the qualitatively different feedback generated by radiative BSF~\cite{Flores:2026SIDMbsf}. The numerical method, code, and data used for the baseline Yukawa problem are presented in Ref.~\cite{Flores:2026num}.

This paper is organized as follows. \cref{Sec:Unitarity} reviews the unitarity bounds and unitarization framework, while \cref{Sec:Model} introduces the model and the baseline Yukawa scattering problem. \Cref{Sec:AnnihilationUnreg} studies annihilation without feedback. \Cref{Sec:UnitarityRestoration} develops the renormalized treatment of annihilation feedback and unitarity restoration, while \cref{Sec:AnnFeedback} applies it to DM self-interactions. We conclude in \cref{Sec:Conclusions}.

\clearpage
\section{Unitarity}\label{Sec:Unitarity}
 
\subsection{Constraints on partial-wave cross sections}
\label{Sec:Unitarity_Constraints}

The unitarity of the $\mathbbm{S}$ matrix, $\mathbbm{S}\mathbbm{S}^\dagger=\ID$, implies the optical theorem, which, upon partial-wave decomposition, yields coupled constraints on the partial-wave elastic and inclusive inelastic cross sections. An early discussion of these bounds was given in Ref.~\cite{Hui:2001wy}, while recent derivations, which also account for identical-particle symmetry factors, can be found in Refs.~\cite{Flores:2024sfy,Flores:2025uoh}. The combined constraints read
\begin{equation}
\qty(
\dfrac{\sigma^\elas_\ell+\sigma^\inel_\ell}{\sigma_\ell^\U}
)^2
\leqslant
\dfrac{\sigma^\elas_\ell}{\sigma_\ell^\U},
\label{eq:UnitarityConstraint_Combined}
\end{equation}
where $\sigma^\inel_\ell$ denotes the sum over all inelastic processes, and
\begin{equation}
\sigma_\ell^\U(k)
\equiv
\sym_\ell\frac{4\pi}{k^2}(2\ell+1).
\label{eq:sigmaU}
\end{equation}
Here, $k=|\vb{k}|$ is the magnitude of the momentum of either incoming particle in the CM frame, while $\sym_\ell$ is an $\ell$-dependent symmetry factor, defined as
\begin{align}
\label{eq:SymmetryFactors}
\sym_\ell \equiv 
\begin{cases}
1, &\text{distinguishable particles},
\\
2, &\text{identical particles, surviving modes}.   
\end{cases}
\end{align}
For identical particles, partial waves forbidden by exchange symmetry are omitted throughout. The inequality~\eqref{eq:UnitarityConstraint_Combined} encompasses the global bounds
\begin{equation}
\sigma^\elas_\ell 
\leqslant 
\sigma_\ell^\U,
\qquad
\sigma^\inel_\ell
\leqslant
\sigma_\ell^\U/4.
\label{eq:UnitarityConstraint_Global}
\end{equation}
Importantly, \cref{eq:UnitarityConstraint_Combined} shows that inelasticity narrows the allowed range of the elastic cross section: increasing $\sigma_\ell^\inel$ lowers the upper bound and raises the lower bound on $\sigma_\ell^\elas$. In particular, a nonzero inelastic cross section requires a nonzero elastic one.

For convenience in the following, we define
\begin{align}
\label{eq:NormalizedCrossSections_def}
x_{\ell}
\equiv
\sigma_{\ell}^{\elas} / 
\sigma_\ell^\U
\qquad \text{and} \qquad
y_{\ell}
\equiv
\sigma_{\ell}^{\inel} / \sigma_\ell^\U,
\end{align}
which allows us to recast the unitarity constraints \eqref{eq:UnitarityConstraint_Combined} and \eqref{eq:UnitarityConstraint_Global} as follows
\begin{subequations}
\label{eq:UnitarityConstraint_xy}
\label[pluralequation]{eqs:UnitarityConstraint_xy}
\begin{gather}
\label{eq:UnitarityConstraint_Combined_xy}
\dfrac{1}{2} 
\qty(1-2y_\ell -\sqrt{1-4y_\ell} )
~\leqslant~
x_\ell
~\leqslant~
\dfrac{1}{2} 
\qty(1-2y_\ell +\sqrt{1-4y_\ell} ),
\\[1ex]
\label{eq:UnitarityConstraint_Global_xy}
x_\ell \leqslant 1, 
\quad 
y_\ell \leqslant 1/4.
\end{gather}
\end{subequations}
At the global inelastic maximum, $y_\ell=1/4$, the elastic cross section is
fixed to $x_\ell=1/4$.

\subsection{Non-Hermitian kernel}\label{Sec:Unitarity_NonHermitianKernel}

Because the optical theorem is nonlinear in the scattering amplitudes, it is not satisfied exactly by a calculation truncated at any finite order in perturbation theory.\footnote{
The perturbative expansion of the optical theorem is nevertheless satisfied order by order, provided that all contributions at the corresponding order are included.} 
Consistency with the unitarity constraints therefore requires resumming the contributions to the self-energy kernel of the incoming two-particle state. Since the optical theorem couples elastic and inelastic amplitudes, this resummation must include both the purely elastic interactions and the contributions induced by inelastic vertices~\cite{Flores:2024sfy,Flores:2025uoh}.

In the nonrelativistic regime, under the instantaneous approximation, this resummation is implemented by solving the Schrödinger equation with a generally nonlocal and non-Hermitian potential,
\begin{equation}
-\dfrac{\vb{\nabla}^2}{2\mu}\,
\psi_{\vb{k}}(\vb{r})
+
\int \dd^3\vb{r}'
\, {\cal V}(\vb{r}',\vb{r})
\, \psi_{\vb{k}}(\vb{r}')
=
{\cal E}_{\bf k} 
\, \psi_{\vb{k}}(\vb{r}),
\label{eq:SchroedingerEquation}
\end{equation}
where $\mu$ and $\mT$ are the reduced and total masses of the interacting particles, respectively, and
\begin{equation}
{\cal E}_{\bf k} \equiv \sqrt{s} -\mT 
\simeq \vb{k}^2/(2\mu)
\label{eq:EnergyEigenvalue_ScattStates}
\end{equation}
for a state of energy $\sqrt{s}$ in the CM frame; $s$ is the first Mandelstam invariant, and the last part of \cref{eq:EnergyEigenvalue_ScattStates} is at leading nonrelativistic order. ${\cal V}(\vb{r'},\vb{r})$ is the Fourier transform of the nonrelativistic limit of the two-particle-irreducible (2PI) kernel, ${\cal K}^{\twoPI}(\vb{p}',\vb{p})$, 
\begin{align}
{\cal V}(\vb{r'},\vb{r})
&=
-
\dfrac{1}{4\mT\mu}
\int
\dfrac{\dd^3 p'}{(2\pi)^3}
\dfrac{\dd^3 p}{(2\pi)^3} 
\, e^{+\im \vb{p}\cdot \vb{r}}
\, {\cal K}^{\twoPI} (\vb{p'},\vb{p})
\, e^{-\im \vb{p'}\cdot \vb{r'}}
.
\label{eq:VfromK}
\end{align}
${\cal K}^{\twoPI}(\vb{p}',\vb{p})$ consists of four-point diagrams irreducible with respect to cutting the two propagators of the interacting pair. Since the kernel is iterated,  the incoming and outgoing three-momenta in the CM frame, $\vb{p}'$ and $\vb{p}$, respectively, are in general off-shell, i.e.~they are not restricted to the physical momentum $\vb{k}$, or by the energy $\sqrt{s}$. The instantaneous approximation neglects any dependence of the kernel on $p^{\prime 0}$ and $p^0$ that parametrize the relative energies\footnote{
In practice, this can be done by evaluating them at their leading nonrelativistic on-shell values. For two particles of equal mass, this is $p^0_{\text{on-shell}} = p'^0_{\text{on-shell}} = 0$.}. 
The kernel may nevertheless depend on $\sqrt{s}$. Away from threshold crossings, this dependence can be evaluated at $\sqrt{s}\simeq\mT$ at leading nonrelativistic order; threshold dependence must instead be kept explicit~\cite{Flores:2025uoh}. Inserting ${\cal V}(\vb{r'},\vb{r})$, computed from \cref{eq:VfromK}, in the Schr\"odinger \cref{eq:SchroedingerEquation} resums the corresponding two-particle-reducible diagrams (see e.g.~\cite{Petraki:2015hla}).

The 2PI kernel can be decomposed in Hermitian and anti-Hermitian components,
\begin{subequations}
\label{eq:Kernel_Components}
\label[pluralequation]{eqs:Kernel_Components}
\begin{gather}
{\cal K}^{\twoPI} (\vb{p'},\vb{p})
=
{\cal K}^{\twoPI}_H (\vb{p'},\vb{p})
+
{\cal K}^{\twoPI}_A (\vb{p'},\vb{p}),
\label{eq:Kernel_Components_Decomposition}
\\
[{\cal K}^{\twoPI}_H (\vb{p'},\vb{p})]^\dagger
=
{\cal K}^{\twoPI}_H  (\vb{p'},\vb{p})
\quad
\text{and}
\quad
[{\cal K}^{\twoPI}_A  (\vb{p'},\vb{p})]^\dagger
=
-{\cal K}^{\twoPI}_A  (\vb{p'},\vb{p}),
\label{eq:Kernel_Components_AandH}
\end{gather}
\end{subequations}
where, in general,
\begin{equation}
\left[
{\cal K}^{\twoPI}(\vb{p}',\vb{p})
\right]^\dagger
\equiv
\left[
{\cal K}^{\twoPI}(\vb{p},\vb{p}')
\right]^*.
\label{eq:Kernel_Adjoint}
\end{equation}
Purely elastic irreducible interactions, such as the one-boson-exchange diagrams that generate the long-range potentials relevant for SIDM, contribute to ${\cal K}^{\twoPI}_H$. Open inelastic channels generate anti-Hermitian contributions through on-shell intermediate states, and may also induce Hermitian, dispersive contributions~\cite{Flores:2025uoh}. The anti-Hermitian component encodes probability flux out of the elastic channel and must be resummed when the inelastic amplitudes become sizable.

The form of the anti-Hermitian kernel, and consequently of the anti-Hermitian potential, can be deduced in three ways~\cite{Flores:2024sfy,Flores:2025uoh}: (i) from the unitarity relation underpinning the optical theorem, (ii) by integrating out the inelastic channels using the Feshbach projection formalism~\cite{Feshbach:1958nx,Feshbach:1962ut}, and (iii) from the continuity equation together with LSZ reduction. The partial-wave contributions of open inelastic channels with two particles in the final state are
\begin{align}
{\cal K}_{A,\ell}^{\twoPI} (p',p) 
\supset
\sum_{j} 
\dfrac{\im \, 2k^j(s)}{\sym^j \sqrt{s}}
\ {\cal A}_\ell^{\inel, j\,*} (p, k^j)
\ {\cal A}_\ell^{\inel, j} (p', k^j) , 
\label{eq:Kernel2PI_AA}
\end{align}
where ${\cal A}_{\ell}^{\inel, j} (p,k^j)$ is the $\ell$-wave irreducible inelastic amplitude of the $j$th inelastic channel, i.e.~the partial-wave inelastic amplitude with all initial-state 2PI factors amputated. The momenta $p=\abs{\vb{p}}$ and $p'=\abs{\vb{p}'}$ remain arbitrary, whereas the particles produced in the inelastic interaction are on shell; their momentum, $k^j$, is thus fixed by $s$. $\sym^j$ is the final-state symmetry factor. We refer to \cite{Flores:2025uoh} for the generalization of \cref{eq:Kernel2PI_AA} to inelastic processes with a different number of final-state particles.

The factorized dependence on $p$ and $p'$ in \cref{eq:Kernel2PI_AA} shows that each open two-body inelastic channel generates a nonlocal but separable contribution to the incoming two-particle self-energy kernel. This separability makes the resummation analytically tractable~\cite{Flores:2024sfy,Flores:2025uoh}, as reviewed below. It also implies that $-\im{\cal K}^{\twoPI}_{A,\ell}$ is a positive-semidefinite operator. Together with the overall minus sign in \cref{eq:VfromK}, this yields the absorptive sign required for probability flux to flow out of the elastic channel~\cite{Flores:2024sfy}.

\subsection{Unitarization}
\label{Sec:Unitarity_Regularization}

Following Refs.~\cite{Flores:2024sfy,Flores:2025uoh}, we now summarize how solving the Schrödinger \cref{eq:SchroedingerEquation} with the full non-Hermitian potential relates the resulting elastic and inelastic cross sections to the amplitudes computed in the baseline Hermitian problem. These relations make explicit how inelastic channels feed  back into elastic scattering. 

Quantities labeled ``unregulated'' are obtained by solving the Schrödinger equation with the chosen baseline Hermitian potential, which in the applications below contains the long-range interactions. Quantities labeled ``regulated” are instead obtained using the full effective potential that incorporates the feedback of inelastic channels, which is generally non-Hermitian. In addition to the absorptive (anti-Hermitian) contribution arising from on-shell inelastic states, off-shell inelastic interactions give rise to dispersive (Hermitian) contributions.  For contact-like interactions, the absorptive and dispersive contributions share the same separable structure~\cite[Sec.~3]{Flores:2025uoh}. Additional contact-like Hermitian interactions of this form, arising from other couplings in the Lagrangian, can be incorporated into the same optical term and are therefore included in the regulated results.

\subsubsection*{Phase shifts}
For finite-range potentials, the scattering solution of \cref{eq:SchroedingerEquation} has the asymptotic form
\begin{equation}
\label{eq:WaveFunction_AsymptoteInfinity}
\psi_{\vb{k}}(\vb{r})
\ \underset{r\to\infty}{\simeq} \
\psi_{\vb{k}}^{\rm in} (\vb{r})
+
f_{\vb{k}}(\vartheta)\,
\frac{e^{\im kr}}{r},
\qquad
\cos\vartheta
\equiv
\hat{\vb{k}}\cdot\hat{\vb{r}},
\end{equation}
where the incoming wave is
\begin{align}
\label{eq:WaveFunction_IncomingWave}
\psi_{\vb{k}}^{\rm in} (\vb{r}) = 
\begin{cases}
e^{\im \vb{k}\cdot \vb{r}},
&\text{distinguishable particles},
\\
\dfrac{1}{\sqrt{2}}
\qty(
e^{\im \vb{k}\cdot \vb{r}} \pm 
e^{-\im \vb{k}\cdot \vb{r}}),
&\text{identical particles}. 
\end{cases}
\end{align}
For identical particles, the sign is fixed by the requirement that the complete two-particle state obey the exchange symmetry dictated by the particle statistics. Consequently, it is determined by the symmetry of the spin state and any other internal degrees of freedom. The elastic scattering amplitude, $f_{\vb{k}}(\vartheta)$, which is understood to be properly symmetrized or antisymmetrized for identical particles, can be analyzed in terms of partial waves, 
\begin{equation}
f_{\vb{k}}(\vartheta)
=
\sum_{\ell=0}^{\infty}
(2\ell+1)\,
f_{k,\ell}\,
P_\ell(\cos\vartheta).
\label{eq:fk_PW}
\end{equation}
For the unregulated and regulated cases, the partial-wave elastic amplitudes can be written in terms of phase shifts as follows
\begin{subequations}
\label{eq:f->PhaseShifts}
\label[pluralequation]{eqs:f->PhaseShifts}
\begin{align}
1+ \dfrac{2\im k}{\sqrt{\sym_\ell}} f_{k,\ell}^{\unreg}
&= e^{2\im\theta_\ell(k)},
\label{eq:f->PhaseShift_Hermitian}
\\    
1+ \dfrac{2\im k}{\sqrt{\sym_\ell}} f_{k,\ell}^{\reg}
&= e^{2\im\Delta_\ell(k)},
\label{eq:f->PhaseShift_Full}
\end{align}
\end{subequations}
with
\begin{equation}
\Delta_\ell(k)=\theta_\ell(k)+\delta_\ell(k).
\end{equation}
$\theta_\ell \in \RR$ is the phase shift arising from the baseline Hermitian potential, while $\delta_\ell \in \CC$ is generated from the resummation of (on-shell and off-shell) inelastic interactions.

\subsubsection*{Inelastic amplitudes}
To present the general result for $\delta_\ell$, we introduce two related normalizations of the partial-wave inelastic amplitudes. For a two-body inelastic channel $j$, we write
\begin{subequations}
\label{eq:InelasticAmplitudes_defs}
\label[pluralequation]{eqs:InelasticAmplitudes_defs}
\begin{align}
{\cal A}_\ell^{\inel, j} (p,k^j(s))
&= \eta_{A,\ell}^j \,
\hat{\cal A}_\ell^{j} (p) ,
\label{eq:InelasticAmplitude_Tree_CouplingFactorization}
\\
{\cal M}_{\ell,\UNreg}^{\inel,j}(q)
&=
\eta_{A,\ell}^{j}\,
\hat{\cal M}_{\ell,\UNreg}^{j}(q),
\label{eq:InelasticAmplitude_Full_CouplingFactorization}
\\
M_{\ell,\UNreg}^{\inel,j}(q)
&=
\eta_{A,\ell}^{j}\,
\hat M_{\ell,\UNreg}^{j}(q),
\label{eq:InelasticAmplitude_MomentumRescaled_CouplingFactorization}
\\
M_{\ell,\UNreg}^{\inel,j}(q)
&\equiv
\sqrt{
\frac{4\,q\,k^j(s)}
{\sym_\ell\sym^j s}
}\,
{\cal M}_{\ell,\UNreg}^{\inel,j}(q) ,
\label{eq:InelasticAmplitude_MomentumRescaled_def}
\end{align}
\end{subequations}
where $\eta_{A,\ell}^{j}\in\RR$ is the absorptive coupling associated with channel $j$. The magnitude of the incoming relative momentum,  $q$, is in general off shell, whereas $k^j(s)$ is the on-shell momentum of the particles produced in channel $j$, fixed by the total CM energy. 

${\cal M}_{\ell,\UNreg}^{\inel,j}$ are the full inelastic amplitudes, obtained from the irreducible inelastic amplitudes, ${\cal A}_\ell^{\inel, j}$, by dressing the latter with the wavefunction, which incorporates the self-energy corrections of the incoming state,
\begin{align}
{\cal M}_{\ell,\UNreg}^{\inel, j}(q)
&\equiv
\dfrac{1}{2\pi^2}
\int_0^\infty \! \dd p \, p^2 
\, \tilde\psi_{q,\ell}^{\UNreg} (p)
\, {\cal A}_\ell^{\inel, j} (p,k^j(s)) .
\label{eq:InelasticAmplitude_M-Arelation}
\end{align}
$M_{\ell,\UNreg}^{\inel,j}$ are their momentum-rescaled counterparts.  

As we will see below, the analytic properties of $\hat{\cal M}_{\ell,\unreg}^{j}$ in the complex-momentum plane are relevant for determining the solution of the Schr\"odinger equation with the full, non-Hermitian potential. $\hat M_{\ell,\UNreg}^{j}$ will be used on-shell to express the normalized cross sections. The omission of the label `$\inel$' from $\hat{\cal M}_{\ell,\unreg}^j$ and $\hat{\cal A}_{\ell}^j$ has been intentional: these quantities will be used to parametrize both the absorptive and dispersive contributions to the 2PI kernel and, consequently, to the regulated cross sections. 

In ${\cal M}_{\ell,\UNreg}^{j}$, $M_{\ell,\UNreg}^{j}$, $\hat{\cal M}_{\ell,\UNreg}^{j}$, $\hat{M}_{\ell,\UNreg}^{j}$ and $\hat{\cal A}_\ell^{j}$, we have suppressed the dependence on the final-state momenta. Any smooth dependence on $s$ can be evaluated at $s= \mT^2$; threshold dependence must be retained explicitly and can be included in $\eta_{A,\ell}^j$.

\subsubsection*{Elastic amplitudes}

The elastic amplitudes and their momentum-rescaled counterparts are
\begin{subequations}
\label{eq:ElasticAmplitudes_f}
\label[pluralequation]{eqs:ElasticAmplitudes_f}
\begin{align}
{\cal M}_{\ell,\UNreg}^{\elas}(k)
&\equiv
\dfrac{\sqrt{\sym_\ell s}}{2} \, f_{k,\ell}^\UNreg ,
\label{eq:ElasticAmplitudes_calM_f}
\\
M_{\ell,\UNreg}^{\elas}(k)
&\equiv
\dfrac{2k}{\sym_\ell \sqrt{s}}
{\cal M}_{\ell,\UNreg}^{\elas}(k)
=
\dfrac{k f_{k,\ell}^\UNreg}{\sqrt{\sym_\ell}},
\label{eq:ElasticAmplitudes_M_f}
\end{align}
\end{subequations}
where we note for clarity that in \cref{eq:ElasticAmplitudes_calM_f}, the $\sqrt{\sym_\ell}$ symmetry factor is due to the final state, which is, however, the same as the initial state in the case of elastic scattering.

\subsubsection*{Non-Hermitian kernel}
In addition to the anti-Hermitian contributions, inelastic interactions also generate Hermitian terms in the potential. For contact-type inelastic interactions, such as those studied in \cref{Sec:AnnihilationUnreg,Sec:UnitarityRestoration}, the absorptive and dispersive contributions share the same separable form~\cite[Sec.~3]{Flores:2025uoh}. The contribution to the 2PI kernel may be parametrized as
\begin{align}
{\cal K}_{\ell}^{\twoPI} (p',p) 
\supset
\sum_{j}
\qty[\eta_{H,\ell}^j+\im (\eta_{A,\ell}^j)^2]
\ \dfrac{2k^j(s)}{\sym^j \sqrt{s}}
\ \hat{\cal A}_\ell^{j\,*} (p, k^j)
\ \hat{\cal A}_\ell^j (p', k^j) , 
\label{eq:Kernel2PI_AA_ComplexCoupling}
\end{align}
where the sign of $\eta_{H,\ell}^j\in\mathbb{R}$ is unrestricted. We note that some $\eta_{H,\ell}^j$ may receive contributions from interactions other than those responsible for the inelastic processes; we will encounter such an example in \cref{Sec:UnitarityRestoration}. We also recall that $\eta_{A,\ell}^j\in\mathbb{R}$. The feedback of the interactions generating \cref{eq:Kernel2PI_AA_ComplexCoupling} on elastic and inelastic scattering will be encoded in the amplitudes $\hat{\cal M}_{\ell,\unreg}^{j}$.

For multiple inelastic channels, we collect the couplings into the diagonal matrices
\begin{align}
\label{eq:EtaMatrices}
(\upeta_{A,\ell})^{ij}
\equiv
\eta_{A,\ell}^{i} \, \delta^{ij},
\qquad
(\upeta_{H,\ell})^{ij}
&\equiv
\eta_{H,\ell}^{i} \, \delta^{ij},
\qquad
\upeta_{\ell}^{\,2}
\equiv
\upeta_{H,\ell}
+
\im \, \upeta_{A,\ell}^{\,2} .
\end{align}
Here, $\upeta_\ell$ is the diagonal matrix with elements $\eta_\ell^j$ whose square is given in \cref{eq:EtaMatrices}, and we restrict to channels for which $\eta_\ell^j\neq0$.

\subsubsection*{Projected Green's function}

For the solution of the Schr\"odinger \cref{eq:SchroedingerEquation} with the full non-Hermitian potential, we introduce the Hermitian matrix, $\WW = \WW^\dagger$, with
\begin{align}
\WW_\ell^{ij}(k)
&\equiv
\dfrac{4}{\sym_\ell\mT^2}
\sqrt{
\frac{k^i(s)k^j(s)}
{\sym^i\sym^j}
}
\label{eq:Wmatrix_def}
\\
&\times
\Bigg[
-\dfrac{2}{\pi} \, \PV
\int_{0}^{+\infty}
\dd q \, q^2
\dfrac{
\hat{\cal M}_{\ell,\unreg}^{i}(q)
\hat{\cal M}_{\ell,\unreg}^{j*}(q)
}
{q^2-k^2}
+
2\mu
\sum_n
\dfrac{
\hat{\cal M}_{n\ell,\unreg}^{\BSD,i}
\hat{\cal M}_{n\ell,\unreg}^{\BSD,j*}
}
{\kappa_{n\ell}^2+k^2}
\Bigg].
\nonumber
\end{align}
As defined in \cref{eq:InelasticAmplitude_Full_CouplingFactorization}, $\hat{\cal M}_{\ell,\unreg}^{i}(q)$ are the unregulated inelastic amplitudes obtained from an (off-shell) initial scattering state of CM momentum $q$ and CM energy $\sqrt{s}$, with the absorptive coupling factored out. If the baseline Hermitian potential supports bound states, the amplitudes $\hat{\cal M}_{n\ell,\unreg}^{\BSD,i}$ describe the decay into channel $i$ of a bound state with binding energy ${\cal E}_{n\ell} = -\kappa_{n\ell}^2/(2\mu)$, where $\kappa_{n\ell}>0$ is the binding momentum of the bound state and $n$ stands for all discrete indices (except $\ell$) needed to enumerate them. 
The final-state momenta $k^i(s)$ and $k^j(s)$ in the prefactor of \cref{eq:Wmatrix_def} are on shell and fixed by $s$, whereas the initial-state momentum $q$ appearing in the continuum integral is off-shell. The principal-value prescription treats the poles at $q=\pm k$; their on-shell absorptive contribution is taken into account separately.

The two terms in \cref{eq:Wmatrix_def} represent the contributions from the continuum and discrete spectra, respectively. The integral contribution can be evaluated in two equivalent ways: by integrating along the positive real axis, or by analytically continuing the integrand to the complex $q$ plane via Schwarz reflection and performing a contour integration. In the latter approach, the following two features become manifestly important for the feedback of BSF and annihilation processes into elastic scattering:
\begin{itemize}
\item 
The continuum integral can receive contributions from poles and branch cuts of the unregulated inelastic amplitudes. Bound-state poles of the amplitudes may partially cancel the explicit bound-state term in \cref{eq:Wmatrix_def}. 
\item
The continuum integral can receive contributions from the arc at infinity. If $\hat{\cal M}_{\ell,\unreg}^{j}(q)\propto q^\gamma$ at large $\abs{q}$, the contribution from the arc vanishes for $\gamma<-1/2$. Amplitudes that decrease more slowly, or grow with $\abs{q}$, generally require renormalization. In particular, contact interactions with $\gamma\in\ZZ_{\geqslant0}$ require Hermitian
counterterms in addition to the absorptive interaction; these are expected to arise from the off-shell contributions of the inelastic interactions and/or other purely elastic (e.g. quartic) couplings, as we further discuss in \cref{Sec:UnitarityRestoration}. 
\end{itemize}

To express the additional phase shift due to inelastic interactions, we define
\begin{subequations}
\label{eq:w-wtilde_def}
\label[pluralequation]{eqs:w-wtilde_def}
\begin{align}
w_\ell &\equiv 
-\dfrac{\im}{2}
\hat{M}_{\ell,\unreg}^\dagger
\Big[
\upeta_\ell 
\left( \ID+\upeta_\ell\WW_\ell\upeta_\ell \right)^{-1} 
\upeta_\ell
-{\rm h.c.} 
\Big]
\hat{M}_{\ell,\unreg} ,
\label{eq:w_def}
\\
\tilde{w}_\ell &\equiv 
+\dfrac{1}{2}
\hat{M}_{\ell,\unreg}^\dagger
\Big[
\upeta_\ell 
\left( \ID+\upeta_\ell\WW_\ell\upeta_\ell \right)^{-1} 
\upeta_\ell
+{\rm h.c.} 
\Big]
\hat{M}_{\ell,\unreg} ,
\label{eq:wtilde_def}
\end{align}
\end{subequations}
where we have suppressed the dependence on the physical momentum $k$. With this,
\begin{align}
\label{eq:PhaseShift_Solution}
e^{2\im \delta_\ell} =  
\dfrac
{1-(w_\ell-\im \tilde{w}_\ell)}
{1+(w_\ell-\im \tilde{w}_\ell)} .
\end{align}

\subsubsection*{Regulated cross sections}

\begin{figure}[t!]
\centering
\includegraphics[width=0.92\linewidth]{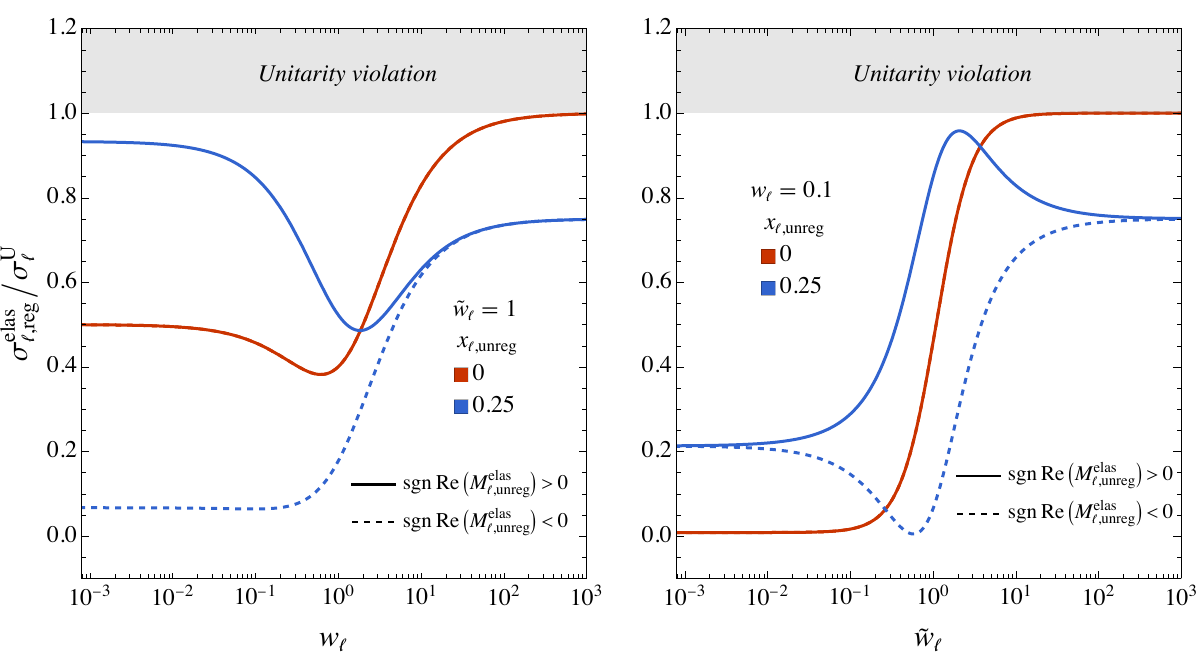}
\caption{Regulated elastic cross section, normalized to the unitarity limit, $x_{\ell,\reg}     =\sigma^{\elas}_{\ell,\reg}/\sigma_\ell^\U$, for representative fixed values of $x_{\ell,\unreg}$. The two branches correspond to the two signs of $\sin(2\theta_\ell)$ in \cref{eq:xReg_General} and coincide for $x_{\ell,\unreg}=0$. 
\emph{Left}: $w_\ell$ is varied at fixed $\tilde{w}_\ell$.  
\emph{Right}: $\tilde{w}_\ell$ is varied at fixed $w_\ell$. For $\abs{\tilde{w}_\ell}\gg\max(1,w_\ell)$, $x_{\ell,\reg}\to1-x_{\ell,\unreg}$.
}
\label{fig:RegulatedCrossSections}
\end{figure}

It is convenient to use the normalized partial-wave cross sections defined in \cref{eq:NormalizedCrossSections_def}. We introduce two versions of these quantities, corresponding to the regulated and unregulated cross sections.  In terms of the momentum-rescaled amplitudes defined in \cref{eq:InelasticAmplitudes_defs,eq:ElasticAmplitudes_f}, the unregulated normalized cross sections are
\begin{subequations}
\label{eq:UnregulatedCrossSections}
\label[pluralequation]{eqs:UnregulatedCrossSections}
\begin{align}
x_{\ell,\unreg}
&= \abs{M_{\ell,\unreg}^{\elas}}^2
 = \sin^2\theta_\ell,
\label{eq:xUnreg}
\\
y_{\ell,\unreg}
&= \sum_j \abs{M_{\ell,\unreg}^{\inel,j}(k)}^2
= 
\hat M_{\ell,\unreg}^\dagger
\, \upeta_{A,\ell}^{\,2}
\, \hat M_{\ell,\unreg}.
\label{eq:yUnreg}
\end{align}
\end{subequations}
The regulated elastic and inelastic cross sections are then
\begin{subequations}
\label{eq:RegulatedCrossSections}
\label[pluralequation]{eqs:RegulatedCrossSections}
\begin{align}
x_{\ell,\reg}
&=
\frac{
\left(
\sqrt{x_{\ell,\unreg}}
\pm
\tilde{w}_\ell
\sqrt{1-x_{\ell,\unreg}}
\right)^2
+
\left(1-x_{\ell,\unreg}\right)w_\ell^2
}{
(1+w_\ell)^2+\tilde{w}_\ell^2
},
\label{eq:xReg_General}
\\
y_{\ell,\reg}
&=
\frac{
w_\ell
}{
(1+w_\ell)^2+\tilde{w}_\ell^2
}.
\label{eq:yReg_General}
\end{align}
\end{subequations}
The upper and lower signs in \cref{eq:xReg_General} apply when $\sin(2\theta_\ell) > 0$ and $\sin(2\theta_\ell) < 0$, respectively,\footnote{
Since the phase shift depends on the momentum $k$, so does ${\rm sgn}\sin(2\theta_\ell)$.} 
and cannot be determined from $x_{\ell,\unreg}$ alone. \Cref{eqs:RegulatedCrossSections} satisfy the coupled unitarity constraint \eqref{eq:UnitarityConstraint_Combined} by construction. The behavior of \cref{eqs:RegulatedCrossSections}, shown in \cref{fig:RegulatedCrossSections}, demonstrates that the feedback from inelastic interactions may either decrease or enhance the elastic cross section.

\subsubsection*{Simple limit}

A particularly simple limit arises if $\WW_\ell \to 0$.  This occurs when the pole, branch-cut and explicit bound-state contributions to $\WW_\ell$ vanish or cancel, and the inelastic amplitudes decrease sufficiently rapidly at large $|q|$ such that the contribution from the arc at infinity vanishes. In this limit, $w_\ell \to y_{\ell,\unreg}$; neglecting also any dispersive separable contributions, $\tilde{w}_\ell \to 0$. Then, the regulated cross sections reduce to~\cite{Flores:2024sfy}
\begin{align}
\label{eq:RegulatedCrossSections_Simple}
x_{\ell,\reg} =
\dfrac{ x_{\ell,\unreg}
+\left(1-x_{\ell,\unreg}\right)
y_{\ell,\unreg}^{\,2}}
{\left(1+y_{\ell,\unreg}\right)^2}
\quad \text{and} \quad
y_{\ell,\reg} =
\dfrac{y_{\ell,\unreg}}
{\left(1+y_{\ell,\unreg}\right)^2}.
\end{align}
The form of $x_{\ell,\reg}$ above admits a simple physical interpretation: the overall suppression factor $(1+y_{\ell,\unreg})^{-2}$ reflects the loss of flux into inelastic channels, the contribution proportional to $x_{\ell,\unreg}$ represents the original elastic-scattering probability, and the contribution proportional to $(1-x_{\ell,\unreg})y_{\ell,\unreg}^2$ describes elastic scattering mediated by intermediate inelastic states~\cite{Flores:2024sfy}. By contrast, $w_\ell \neq y_{\ell,\unreg}$ and $\tilde{w}_\ell \neq 0$ in \cref{eq:xReg_General} introduce further interference.  

Equations~\eqref{eq:RegulatedCrossSections_Simple} are exact under the conditions stated above, and provide a useful approximation in certain cases. Contact-like annihilation amplitudes do not generally satisfy the large-momentum convergence condition assumed by \eqref{eq:RegulatedCrossSections_Simple}. Their resummation therefore requires the renormalized treatment discussed in \cref{Sec:UnitarityRestoration}.

\subsection{Angularly weighted cross sections for elastic scattering}
\label{Sec:Unitarity_WeightedCrossSections}

The elastic cross section counts all scattering events equally, whereas their efficiency in redistributing momentum and energy depends on the scattering angle. It is therefore useful to define the transfer and viscosity cross sections in addition to the elastic one. The transfer weight suppresses forward scattering, while the viscosity weight suppresses both forward and backward scattering, emphasizing transverse deflections,
\begin{subequations}
\label{eq:CrossSections_ETV_def}
\label[pluralequation]{eqs:CrossSections_ETV_def}
\begin{align}
 \sigma^{\elas} (k)
 &\equiv
 \int \dd\Omega
 \,\frac{\dd\sigma (k)}{\dd\Omega},
\label{eq:ElasticCrossSectionDefinition}
\\
 \sigma^\T (k)
 &\equiv
 \int \dd\Omega
 \, (1-\cos\vartheta)
 \, \frac{\dd\sigma (k)}{\dd\Omega},
\label{eq:TransferCrossSectionDefinition}
\\
 \sigma^\V (k)
 &\equiv
 \int \dd \Omega
 \, \sin^2\vartheta
 \, \frac{\dd\sigma (k)}{\dd\Omega},
\label{eq:ViscosityCrossSectionDefinition}
\end{align}
\end{subequations}
where $d\sigma/d\Omega=|f_k(\vartheta)|^2$ for both distinguishable and identical particles, since in the latter case the amplitude is understood to be symmetrized or antisymmetrized, as appropriate. Using the partial-wave decomposition presented in
\cref{App:PWanalysis}, we find
\begin{subequations}
\label{eq:CrossSections_PW}
\label[pluralequation]{eqs:CrossSections_PW}
\begin{align}
\sigma_\UNreg^{\elas}(k) &=
\sum_{\ell=0}^{\infty}
\sigma_\ell^\U(k)
\, x_{\ell,\UNreg}^{\elas}(k),
\label{eq:CrossSection_PW_Elastic}
\\
\sigma_\UNreg^\T(k)
&=
\sum_{\ell=0}^{\infty}
\sigma_\ell^\UT(k)
\, x_{\ell,\UNreg}^\T(k),
\quad
\text{for distinguishable particles,}
\label{eq:CrossSection_PW_Transfer}
\\
\sigma_\UNreg^\V(k)
&=
\sum_{\ell=0}^{\infty}
\sigma_\ell^\UV(k)
\, x_{\ell,\UNreg}^\V(k),
\label{eq:CrossSection_PW_Viscosity}
\end{align}
\end{subequations}
where $\sigma_\ell^\U(k)$ is given in \cref{eq:sigmaU}, and
\begin{align}
\sigma_\ell^\UT(k)
\equiv
\frac{4\pi}{k^2}(\ell+1)
\qquad \text{and} \qquad
\sigma_\ell^\UV(k)
\equiv
\sym_\ell
\frac{4\pi}{k^2}
\frac{(\ell+1)(\ell+2)}{2\ell+3}.
\label{eq:sigmaUs_TandV}
\end{align}
The unregulated normalized cross sections are
\begin{align}
x_{\ell,\unreg}^{\elas}
=\sin^2\theta_\ell,
\qquad
x_{\ell,\unreg}^\T 
=\sin^2(\theta_{\ell+1}-\theta_\ell),
\qquad
x_{\ell,\unreg}^\V
&=\sin^2(\theta_{\ell+2}-\theta_\ell).
\label{eq:xUnreg_ETV}
\end{align}

$\sigma_\ell^\UT$ and $\sigma_\ell^\UV$ are term-by-term upper bounds on the corresponding contributions of the transfer and viscosity cross sections. 
The regulated versions of the cross sections can be found by propagating the regularization of the elastic cross section into the transfer and viscosity ones (cf.~\cref{App:PWanalysis}). The result can be summarized as follows:
\begin{subequations}
\label{eq:xregulated_Main}
\label[pluralequation]{eqs:xregulated_Main}
\begin{align}
\label{eq:xregulated_Elastic_Main}
x_{\ell,\reg}^{\elas} 
&=
{\cal U} \big(
x_{\ell,\unreg}^{\elas};~
\{w_\ell,\tilde{w}_\ell\},\{0,0\}
\big),
\\
\label{eq:xregulated_Transfer_Main}
x_{\ell,\reg}^\T 
&=
{\cal U} \big(
x_{\ell,\unreg}^\T;~
\{w_\ell,\tilde{w}_\ell\},
\{w_{\ell+1},\tilde{w}_{\ell+1}\}
\big),
\quad \text{for distinguishable particles,}
\\
\label{eq:xregulated_Viscosity_Main}
x_{\ell,\reg}^\V 
&=
{\cal U} \big(
x_{\ell,\unreg}^\V;~
\{w_\ell,\tilde{w}_\ell\},
\{w_{\ell+2},\tilde{w}_{\ell+2}\}
\big) ,
\end{align}
\end{subequations}
where we define the function
\begin{align}
{\cal U}
\big(
X;~\{w,\tilde w\},\{w',\tilde w'\}
\big)
&\equiv
\dfrac{\left[
\sqrt{X}\,(1-ww'+\tilde w\,\tilde w')
\pm
\sqrt{1-X}\,(\tilde w-\tilde w')
\right]^2}
{\left[(1+w)^2+\tilde w^2\right]
\left[(1+w')^2+\tilde w'^2\right]}
\nonumber
\\
&+
\frac{\left[
\sqrt{X}\,(w\tilde w'+\tilde w\,w')
\pm
\sqrt{1-X}\,(w-w')
\right]^2}
{\left[(1+w)^2+\tilde w^2\right]
\left[(1+w')^2+\tilde w'^2\right]}.
\label{eq:RegularizationFunction_Main}
\end{align}
The signs in \cref{eq:RegularizationFunction_Main}, when inserted in \cref{eqs:xregulated_Main}, are chosen according to the sign of
\begin{equation}
\sin(2\theta_\ell),
\qquad
\sin\left[2(\theta_\ell-\theta_{\ell+1})\right],
\qquad
\sin\left[2(\theta_\ell-\theta_{\ell+2})\right],
\end{equation}
respectively. If one of these quantities vanishes, the corresponding sign-dependent term vanishes. The transfer cross section formulas above apply to distinguishable particles only (except for the definition). For identical particles $\sigma^\T=\sigma^{\elas}$ (cf.~\cref{eq:IdenticalParticles_Elas=Transfer}). In \cref{Sec:UnitarityRestoration}, we apply the regularization formulas \eqref{eq:xregulated_Main} to annihilation processes.

\subsection{Halo-averaged cross sections}
\label{Sec:Unitarity_AveragedCrossSections}

To study the heat transport inside halos due to DM self-scattering, the cross sections must be averaged over the DM velocity distribution in the halo. Following Ref.~\cite{Colquhoun:2020adl}, we define the scattering, momentum-transfer, and transverse-energy-transfer rates as
\begin{align}
\Gamma_{\rm scat}
&\equiv
\nDM \,
\overline{\sigma}^{\elas}\average{\vrel},
\quad
\Gamma_p 
\equiv
\dfrac{\average{\dot{p}}}{\average{p}} 
=
\nDM \,
\overline{\sigma}^{\T}\average{\vrel},
\quad
\Gamma_{E_\perp} 
\equiv
\dfrac{\langle\dot{E}_\perp\rangle}{\average{E}}
=
\nDM \,
\overline{\sigma}^{\V}\average{\vrel},
\label{eq:SIDM_HaloRates}
\end{align}
where $n_{\DM}$ and $\rho_{\DM}$ are the local DM number and mass densities, respectively, with $n_{\DM}=\rho_{\DM}/\mX$. $p$ and $E$ denote the momentum and kinetic energy of an individual DM particle in the halo frame, while $\dot{E}_\perp$ denotes the rate of transverse-energy transfer induced by collisions. The corresponding velocity-weighted cross sections are\footnote{
For the viscosity cross section, our normalization follows directly from the transverse-energy-transfer rate in Eq.~(C6) of Ref.~\cite{Colquhoun:2020adl}. This yields a factor two difference relative to Eq.~(C10) of that reference.}
%
\begin{align}
\overline{\sigma}^{\elas}
&\equiv
\dfrac
{\average{\sigma^{\elas}\vrel}}
{\average{\vrel}},
\qquad
\overline{\sigma}^{\T}
\equiv
\dfrac
{\average{\sigma^{\T}\vrel^2}}
{2\average{v_{\DM}}\average{\vrel}},
\qquad
\overline{\sigma}^{\V}
\equiv
\dfrac{\average{\sigma^{\V}\vrel^3}}
{4\average{v_{\DM}^2}\average{\vrel}},
\label{eq:SIDM_AveragedCrossSections}
\end{align}
where the brackets indicate averages over the velocity distribution, and $v_{\DM}$ denotes the speed of an individual DM particle in the halo frame. The three quantities characterize different aspects of halo transport: $\overline{\sigma}^{\elas}$ controls the total scattering rate, while $\overline{\sigma}^{\T}$ and $\overline{\sigma}^{\V}$ characterize momentum and transverse-energy transfer, respectively. Different velocity moments may be appropriate for other transport observables and regimes~\cite{Fischer:2024dte,Yang:2022hkm,Outmezguine:2022bhq}. In the phenomenological analysis below, we use $\overline{\sigma}^{\V}$. The definitions of the velocity-weighted cross sections in \cref{eq:SIDM_AveragedCrossSections} generalize those of Ref.~\cite{Colquhoun:2020adl}, formulated there for an untruncated Maxwellian distribution, to a general isotropic velocity distribution.

We evaluate the averages assuming an isotropic Maxwellian velocity distribution truncated at the escape speed $\vesc$~\cite{Chu:2019awd}, as detailed in \cref{App:VelocityAverage}. We adopt $\vesc\simeq\sqrt{2}\,v_c$, where $v_c/\sqrt{2}$ is the one-dimensional velocity-dispersion parameter of the corresponding untruncated Maxwellian. We note, however, that many analyses instead use an untruncated Maxwellian distribution, corresponding formally to $\vesc\to\infty$.
\clearpage
\section{The model}
\label{Sec:Model}

In this section, we specify the dark-sector model, derive the long-range potentials that define the baseline scattering problem, and identify the inelastic processes whose feedback modifies elastic scattering.

\subsection{The dark sector}
\label{Sec:Model_Lagrangian}

\subsubsection*{Lagrangian}

We consider a simple scalar dark sector containing a complex scalar DM field $X$ and a complex scalar mediator $\varrho$, charged under a dark Abelian symmetry, $U(1)_D$, associated with the dark gauge field $V_\mu$~\cite{Oncala:2019yvj}. The charge assignments are shown in \cref{tab:Model_Charges}. 
%
\begin{table}[h!]
\centering
\begin{tabular}{|c||c|c|}
\hline
& $U(1)_D$ & Type of field\\
\hline\hline
$X$ & $1$ & complex scalar\\
\hline
$\varrho$ & $2$ & complex scalar\\
\hline
$V_\mu$ & -- & gauge field\\
\hline
\end{tabular}
\caption{Charge assignments of the dark-sector fields.}
\label{tab:Model_Charges}
\end{table}

The scalar $\varrho$ carries twice the charge of $X$, so that the trilinear interaction $X^2\varrho^\dagger$ is allowed. Conservation of the $U(1)_D$ charge stabilizes the lightest particle with unit charge, which we take to be $X$. The model can be augmented with additional charged or uncharged scalar mediators~\cite{Flores:2026yay}, which could result in richer phenomenology, but we do not consider this possibility here. The terms relevant for the long-range dynamics and inelastic processes considered in this work are
\begin{equation}
\begin{aligned}
\mathcal{L}
=&
-\frac14 F_{\mu\nu}F^{\mu\nu}
+\frac12 \mV^2 V_\mu V^\mu
+\abs{D_\mu X}^2-\mX^2\abs{X}^2
+\abs{D_\mu\varrho}^2-\mrho^2\abs{\varrho}^2
\\
&-\frac{\y\,\mX}{2}
\left[X^2\varrho^\dagger+(X^\dagger)^2\varrho\right]
+\mathcal{L}_{\rm quartic},
\end{aligned}
\label{eq:Model_Lagrangian}
\end{equation}
where $F_{\mu\nu} \equiv \partial_\mu V_\nu- \partial_\nu V_\mu$ is the gauge field strength, and $D_\mu\equiv\partial_\mu+iqgV_\mu$, with $q$ the $U(1)_D$ charge of the field on which the covariant derivative acts. The vector mass $\mV$ should be understood as arising from a Higgs or Stueckelberg sector; for an unbroken gauge symmetry we set $\mV=0$.  In the applications below we assume
\begin{equation}
\mrho \ll \mX.
\label{eq:Model_MassHierarchy}
\end{equation}
This hierarchy allows $\varrho$ exchange to mediate a long-range force on halo scales. We define
\begin{equation}
\aV\equiv \dfrac{g^2}{4\pi}
\qquad \text{and} \qquad
\arho\equiv \dfrac{\y^2}{16\pi}.
\label{eq:Model_Couplings}
\end{equation}
$\mathcal{L}_{\rm quartic}$ encapsulates the renormalizable quartic scalar self-interactions allowed by the dark symmetry,
\begin{equation}
\mathcal{L}_{\rm quartic}
=
-\frac{\lambda_X}{4}\abs{X}^4
-\frac{\lambda_\varrho}{4}\abs{\varrho}^4
-\lambda_{X\varrho}\abs{X}^2\abs{\varrho}^2 .
\label{eq:Model_SelfInteractions}
\end{equation}
We require the scalar potential to be bounded from below,
\begin{subequations}
\label{eq:Model_QuarticCouplingConditions}
\label[pluralequation]{eqs:Model_QuarticCouplingConditions}
\begin{align}
\lambda_X>0,
\qquad
\lambda_\varrho>0,
\qquad
\lambda_{X\varrho}>
-\frac{1}{2}\sqrt{\lambda_X\lambda_\varrho},
\label{eq:Model_QuarticBoundedness}
\end{align}
and take all couplings to remain within the perturbative regime,
\begin{align}
\arho,\ \aV,\ 
\frac{\lambda_X}{16\pi},\
\frac{\lambda_\varrho}{16\pi},\
\frac{|\lambda_{X\varrho}|}{16\pi}
\ll 1.
\label{eq:Model_CouplingsPerturbative}
\end{align}
\end{subequations}
We do not assume a hierarchy between the quartic and trilinear or gauge interactions. The quartic couplings generate short-distance contributions to elastic and inelastic scattering, which will be incorporated in \cref{Sec:UnitarityRestoration} through the corresponding renormalized contact coefficients.

\subsubsection*{Long-range potentials}

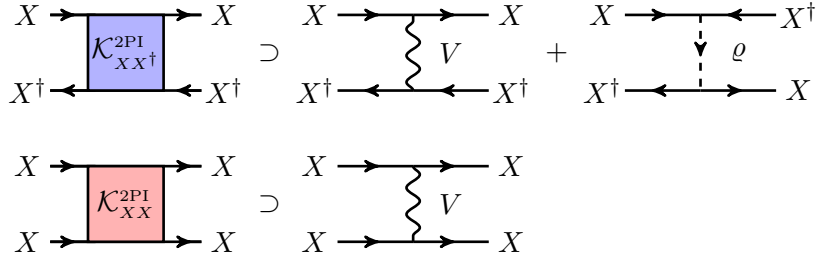
\begin{figure}[h!]
\centering
\begin{tikzpicture}[line width=1pt, scale=1]
\begin{scope}[shift={(0,1)}]
\begin{scope}[shift={(-1.9,0)}]
\node at (-1.3,1){$X$};
\node at (-1.3,0){$X^{\dagger}$};
\draw (-1,1) -- (1,1);\draw[fermion] 	(-1,1) -- (-0.4,1);\draw[fermion] 	(0.7,1) -- (1,1);
\draw (-1,0) -- (1,0);\draw[fermionbar] (-1,0) -- (-0.4,0);\draw[fermionbar](0.7,0) -- (1,0);
\node at (1.3,1){$X$};
\node at (1.3,0){$X^{\dagger}$};
\draw[fill=lightblue,shift={(0,0.5)}] (-0.5,-0.5) rectangle (0.5,0.5);
\node at (0,0.5){${\cal K}_{\scriptscriptstyle{XX^{\dagger}}}^{\twoPI}$};
\end{scope}
\node at (0,0.5){$\supset$};
\begin{scope}[shift={(1.9,0)}]
\node at (-1.3,1){$X$};
\node at (-1.3,0){$X^{\dagger}$};
\draw[fermion] 		(-1,1) -- (0,1);
\draw[fermionbar] 	(-1,0) -- (0,0);	
\draw[vector]	(0,0) -- (0,1);
\node at (0.5,0.5){$V$};
\draw[fermion] 		(0,1) -- (1,1);
\draw[fermionbar] 	(0,0) -- (1,0);	
\node at (1.3,1){$X$};
\node at (1.3,0){$X^{\dagger}$};
\end{scope}
\node at (3.8,0.5){$+$};
\begin{scope}[shift={(5.7,0)}]
\node at (-1.3,1){$X$};
\node at (-1.3,0){$X^{\dagger}$};
\draw[fermion] 		(-1,1) -- (0,1);
\draw[fermionbar] 	(-1,0) -- (0,0);	
\draw[scalar]	(0,1) -- (0,0);
\node at (0.5,0.5){$\varrho$};
\draw[fermionbar] 	(0,1) -- (1,1);
\draw[fermion] 		(0,0) -- (1,0);	
\node at (1.3,1){$X^{\dagger}$};
\node at (1.3,0){$X$};
\end{scope}
\end{scope}
%
%
\begin{scope}[shift={(0,-1)}]
\begin{scope}[shift={(-1.9,0)}]
\node at (-1.3,1){$X$};
\node at (-1.3,0){$X$};
\draw (-1,1) -- (1,1);\draw[fermion] (-1,1) -- (-0.4,1);\draw[fermion] (0.7,1) -- (1,1);
\draw (-1,0) -- (1,0);\draw[fermion] (-1,0) -- (-0.4,0);\draw[fermion] (0.7,0) -- (1,0);
\node at (1.3,1){$X$};
\node at (1.3,0){$X$};
\draw[fill=lightred,shift={(0,0.5)}] (-0.5,-0.5) rectangle (0.5,0.5);
\node at (0,0.5){${\cal K}_{\scriptscriptstyle{XX}}^{\twoPI}$};
\end{scope}
\node at (0,0.5){$\supset$};
\begin{scope}[shift={(1.9,0)}]
\node at (-1.3,1){$X$};
\node at (-1.3,0){$X$};
\draw[fermion]	(-1,1) -- (0,1);
\draw[fermion]	(-1,0) -- (0,0);	
\draw[vector]	(0,1) -- (0,0);
\node at (0.5,0.5){$V$};
\draw[fermion] 	(0,1) -- (1,1);
\draw[fermion] 	(0,0) -- (1,0);	
\node at (1.3,1){$X$};
\node at (1.3,0){$X$};
\end{scope}
\end{scope}
\end{tikzpicture}
\caption{One-boson-exchange kernels generating long-range potentials in the model.}
\label{fig:Model_2PI_Diagrams}
\end{figure}

\noindent
The exchange of $V^\mu$ and $\varrho$ bosons generates 2PI Hermitian kernels, shown in \cref{fig:Model_2PI_Diagrams}. Using standard methods and projecting \cref{eq:VfromK} onto partial waves, we find the corresponding potentials to be ${\cal V}_\ell (r',r) = V(r) \delta(r-r') / r^2$, with 
\begin{subequations}
\label{eq:Model_OBEpotentials}
\label[pluralequation]{eqs:Model_OBEpotentials}
\begin{align}
V_{XX^\dagger,\ell}(r)
&=
-\frac{\aV e^{-\mV r}}{r}
-\frac{(-1)^\ell\arho e^{-\mrho r}}{r},
\label{eq:Model_OBEpotential_XXdagger}
\\
V_{XX,\ell}(r)
=
V_{X^\dagger X^\dagger,\ell}(r)
&=
+\frac{1+(-1)^\ell}{2}\,
\frac{\aV e^{-\mV r}}{r}.
\label{eq:Model_OBEpotential_XX}
\end{align}
\end{subequations}
The $u$-channel $\varrho$ exchange contributes only to the $XX^\dagger$ kernel. Its coordinate-space potential is formally nonlocal, $\propto\delta^3(\vb r+\vb r')$, but becomes a local Yukawa potential multiplied by $(-1)^\ell$ after partial-wave projection~\cite[Appendix~A]{Oncala:2019yvj}. Vector exchange contributes to both opposite- and same-charge pairs, with opposite signs. For identical pairs, symmetrization of the $t$- and $u$-channel kernels gives the factor $[1+(-1)^\ell]/2$, projecting out the odd partial waves~\cite[Appendix~A]{Oncala:2021tkz}.

\subsubsection*{Global $U(1)_D$ limit}

For simplicity, we neglect interactions involving $V^\mu$ and work in the global-$U(1)_D$ limit. This may be realized either by taking $\aV\to0$, or by breaking the gauge symmetry such that an accidental global $U(1)_D$ remains in the $X$--$\varrho$ sector while $V^\mu$ is heavy. For $\mV$ well above the momenta relevant for halo scattering and $\mrho$, vector exchange is short ranged and can be integrated out. The long-range dynamics is then governed solely by $\varrho$ exchange,
\begin{align}
V_{XX^\dagger,\ell}(r) =
-\frac{(-1)^\ell\arho e^{-\mrho r}}{r}
\qquad \text{and} \qquad
V_{XX,\ell}(r) =
V_{X^\dagger X^\dagger,\ell}(r)
=0.
\label{eq:Model_OBEpotential_GlobalU1}
\end{align}

\subsubsection*{Particle-physics context}

The trilinear interaction $X^2\varrho^\dagger$ considered here has analogues in a variety of extensions of the Standard Model, motivated by different particle-physics and cosmological considerations. In particular, couplings of a scalar to two singlet fermions appear in extensions of the $\nu$MSM and related sterile-neutrino models, where they play a role in the generation of neutrino masses, sterile-neutrino dark-matter production, and inflation~\cite{Shaposhnikov:2006xi,Petraki:2007gq,Allison:2012qn,Bringmann:2022aim}. Such interactions also arise in Majoron models, while the same cubic scalar structure appears in radiative neutrino-mass models, notably the Zee--Babu model~\cite{Zee:1985id,Babu:1988ki}. Moreover, the trilinear interaction between singly and doubly charged complex scalars has been considered in dark-sector models motivated by lepton-number conservation and neutrino phenomenology~\cite{Berryman:2018ogk,Ma:2018pft}. Although these constructions often involve different mass hierarchies or symmetry-breaking patterns, they provide independent theoretical motivations for the interaction studied here.

\subsubsection*{Cosmological considerations}

While the positive $X$ and $\varrho$ mass-squared parameters ensure that $X=\varrho=0$ is a local minimum, additional conditions may be needed to ensure that it is either the global minimum or sufficiently long-lived on cosmological timescales. Additional cosmological and astrophysical considerations relevant to these model realizations, including the relic abundance and the fate of other dark-sector states, are discussed in our companion paper~\cite{Flores:2026SIDMbsf}.

\subsection{Yukawa scattering}
\label{Sec:Model_YukawaScattering}

Having identified the long-range interaction, we define the baseline scattering problem by retaining only the Yukawa potential generated by light-mediator exchange,
\begin{equation}
V(r) =
-\dfrac{\alpha\,e^{-m_{\med}r}}{r}.
\label{eq:Model_ScatteringPotential}
\end{equation}
For the $XX^\dagger$ potential in \cref{eq:Model_OBEpotential_GlobalU1}, $\alpha=(-1)^\ell\arho$ and $m_{\med}=\mrho$. Short-distance quartic contributions are incorporated through the renormalized contact coefficients in \cref{Sec:UnitarityRestoration}.

The normalized elastic and viscosity cross sections due to the potential \eqref{eq:Model_ScatteringPotential} are
\begin{subequations}
\label{eq:Model_UnregulatedCrossSections}
\begin{align}
x_{\ell,\unreg}
&\equiv
\sigma^{\elas}_{\ell,\unreg} / 
\sigma^\U_\ell
= \sin^2\theta_\ell,
\label{eq:Model_UnregElastic}
\\
x_{\ell,\unreg}^\V
&\equiv
\sigma^\V_{\ell,\unreg} / 
\sigma^\UV_\ell
= \sin^2(\theta_{\ell+2}-\theta_\ell).
\label{eq:Model_UnregViscosity}
\end{align}
\end{subequations}
The phase shifts $\theta_\ell\in\RR$
are obtained by solving the radial Schr\"odinger equation for the potential \eqref{eq:Model_ScatteringPotential}, using the variable-phase method~\cite{Calogero:1967var}, described in Ref.~\cite{Flores:2026num}. The result depends on the dimensionless variables
\begin{subequations}
\label{eq:Model_DimensionlessParameters}
\begin{alignat}{4}
\zeta 
&\equiv
\dfrac{\mu\alpha}{k} = 
\dfrac{\alpha}{\vrel}
&
&\qquad
\stackrel{XX^\dagger}{\longrightarrow}~~
(-1)^\ell \dfrac{\arho}{\vrel},
\label{eq:zeta_def}
\\
\xi 
&\equiv
\dfrac{\mu|\alpha|}{m_{\med}}
=\dfrac{|\alpha|}{\tau}
&
&\qquad
\stackrel{XX^\dagger}{\longrightarrow}~~
\dfrac{\mX\arho}{2\mrho}
,
\label{eq:xi_def}
\end{alignat}
where for later convenience we also define the parameter
\begin{align}
\tau\equiv\frac{m_{\med}}{\mu},
\label{eq:tau_def}
\end{align}
\end{subequations}
which does not enter the phase shifts and Sommerfeld factors but will appear in the expressions for the scattering cross sections in \cref{Sec:AnnihilationUnreg,Sec:UnitarityRestoration}.

Note that $\zeta$ may be positive, negative or zero, for attractive, repulsive and vanishing potentials, while $\xi \geqslant 0$. The largest angular momentum that contributes significantly to elastic scattering can be estimated classically as
\begin{align}
\ell_{\rm clas} 
\sim r_{\rm force} \,  k    
\sim \mu \vrel / m_{\med} 
= \xi / |\zeta| .
\label{eq:ell_clas}
\end{align}

The Yukawa problem exhibits the familiar Born, resonant, classical, and Coulomb regimes discussed extensively in the DM literature (see e.g.~\cite{Tulin:2013teo,Tulin:2017ara}). 
\begin{description}
\item[Born regime, $\bm{\xi\ll 1}$:] 
The potential is weak over its range, the phase shifts are small, and tree-level exchange gives an adequate approximation. The normalized partial-wave cross sections are well below their unitarity limits.

\item[Resonant regime, $\bm{0.84 \simeq \xi_{1,0} \lesssim \xi \lesssim |\zeta|}$:]
Here, $\ell_{\rm clas}\lesssim1$, so only a few low partial waves are significant. Attractive partial waves support bound states for sufficiently large $\xi$. Near the threshold $\xi$ values for the existence of bound levels, $\xi_{n,\ell}$, the corresponding partial-wave cross sections are resonantly enhanced. On $s$-wave resonances, $\theta_0\to\pi/2\pmod{\pi}$ as the velocity vanishes, and the elastic $s$-wave cross section approaches its unitarity limit. Higher partial waves also exhibit resonant enhancement, but their low-velocity behavior is modified by the centrifugal barrier. Repulsive partial waves do not support bound states and thus do not exhibit resonant features.

\item[Classical regime, $\bm{|\zeta|\lesssim 1 \ll \xi}$:] 
Here, $\ell_{\rm clas}\gg1$, many partial waves contribute and the scattering becomes semiclassical. Since $|\zeta|\lesssim1$, individual partial-wave phase shifts are typically small, although the cumulative elastic cross section may be sizable because many partial waves contribute.

\item[Coulomb regime, $\bm{1\ll|\zeta| \ll \xi}$:] 
Scattering is nonperturbative and resembles scattering in a Coulomb potential. Many partial waves contribute,  $\ell_{\rm clas}\gg1$. In the Coulomb approximation, the sum over partial waves diverges; however, the finite Yukawa mass screens the interaction and renders the partial-wave sums finite.
\end{description}

\begin{figure}[p!]
\centering
\includegraphics[width=0.9\linewidth]{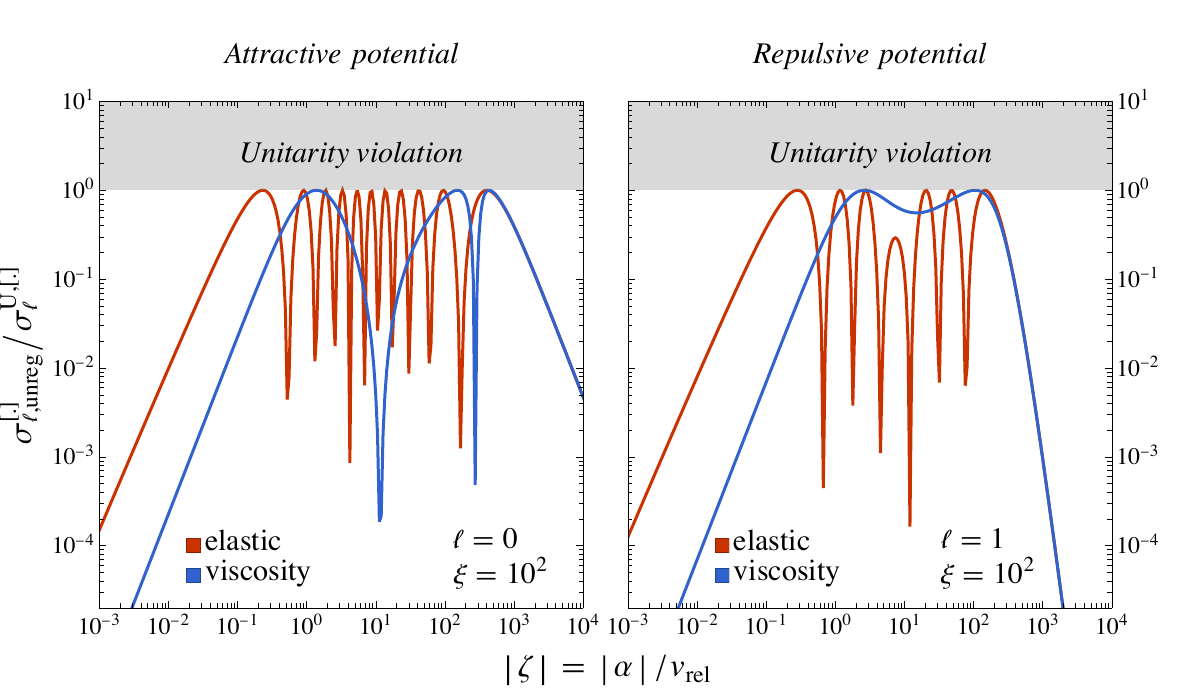}
\caption{Unregulated normalized partial-wave cross sections, $x_{\ell,\unreg}^\elas$ and $x_{\ell,\unreg}^\V$, generated by the Yukawa potential \eqref{eq:Model_ScatteringPotential}, for fixed $\ell$ and fixed off-resonance $\xi\equiv\mu|\alpha|/m_{\med}$. The elastic partial-wave cross sections exhibit oscillations in the Coulomb-like regime ($|\zeta| \lesssim \xi$), for both the attractive $\ell=0$ mode (\emph{left}) and the repulsive $\ell=1$ mode (\emph{right}). 
The velocity dependence of a resonant point is shown in \cref{fig:ElasticReg_FirstRes_vsZeta}.
}
\label{fig:YukawaCrossSection_Fixed-ell+xi}

\centering
\includegraphics[width=0.9\linewidth]{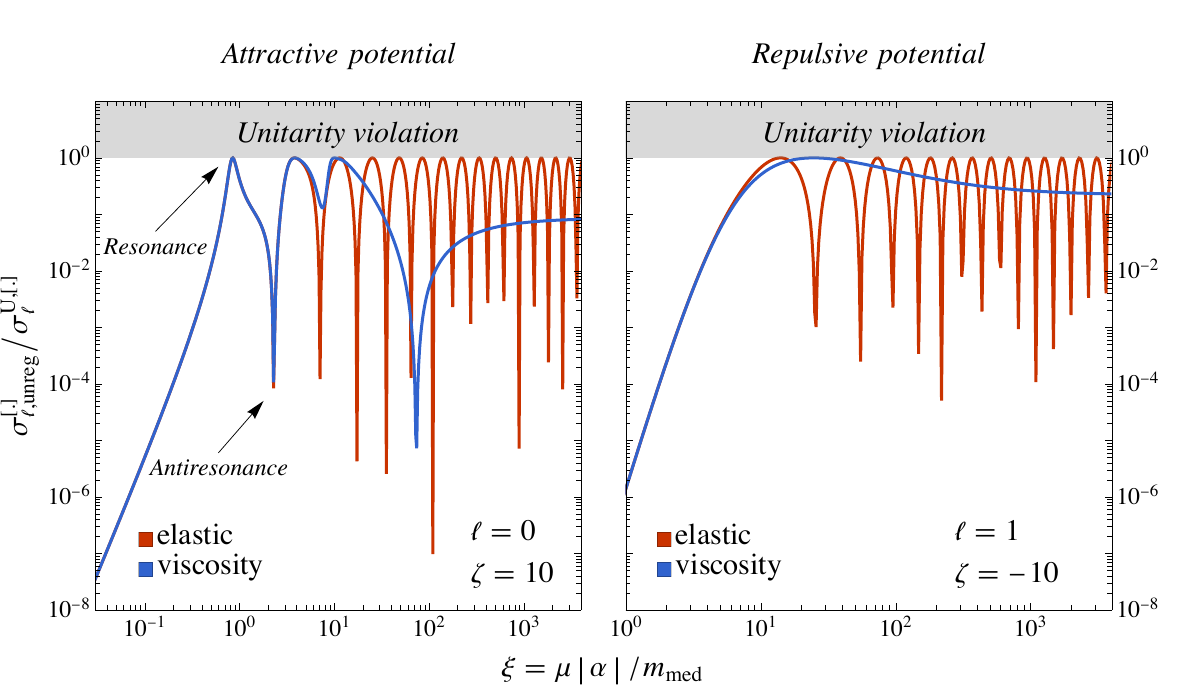}
\caption{As in \cref{fig:YukawaCrossSection_Fixed-ell+xi}, for fixed $\ell$ and $\zeta=\alpha/\vrel$. 
Small $\xi$ corresponds to the Born regime, while large $\xi$ corresponds to the Coulomb regime. The elastic partial-wave cross sections exhibit oscillations as $\xi$ is varied, even deep in the Coulomb regime, $\xi\gg (1+\ell)|\zeta|$. The oscillations are smoothed out in the viscosity cross sections, whose asymptotic values at $\xi \gg (1+\ell) |\zeta|$ are well described by the analytic Coulomb limit \eqref{eq:Coulomb_xV}. Between the Born and Coulomb regimes, attractive partial waves exhibit resonances and antiresonances.
}
\label{fig:YukawaCrossSection_Fixed-ell+zeta}
\end{figure}

The numerical results in \cref{fig:YukawaCrossSection_Fixed-ell+xi,fig:YukawaCrossSection_Fixed-ell+zeta} illustrate these expectations. At fixed $\xi$ away from threshold resonances, the normalized partial-wave cross sections become small both for $|\zeta| \ll 1+\ell$, where the interaction is perturbative, and for $|\zeta| \gg \xi$, where only the lowest partial waves remain important. In the intermediate range, nonperturbative phase shifts can drive the elastic partial-wave cross sections close to their unitarity limits. At fixed $\zeta$, increasing $\xi$ takes the system from the Born regime into the Coulomb regime. In between, attractive partial waves show threshold resonances at the discrete values of $\xi$ where bound levels appear, as well as antiresonances. Repulsive partial waves vary smoothly. The oscillations of the elastic cross sections in the Coulomb regime are smoothed out in angularly weighted cross sections, which suppress forward (and backward) scattering and depend on phase-shift differences rather than on individual phase shifts; signatures of resonances and antiresonances nevertheless persist.

\subsubsection*{Analytic Coulomb limit}

It is useful to consider the Coulomb limit analytically. The Coulomb phase shifts are
\begin{align}
\theta_\ell^C 
= \arg \Gamma(1+\ell-\im\zeta)  
= \dfrac{1}{2\im} \ln \left[ 
\dfrac{\Gamma(1+\ell-\im\zeta)}{\Gamma(1+\ell+\im\zeta)}
\right] .
\label{eq:Coulomb_PhaseShift}
\end{align}
This implies
\begin{align}
(\sigma_\ell^{\elas})^C \simeq 
\sym_\ell (2\ell+1) \times
\begin{cases}
[\psi^{(0)} (1+\ell)]^2
\, \dfrac{4\pi \alpha^2}{\mu^2 \vrel^4}, 
&|\zeta| \ll
\min\qty(1+\ell,\dfrac{1}{|\psi^{(0)} (1+\ell)|} ) ,   
\\
\dfrac{4\pi}{\mu^2 \vrel^2} 
\, \sin^2\theta_\ell^C,
&|\zeta| \gg 1+\ell ,
\end{cases}
\label{eq:Coulomb_Elastic_Limits}
\end{align}
where $\psi^{(0)}$ is the digamma function. That is, at high velocities, the partial-wave elastic cross section exhibits the Rutherford scaling: it is coupling-suppressed but grows rapidly as the velocity decreases. At sufficiently low velocities, its envelope reaches the unitarity limit, while the cross section retains a rapidly oscillatory factor, for \emph{any nonzero coupling}; the coupling only sets the velocity scale at which $\theta_\ell^C$ becomes of order unity. Clearly, the sum over partial waves diverges.

Turning to the viscosity cross section, from \cref{eq:Coulomb_PhaseShift}, we deduce 
\begin{align}
(x_\ell^\V)^C 
= 
\dfrac
{(3+2\ell)^2\zeta^2}
{[(2+\ell)(1+\ell)+\zeta^2]^2+\zeta^2},
\label{eq:Coulomb_xV}
\end{align}
which implies
\begin{align}
(\sigma_\ell^\V)^C \approx
\sym_\ell \times    
\begin{cases}
\dfrac{3+2\ell}{(2+\ell)(1+\ell)}
\ \dfrac{4\pi \alpha^2}{\mu^2 \vrel^4}, 
&|\zeta| \ll 1+\ell ,   
\\[1em]
(\ell+1)(\ell+2)(2\ell+3)
\ \dfrac{4\pi}{\mu^2 \alpha^2} ,
&|\zeta| \gg 1+\ell .
\end{cases}
\label{eq:Coulomb_Viscosity_Limits}
\end{align}
Thus each mode of the viscosity cross section grows rapidly with decreasing velocity until it saturates to a velocity-independent value proportional to $\alpha^{-2}$. The phase-shift difference entering $x_\ell^\V$ eliminates the rapid oscillations present in the elastic partial-wave cross sections. The partial-wave sum is still divergent in the strict Coulomb limit, although more mildly than for the total elastic cross section. 
For $|\zeta| \gg 1$, we find
\begin{align}
\sum_{\ell=0}^{\ell_{\rm max}}
(\sigma_\ell^{\V})^C 
&\sim 
\dfrac{8\pi}{\mu^2 \alpha^2} 
\qty[
\int_0^{\min(\ell_{\max},|\zeta|)} 
\dd\ell \, \sym_\ell \, \ell^3
+
\Theta(\ell_{\max}-|\zeta|)
\, |\zeta|^4
\int_{|\zeta|}^{\ell_{\rm max}} 
\dd\ell
\, \dfrac{\sym_\ell}{\ell}
]
\nonumber \\
&\sim
\begin{cases}
\dfrac{2\pi}{\mu^2 \alpha^2} \, \ell_{\max}^4, 
& \ell_{\rm max} \lesssim |\zeta|,
\\[1em]
\dfrac{8\pi\alpha^2}{\mu^2 \vrel^4}
\, \ln (\ell_{\rm max}/|\zeta|), 
& \ell_{\rm max} \gg |\zeta|,
\end{cases}
\label{eq:Coulomb_Viscosity_SumApprox}
\end{align}
where, for identical particles, the reduced number of contributing partial waves is offset by the enhanced weight of the surviving modes. Setting the partial-wave cutoff according to \cref{eq:ell_clas}, $\ell_{\rm max}\sim \ell_{\rm clas} = \mu\vrel/m_{\med}$, we may obtain from the above an estimate for Yukawa scattering. For more detailed estimates, we refer to \cite{Colquhoun:2020adl,Flores:2026num}.

\subsection{Inelastic processes}
\label{Sec:Model_InelasticProcesses}

The baseline Yukawa scattering described above is modified by inelastic processes
through the feedback mechanism discussed in \cref{Sec:Unitarity}. We restrict to $2\to2$ inelastic channels and identify the following two inelastic processes.
\begin{description}
\item[Annihilation.] 
$XX^\dagger$ pairs can annihilate into scalar radiation,
\begin{equation}
X+X^\dagger \to \varrho+\varrho^\dagger .
\label{eq:Process_Annihilation}
\end{equation}
The unregulated annihilation amplitudes are affected by the Sommerfeld effect due to the $XX^\dagger$ potential of \cref{eq:Model_OBEpotential_GlobalU1}. Their feedback on elastic $XX^\dagger$ scattering is discussed in \cref{Sec:UnitarityRestoration,Sec:AnnFeedback}.

\item[Bound-state formation.]
If the scalar mediator is sufficiently light, the attractive even partial waves
of the $XX^\dagger$ potential in
\cref{eq:Model_OBEpotential_GlobalU1} can support bound states. Same-charge
pairs can then form these bound states radiatively,
\begin{equation}
X+X\to \mathcal{B}(XX^\dagger)+\varrho,
\label{eq:Process_BSF}
\end{equation}
and similarly for $X^\dagger X^\dagger$. This process is kinematically allowed only if the emitted scalar can be produced from the sum of the initial kinetic energy and the binding energy of the final state. Thus BSF imposes an additional, and generally stronger, upper limit on $\mrho$ than the existence of the bound state alone.

In the global-$U(1)_D$ limit considered here, same-charge pairs have no long-range Hermitian baseline potential. Nevertheless, the BSF channel provides an absorptive contribution to the same-charge two-particle self-energy, and can therefore generate elastic $XX$ and $X^\dagger X^\dagger$ scattering after unitarization. We discuss BSF and its feedback on same-charge elastic scattering
in a companion paper~\cite{Flores:2026SIDMbsf}.
\end{description}

\clearpage
\section{Annihilation without feedback and unitarity violation}
\label{Sec:AnnihilationUnreg}

$XX^\dagger$ pairs can annihilate into scalar radiation, as shown in \eqref{eq:Process_Annihilation}. The corresponding Feynman diagram is displayed in \cref{fig:DMannihilation}. The full amplitude depends on both the hard-scattering process and the $XX^\dagger$ potential generated by iterating the 2PI kernel. In the baseline \emph{unregulated} calculation discussed in this section, the $\varrho$-exchange kernel generates the Hermitian Yukawa potential of \cref{eq:Model_OBEpotential_GlobalU1}. The \emph{regulated} calculation, discussed in \cref{Sec:UnitarityRestoration}, additionally includes the non-Hermitian kernel associated with annihilation. The contributions to the 2PI $XX^\dagger$ kernel are shown in \cref{fig:Kernel_XXdagger}.

\begin{figure}[b!]
\centering
\begin{tikzpicture}[line width=1pt, scale=1]
\begin{scope}[shift={(0,0.5)}]
\node at (-5.6,1){$X$};
\node at (-5.6,0){$X^\dagger$};
\draw[fermion] 	(-5.3,1) -- (-4.7,1);
\draw[fermionbar] 	(-5.3,0) -- (-4.7,0);
\draw[fermion]	(-3.2,1) -- (-3.1,1);
\draw[fermionbar]	(-3.2,0) -- (-3.1,0);
\draw (-4.8,1) -- (0,1);
\draw (-4.8,0) -- (0,0);
%
\draw[fill=lightblue,shift={(-4.1,0.5)}] (-0.5,-0.5) rectangle (0.5,0.5);
\draw[fill=lightblue,shift={(-2.1,0.5)}] (-0.5,-0.5) rectangle (0.5,0.5);
\node at (-4.1,0.5){${\cal K}_{\scriptscriptstyle{XX^\dagger}}^{\twoPI}$};
\node at (-2.1,0.5){${\cal K}_{\scriptscriptstyle{XX^\dagger}}^{\twoPI}$};
\node at (-3.1,0.5){$\cdots$};
\draw[fermion] 	(-0.8,1) -- (-0.5,1);
\draw[fermionbar] 	(-0.8,0) -- (-0.5,0);
\draw[fermion]	(0,0) -- (0,1);
\node at (1.8,1){$\varrho$};
\node at (1.8,0){$\varrho^{\dagger}$};

\draw[scalar] (0,1) -- (1.6,1);
\draw[scalarbar] (0,0) -- (1.6,0);

\end{scope}

\end{tikzpicture}
\caption{DM annihilation into two scalar mediators, $XX^\dagger\to \varrho\varrho^\dagger$. The hard annihilation process is dressed with the iterated $XX^\dagger$ 2PI kernel.}
\label{fig:DMannihilation}

\vspace{1em}

\begin{tikzpicture}[line width=1pt, scale=.85]


\begin{scope}[shift={(4,3)}]
\draw[fermion] (-0.7,1) -- (0,1);
\draw[fermionbar] (-0.7,0) -- (0,0);
\draw[scalar] (0,1) -- (0,0);
\node at (0.5,0.5){$\varrho$};
\draw[fermionbar] (0,1) -- (0.7,1);
\draw[fermion] (0,0) -- (0.7,0);
\node at (-1.0,1){$X$};
\node at (-1.0,0){$X^{\dagger}$};
\node at (1.0,1){$X^{\dagger}$};
\node at (1.0,0){$X$};
\node at (0,-0.75){$(a)$};
\end{scope}

\begin{scope}[shift={(9.5,3)}]
\draw[fermion] (-0.6,1) -- (0,0.5);
\draw[fermionbar] (-0.6,0) -- (0,0.5);
\draw[fermionbar] (0,0.5) -- (0.6,1);
\draw[fermion] (0,0.5) -- (0.6,0);
\node at (-0.9,1){$X$};
\node at (-0.9,0){$X^{\dagger}$};
\node at (0.9,1){$X^{\dagger}$};
\node at (0.9,0){$X$};
\node at (0,-0.75){$(b)$};
\end{scope}


\begin{scope}[shift={(0,0)}]
\draw[fermion] (-0.7,1) -- (0,1);
\draw[fermionbar] (-0.7,0) -- (0,0);
\draw[fermion] (0,0) -- (0,1);

\draw[scalar] (0,1) -- (1.3,1);
\draw[scalarbar] (0,0) -- (1.3,0);

\node at (0.4,1.3){$\varrho$};
\node at (0.9,-0.3){$\varrho^{\dagger}$};

\draw[red,dashed,line width=1pt]
    (0.65,1.6) -- (0.65,-0.6);

\draw[fermionbar] (1.3,0) -- (1.3,1);
\draw[fermion] (1.3,1) -- (2.0,1);
\draw[fermionbar] (1.3,0) -- (2.0,0);

\node at (-1.0,1){$X$};
\node at (-1.0,0){$X^{\dagger}$};
\node at (2.3,1){$X$};
\node at (2.3,0){$X^{\dagger}$};
\node at (0.65,-1.0){$(c)$};
\end{scope}

\begin{scope}[shift={(4.5,0)}]
\draw[fermion] (-0.7,1) -- (0,1);
\draw[fermionbar] (-0.7,0) -- (0,0);
\draw[fermion] (0,0) -- (0,1);

\draw[scalar]
    (0,1) .. controls (0.45,1.0) and (0.9,0.85) .. (1.3,0.5);
\draw[scalarbar]
    (0,0) .. controls (0.45,0.0) and (0.9,0.15) .. (1.3,0.5);

\draw[fermion] (1.3,0.5) -- (1.9,1);
\draw[fermionbar] (1.3,0.5) -- (1.9,0);

\draw[red,dashed,line width=1pt]
    (0.65,1.5) -- (0.65,-0.5);

\node at (0.45,1.25){$\varrho$};
\node at (0.90,-0.20){$\varrho^{\dagger}$};

\node at (-1.0,1){$X$};
\node at (-1.0,0){$X^{\dagger}$};
\node at (2.2,1){$X$};
\node at (2.2,0){$X^{\dagger}$};
\node at (0.65,-1.0){$(d)$};
\end{scope}

\begin{scope}[shift={(8.5,0)}]
\draw[fermion] (-0.6,1) -- (0,0.5);
\draw[fermionbar] (-0.6,0) -- (0,0.5);

\draw[scalar]
    (0,0.5) .. controls (0.4,0.85) and (0.85,1.0) .. (1.3,1);
\draw[scalarbar]
    (0,0.5) .. controls (0.4,0.15) and (0.85,0.0) .. (1.3,0);

\draw[fermionbar] (1.3,0) -- (1.3,1);
\draw[fermion] (1.3,1) -- (2.0,1);
\draw[fermionbar] (1.3,0) -- (2.0,0);

\draw[red,dashed,line width=1pt]
    (0.65,1.5) -- (0.65,-0.5);

\node at (0.45,1.25){$\varrho$};
\node at (0.90,-0.20){$\varrho^{\dagger}$};

\node at (-0.9,1){$X$};
\node at (-0.9,0){$X^{\dagger}$};
\node at (2.3,1){$X$};
\node at (2.3,0){$X^{\dagger}$};
\node at (0.65,-1.0){$(e)$};
\end{scope}

\begin{scope}[shift={(13,0)}]
\draw[fermion] (-0.6,1) -- (0,0.5);
\draw[fermionbar] (-0.6,0) -- (0,0.5);

\draw[scalar]
    (0,0.5) .. controls (0.4,1.0) and (0.9,1.0) .. (1.3,0.5);
\draw[scalarbar]
    (0,0.5) .. controls (0.4,0.0) and (0.9,0.0) .. (1.3,0.5);

\draw[fermion] (1.3,0.5) -- (1.9,1);
\draw[fermionbar] (1.3,0.5) -- (1.9,0);

\draw[red,dashed,line width=1pt]
    (0.65,1.4) -- (0.65,-0.4);

\node at (0.4,1.1){$\varrho$};
\node at (0.9,-0.1){$\varrho^{\dagger}$};

\node at (-0.9,1){$X$};
\node at (-0.9,0){$X^{\dagger}$};
\node at (2.2,1){$X$};
\node at (2.2,0){$X^{\dagger}$};
\node at (0.65,-1.0){$(f)$};
\end{scope}

\end{tikzpicture}
\caption{Contributions to the $XX^\dagger$ 2PI kernel. 
(a) Dispersive long-range contribution generated by $\varrho$ exchange. 
(b) Dispersive contact contribution generated by the quartic self-coupling.
(c)--(f) Optical-kernel contributions generated by annihilation into $\varrho\varrho^\dagger$. The red dashed lines denote on-shell cuts, which generate absorptive terms, whereas off-shell $\varrho\varrho^\dagger$ intermediate states generate dispersive terms.
For the unregulated cross sections in this section, only the long-range contribution (a) is retained, whereas the contributions (a)–(f) are included in the regulated scattering studied in \cref{Sec:UnitarityRestoration,Sec:AnnFeedback}.
}
\label{fig:Kernel_XXdagger}
\end{figure}

\subsection{Cross section}
\label{Sec:AnnihilationUnreg_CrossSection}

At leading order in $\vrel$ within each partial wave, the irreducible annihilation amplitude can be cast in the form of \cref{eq:InelasticAmplitude_Tree_CouplingFactorization}, with 
\begin{align}
\eta_{A,\ell} 
\simeq    
\dfrac{\arho}{2} \dfrac{\ell!^2}{(2\ell+1)!}
\dfrac
{\qty(1-\tau^2/4)^{\ell/2}}
{\qty(1-\tau^2/8)^{\ell+1}}
- \dfrac{\lambda_{X\varrho}}{16\pi} \, \delta_{\ell,0} 
\qquad \text{and} \qquad
\hat{\cal A}_\ell (p)  \simeq  (p/\mu)^\ell,
\label{eq:AnnihilationAmplitude_Tree_scaling}
\end{align}
where, as previously, $\mu = \mX/2$ is the $XX^\dagger$ reduced mass and $\tau = \mrho / \mu$. The calculation is presented in \cref{App:ShortDistanceAmplitudes}. The velocity-weighted unregulated partial-wave annihilation cross sections, including the Sommerfeld effect due to the Hermitian potential \eqref{eq:Model_OBEpotential_GlobalU1}, are
\begin{align}
\qty(\dfrac{k}{\mu}) \, \sigma_{\ell,\unreg}^{\ann} (k) 
\simeq
\eta_{A,\ell}^2
\, \dfrac{4\pi (2\ell+1)}{\mu^2}
\, \dfrac{(1-\tau^2/4)^{1/2}}{2}
\, \qty(\dfrac{k}{\mu})^{2\ell}
\, S_\ell (k)
, \quad \tau \leqslant 2,
\label{eq:AnnCS_unreg}
\end{align}
where $k=\mu \vrel$ is the momentum in the CM frame and $S_\ell(k)$ are the Sommerfeld factors due to the Yukawa potential. \Cref{eq:AnnCS_unreg} agrees with the result of Ref.~\cite{Oncala:2019yvj} for $\ell=0$ and extends it to higher partial waves. The Sommerfeld factors $S_\ell(k)$ are determined by the dimensionless parameters defined in \cref{eq:Model_DimensionlessParameters}, $\zeta = \alpha/\vrel$ with $\alpha = (-1)^\ell \arho$ as per \cref{eq:Model_OBEpotential_GlobalU1}, and $\xi = \mu |\alpha|/\mrho = \mu \arho/\mrho$. 

\subsection{Sommerfeld factors}
\label{Sec:AnnihilationUnreg_SommerfeldFactors}

In the Coulomb limit, $\xi \to \infty$, the Sommerfeld factors take the analytic form 
\begin{align}
\label{eq:SommerfeldFactors_Coulomb}
S_\ell^C  
=
\dfrac{2\pi\zeta}{1-e^{-2\pi\zeta}}
\prod_{j=1}^\ell\left(1+\dfrac{\zeta^2}{j^2}\right)
\simeq 
\begin{cases}
1, 
&|\zeta| \ll 1,
\\[1ex]
\dfrac{2\pi}{\ell!^2} \,\zeta^{2\ell+1}, 
&\zeta\gg 1+\ell,
\\[1em]
\dfrac{2\pi}{\ell!^2} 
\, |\zeta|^{2\ell+1} 
\, e^{-2\pi |\zeta|},
&\zeta\ll -(1+\ell).
\end{cases}
\end{align}
For even $\ell$ corresponding to an attractive potential, $\zeta>0$, the Sommerfeld enhancement transforms the $\vrel^{2\ell}$ suppression of higher partial waves into a coupling suppression, $\arho^{2\ell}$, in the regime $\vrel \ll \arho$, such that
\begin{align}
\label{eq:AnnCS_unreg_Coul_evenL}
\sigma_{\ell,\unreg}^{\ann}
\ \overset{\zeta \gg 1+\ell }{\simeq} \
\sigma_\ell^{\U}
\, \dfrac{\pi}{\ell!^2}
\, \eta_{A,\ell}^2
\, \arho^{2\ell+1}
,
\qquad \ell = \text{even},
\end{align}
where we also set $\tau\to0$. Thus, the partial-wave annihilation cross sections acquire the same velocity dependence as the unitarity limit \eqref{eq:sigmaU}~\cite{vonHarling:2014kha,Baldes:2017gzw}, and would exceed the global inelastic bound for 
$\eta_{A,\ell}^2 \, \arho^{2\ell+1} \gtrsim \ell!^2 / (4\pi)$. 
Neglecting $\lambda_{X\varrho}$, this corresponds to
\begin{align}
\arho
\gtrsim
\left[
\dfrac{(2\ell+1)!}
{\sqrt{\pi}\,\ell!}
\right]^{2/(2\ell+3)}
\simeq
\begin{cases}
0.68,
& \ell=0,
\\[1ex]
4\ell/e,
& \ell \geqslant 2~\text{and even},
\end{cases}
\label{eq:UnitarityViolation_Coul_CriticalCoupling}
\end{align}
where in the last step we used the Stirling approximation. The critical coupling is therefore already of order unity for the $s$ wave and increases with $\ell$. Although the low-$\ell$ critical values do not violate a naive perturbativity bound ($\arho\lesssim 4\pi$), the hierarchy required for a controlled nonrelativistic instantaneous-potential treatment is not guaranteed for $\arho\gtrsim1$. We therefore do not consider this regime further. For the odd $\ell$ modes corresponding to a repulsive potential, $\zeta<0$, the annihilation cross sections become exponentially suppressed at low velocities, $|\zeta| \gg 1+\ell$.

\begin{figure}[p!]
\centering
\includegraphics[height=0.48\linewidth]{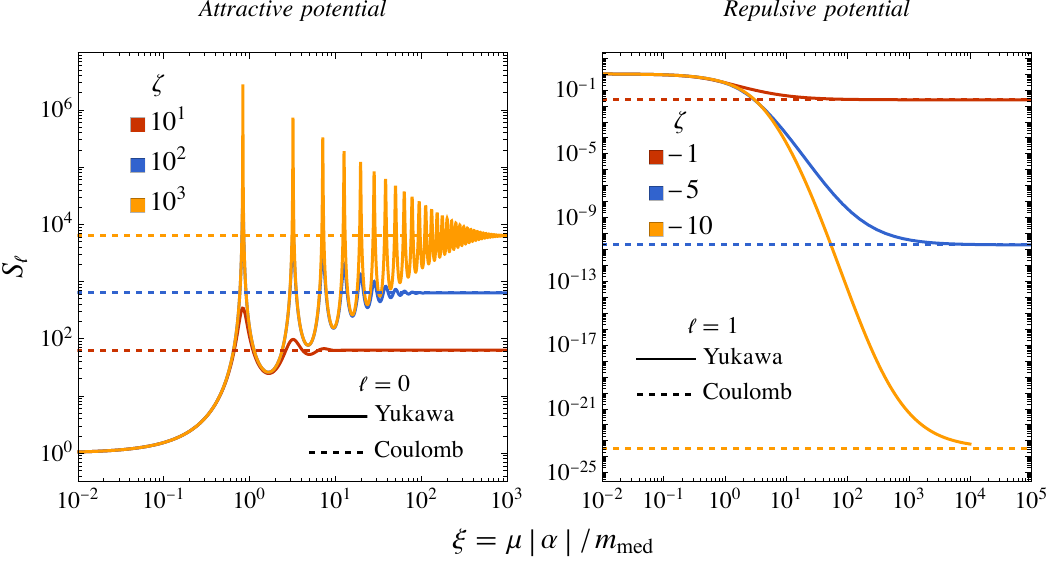}
\caption{
Sommerfeld factors as functions of $\xi \equiv \mu |\alpha|/m_{\med}$, for fixed $\ell$ and $\zeta \equiv \alpha/\vrel$. As $\xi \to \infty$, they approach the corresponding Coulomb values, $S_\ell^C$. \emph{Left}: For $\ell=0$, the potential is attractive, and resonances occur at fixed values of $\xi$ due to bound states at threshold. Right: For $\ell=1$, the potential is repulsive, and the Sommerfeld factor introduces a suppression. There are no bound states, so no resonances occur.
}
\label{fig:SommerfeldFactors_ell=0and1_zetafixed}

\vspace{2em}

\includegraphics[height=0.48\linewidth]{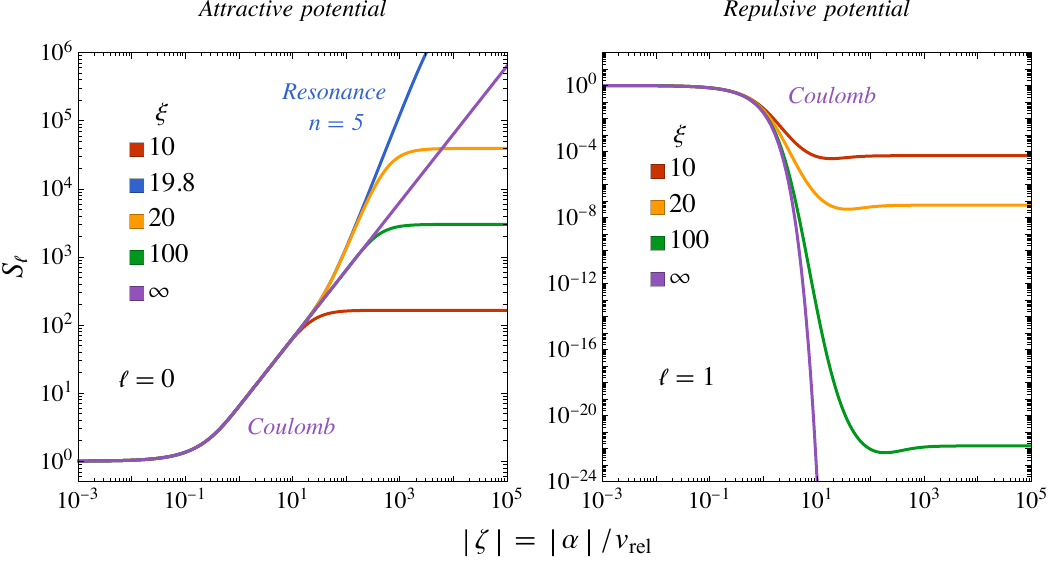}
\caption{
Sommerfeld factors as functions of $|\zeta|=|\alpha|/\vrel$, for fixed $\ell$ and $\xi=\mu|\alpha|/m_{\med}$. For attractive potentials, the Yukawa result follows the Coulomb limit for $|\zeta|\lesssim \xi/(\ell+1)$, while finite-range effects become important at larger $|\zeta|$. For repulsive potentials, the exponential Coulomb suppression is more sensitive to screening, which becomes significant at smaller $|\zeta|$.
See text for further discussion. 
}
\label{fig:SommerfeldFactors_ell=0and1_xifixed}
\end{figure}

The Schrödinger equation for the Yukawa potential has no analytic solution, so we compute the Sommerfeld factors numerically using the method presented in Ref.~\cite{Flores:2026num}. The results are shown in \cref{fig:SommerfeldFactors_ell=0and1_xifixed,fig:SommerfeldFactors_ell=0and1_zetafixed}. Attractive interactions generally enhance the annihilation cross sections, whereas repulsive interactions suppress them, although finite-range effects may produce non-resonant suppression for attractive higher partial waves. We summarize several important features.
\begin{itemize}
\item 
In the Born regime, $\xi\ll1$, the Sommerfeld factors are independent of $\zeta$ at leading order: provided that  $\xi \ll |\zeta|$, $S_\ell = 1 \pm 4\xi/(2\ell+1) + {\cal O}[\xi^2,\xi (\xi/|\zeta|)^2]$, where the upper (lower) sign corresponds to an attractive (repulsive) potential.\footnote{
This follows from the first-order perturbation in $V$ of the zero-energy ($k\to0$) solution (see, e.g., Ref.~\cite[Sections~11h--11i]{Taylor:1972}), which gives
\begin{align*}
S_\ell \simeq 1 
- \dfrac{4\mu}{2\ell+1} 
\int_0^\infty \dd r \, r \, V(r) + {\cal O} (V^2).
\end{align*}
This approximation requires that $1/k$ be much larger than the range of the potential and that the effect of the potential be perturbative. For a Yukawa potential, these conditions become $\xi \ll |\zeta|$ and $\xi \ll 1$.
}

\item
For a repulsive potential, $S_\ell$ is exponentially suppressed at $|\zeta| \gtrsim 1$, unless $\xi \lesssim 1$.

\item
For an attractive potential, resonances occur near the bound-state thresholds $\xi=\xi_{n,\ell}$; a level $\{n,\ell\}$ becomes bound for $\xi>\xi_{n,\ell}$. The associated near-threshold poles also affect elastic scattering~\cite{Kamada:2023iol}. The resonances become pronounced outside the Coulomb regime, $\zeta\gtrsim\xi_{n,\ell}/(\ell+1)$. While $\xi_{n,\ell}$ are fixed, for $\ell\geqslant1$ the finite-velocity peak positions may shift with velocity because of quasi-bound states~\cite{Beneke:2024iev,Petraki:2016cnz}.

\item
For $\xi\gtrsim1$, the long-range effect becomes important at $|\zeta| \gtrsim 1$, where $S_\ell$ departs substantially from unity. For attractive potentials, the Coulomb regime extends up to $|\zeta|\lesssim\xi/(\ell+1)$. For repulsive potentials, screening becomes important already at smaller $\zeta$. 
Outside the Coulomb regime, for non-resonant $\xi$ values, the Sommerfeld factor saturates. At sufficiently low velocities, the annihilation cross section thus recovers its perturbative scaling, $\sigma_\ell^\ann \vrel \propto \vrel^{2\ell}$, with a modified normalization.

\item
For an attractive potential outside the Coulomb regime, $\zeta \gtrsim \xi/(\ell+1)$, and for $\xi$ near a resonant value, $\xi_{n,\ell}$, the behavior depends on the partial wave.

$\ell=0$: The Sommerfeld factor grows above the Coulomb value, scaling as $S_0\propto\zeta^2$. Exactly on resonance, this growth persists down to arbitrarily low velocities. Slightly off resonance, however, $S_0$ eventually saturates, with the saturation point determined by the detuning, $\Delta\xi = \xi-\xi_{n,0}$.

$\ell\geqslant1$: For $\zeta$ moderately outside the Coulomb regime, $S_\ell\propto\zeta^4$. Exactly on resonance, this scaling persists towards arbitrarily low velocities. Slightly off resonance, $S_\ell$ eventually saturates. For $\xi<\xi_{n,\ell}$, an additional narrow peak may occur at finite velocity; such a peak does not appear for $\xi > \xi_{n,\ell}$; see Ref.~\cite{Beneke:2024iev} for details.

\item
Repulsive potentials do not support bound states and do not exhibit resonances.
\end{itemize}

\subsection{Global unitarity bound}
\label{Sec:AnnihilationUnreg_Unitarity}

\begin{figure}[t!]
\centering
\includegraphics[width=0.85\linewidth]{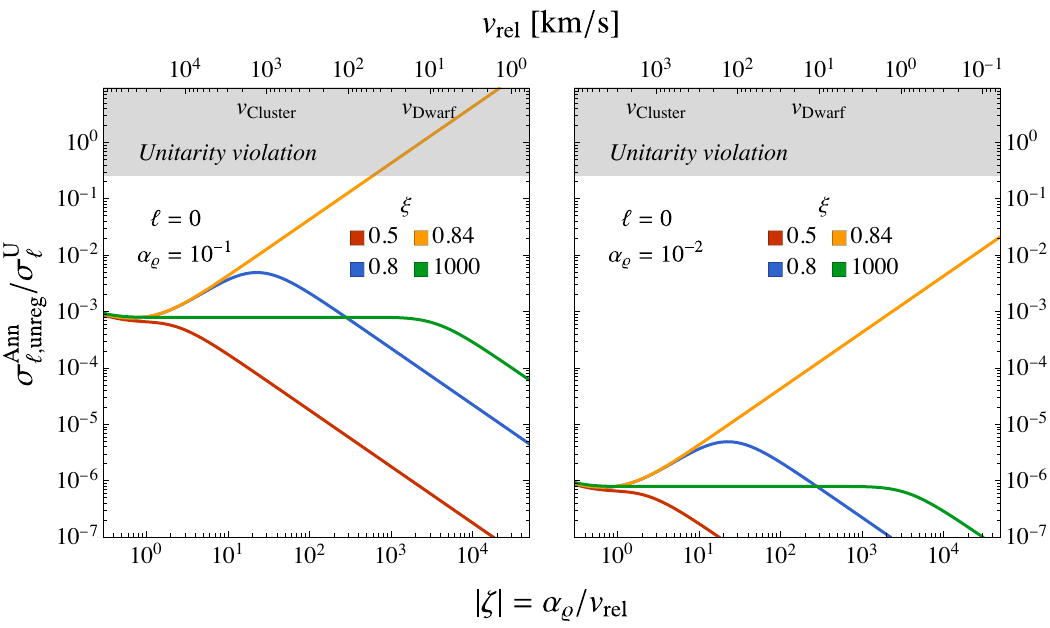}
\caption{
Unregulated $s$-wave annihilation cross sections normalized to the unitarity cross section, as a function of $\zeta$, for $\lambda_{X\varrho}=0$ and different values of $\arho$ and $\xi=\mu\arho/m_{\varrho}$. Away from the discrete resonance values of $\xi$, the cross sections remain well below the global inelastic unitarity bound, $y_{\ell,\unreg}^{\ann}\leqslant 1/4$. Near resonances, they can approach and, at sufficiently small velocities, exceed this bound. For $\xi = \xi_{n=1,\ell=0} \simeq 0.84$ and the larger coupling shown, $\arho = 10^{-1}$ (\emph{left}), the global unitarity bound is exceeded below $\vrel \sim 50~{\rm km\,s^{-1}}$, at velocities characteristic of dwarf galaxies.
For the smaller coupling, $\arho = 10^{-2}$ (\emph{right}), at the same $\xi$, the global unitarity bound is exceeded only at lower velocities.}
\label{fig:AnnCS_xi+alphaFixed}
\end{figure}

Having computed the Sommerfeld factors, we now consider the annihilation cross section of \cref{eq:AnnCS_unreg}. We focus on the $s$ wave. The normalized $s$-wave cross section, $y_{0,\unreg}^{\ann}=\sigma_{0,\unreg}^{\ann}/\sigma_0^{\rm U}$, is shown in \cref{fig:AnnCS_xi+alphaFixed}. Away from resonant $\xi$ values, the normalized unregulated annihilation cross section, $y_{0,\unreg}^{\ann}$, remains far below the global inelastic unitarity bound, unless the coupling is of order unity. As $\xi$ approaches a resonant value, $y_{0,\unreg}^{\ann}$ develops an increasingly pronounced enhancement over a finite velocity range. The closer $\xi$ lies to resonance, the larger this enhancement becomes and the farther it extends toward lower velocities. Exactly on resonance, as illustrated by $\xi=\xi_{n=1,0} \simeq 0.84$ in \cref{fig:AnnCS_xi+alphaFixed}, $y_{0,\unreg}^{\ann}$ continues to grow as $\vrel$ decreases and eventually exceeds the global inelastic unitarity bound. This can occur at velocities within the range of interest for SIDM. For $\arho=0.1$, as in the left panel of \cref{fig:AnnCS_xi+alphaFixed}, the global inelastic unitarity bound is violated below $\vrel\sim50~\mathrm{\rm km\,s^{-1}}$, at velocities characteristic of dwarf galaxies. This behavior occurs at every $s$-wave resonance, although at different velocities. Decreasing $\arho$ suppresses $y_{0,\unreg}^\ann$, shifting the same behavior to lower $\vrel$, as shown in the right panel of  \cref{fig:AnnCS_xi+alphaFixed}.

\clearpage
\section{Unitarity restoration through annihilation feedback} 
\label{Sec:UnitarityRestoration}

Having established that long-range enhancement and resonant structure can cause the unregulated partial-wave annihilation cross sections to approach or exceed their global inelastic unitarity bounds, we now show how resumming the corresponding non-Hermitian contributions to the two-particle self-energy kernel restores unitarity and modifies elastic and inelastic cross sections. Since the annihilation vertex is contact-like in the non-relativistic effective theory, its iteration is ultraviolet (UV) sensitive and requires renormalization. We first discuss the UV behavior and renormalization of the short-distance interaction, following Ref.~\cite{Flores:2025uoh}, and then formulate the benchmark prescription used in the phenomenological analysis of \cref{Sec:AnnFeedback}. (We refer to \cite{Blum:2016nrz,Parikh:2024mwa,Watanabe:2025kgw,Cimring:2026jzn} for other treatments of absorptive-interaction resummation and short-distance renormalization.)

\subsection{Ultraviolet behavior and finite nonanalytic contributions}
\label{Sec:UnitarityRestoration_RenormVsSingularities}

Extrapolating the leading contact approximation to $\hat{\cal A}_\ell$, given in \cref{eq:AnnihilationAmplitude_Tree_scaling}, to arbitrarily large off-shell momenta implies that $\hat{\cal M}_{\ell,\unreg}(q)\propto q^\ell$, since the effect of the baseline long-range potential vanishes at large $q$. Consequently, within this contact approximation, the continuum integral entering $\WW_\ell$ in \cref{eq:Wmatrix_def} is UV divergent.

For $\ell=0$, the divergent part is independent of the external momentum and can be absorbed into a single complex contact coupling, or equivalently fixed by two real renormalization conditions~\cite{Flores:2025uoh,Blum:2016nrz}. In the scalar theory considered here, the independent interactions proportional to $\lambda_X$ and $\lambda_{X\varrho}$ provide the contact terms required for this renormalization.

For $\ell\geqslant1$, extending the leading contact approximation to arbitrarily large momenta generates UV divergences multiplying a polynomial in $k^2$ of degree $\ell$. Complete renormalization within this contact description therefore requires counterterms containing up to $2\ell$ derivatives and $2(\ell+1)$ real renormalization conditions. Alternatively, the derivative expansion may be truncated at a fixed order and regarded as an effective description valid only at sufficiently low momenta, with an accuracy controlled by the omitted higher-derivative terms~\cite{Flores:2025uoh}.

This UV behavior is, however, an artifact of extrapolating the leading-order nonrelativistic contact approximation beyond its regime of validity. Indeed, for $\ell \geqslant 1$, the full irreducible amplitude of \cref{eq:AnnihilationAmplitude_Tree_PW} behaves, for fixed $s$, as
\begin{align}
\hat{\cal A}_{\ell} (p)
\propto p^{-\ell-2},
\qquad p\to\infty.
\label{eq:AnnihilationAmplitude_Tree_LargeMomenta_L>=1}
\end{align}
This UV softening renders the continuum integral in \cref{eq:Wmatrix_def} convergent. The corresponding higher-derivative absorptive and dispersive coefficients are then fixed, and generally correlated, rather than being independent fundamental parameters. Nevertheless, UV convergence does not imply that $\WW_\ell$ vanishes. Poles and branch cuts due to the full momentum dependence of the irreducible annihilation amplitude, as well as those associated with the long-range scattering wavefunction, can generate finite, momentum-dependent contributions to $\WW_\ell$ defined in \cref{eq:Wmatrix_def}.\footnote{
One may ask whether a nontrivial inelastic amplitude can make the continuum contribution to $\WW_\ell$ vanish identically, without cancellations among pole or branch-cut contributions. Considering a single channel for simplicity, define
\[
F(q) \equiv
q^2
\hat{\cal M}_{\ell,\unreg}(q)
\hat{\cal M}_{\ell,\unreg}^\star(q),
\qquad
\hat{\cal M}^\star(q)\equiv[\hat{\cal M}(q^*)]^*.
\]
The continuum integrand is proportional to $F(q)/(q^2-k^2)$. Thus, for the contour integral to receive neither singularity nor arc contributions, $F$ must be analytic in the upper half-plane, extend continuously to the real axis, and grow more slowly than linearly at large $|q|$ in the upper half-plane. Since $F$ is real on the real axis, Schwarz reflection extends it to an entire function on $\CC$, and the above asymptotic bound then implies that $F$ is constant. Thus,
\[
\hat{\cal M}_{\ell,\unreg}(q) = h(q) / q,
\qquad
h(q)h^\star(q)={\rm constant}.
\]
If the amplitude is regular at $q=0$, then $F(0)=0$, and its constancy implies that the amplitude vanishes identically. A nontrivial solution requires a $1/q$ behavior over the full momentum range. This is highly restrictive: it does not resemble the behavior of a contact-like inelastic amplitude, and it is not obvious whether it is realized by any physical amplitude. 
}

For $\ell=0$, the contribution proportional to $\arho$ is likewise UV convergent when its full momentum dependence is retained, whereas the constant annihilation amplitude proportional to $\lambda_{X\varrho}$ remains contact-like and requires
renormalization.\footnote{
For fermionic DM, no renormalizable local four-fermion interaction exists. Instead, a trilinear coupling to a scalar mediator, as in the present scenario, leads to a convergent behavior once the full momentum dependence of the irreducible $s$-wave annihilation amplitude is retained, analogous to the $\ell \geqslant 1$ case discussed above.}
For the $s$-wave analysis below, we adopt the low-energy contact description and renormalize the corresponding short-distance coefficients.

\subsection{Renormalization of contact-like interactions}
\label{Sec:UnitarityRestoration_Renorm}

Annihilation feedback becomes important primarily in the vicinity of resonances and antiresonances. We restrict our discussion to the $s$ wave, which exhibits sustained low-velocity growth near resonances and suppression near antiresonances. For higher partial waves, the centrifugal barrier suppresses the low-velocity growth of the corresponding resonant contributions and, in the present model, the associated annihilation feedback. We discuss this in more detail in \cref{App:UnitHigherWaves}.

For the low-energy $s$-wave analysis that follows, we integrate out the hard momentum dependence and nonanalytic structure of the irreducible annihilation vertex, absorbing their effects into the renormalized contact coefficients. We work at leading order with $\hat{\cal A}_0(p)=1$, so that any momentum dependence in the unregulated annihilation amplitudes arises solely from the wavefunction. Following the renormalization procedure of Ref.~\cite{Flores:2025uoh}, the regulated $s$-wave elastic and annihilation cross sections are found from \cref{eqs:RegulatedCrossSections} with
\begin{subequations}
\label{eq:RegulRenorm_w0s}
\label[pluralequation]{eqs:RegulRenorm_w0s}
\begin{align}
w_0(k;k_\diamond) &=
\dfrac
{y_{0,\unreg}^\ren(k;k_\diamond)}
{\qty[1+\beta_0(k_\diamond) z_0(k;k_\diamond)]^2
+z_0^2(k;k_\diamond)},
\label{eq:RegulRenorm_w0}
\\
\tilde{w}_0 (k;k_\diamond) &=
y_{0,\unreg}^\ren(k;k_\diamond) \ 
\dfrac
{\beta_0(k_\diamond)
+[1+\beta_0^2(k_\diamond)] \, z_0(k;k_\diamond) }
{\qty[1+\beta_0(k_\diamond) z_0(k;k_\diamond)]^2
+z_0^2(k;k_\diamond)}
.
\label{eq:RegulRenorm_wtilde0}
\end{align}
\end{subequations}
In the above, $k_\diamond$ is the renormalization momentum, or subtraction scale, and 
\begin{subequations}
\label{eq:Renorm_yzbeta}
\label[pluralequation]{eqs:Renorm_yzbeta}
\begin{align}
y_{0,\unreg}^\ren (k;k_\diamond)  
&\equiv 
[\eta_{A,0}^{\ren} (k_\diamond)]^2 
\times |\hat{M}_{0,\unreg} (k)|^2 ,  
\label{eq:Renorm_y_unreg}
\\[0.5ex]
z_0 (k;k_\diamond)  
&\equiv 
[\eta_{A,0}^{\ren} (k_\diamond)]^2 
\times 
W_{0}^{\fin} (k;k_\diamond) ,
\label{eq:Renorm_z}
\\[0.5ex]
\beta_0(k_\diamond)
&\equiv
\dfrac
{\eta_{H,0}^{\ren}(k_\diamond)}
{[\eta_{A,0}^{\ren}(k_\diamond)]^2}. 
\label{eq:Renorm_beta}
\end{align}
\end{subequations}
The superscript `$\ren$' denotes renormalized quantities. The absorptive coupling $\eta_{A,\ell}$ and rescaled amplitude $\hat{M}_{\ell,\unreg}$ have been introduced in \cref{Sec:Unitarity_Regularization}. Moreover,
\begin{align}
W_0^\fin(k;k_\diamond) \equiv W_0(k)-W_0(k_\diamond)    
\label{eq:W0fin}
\end{align}
is the finite part of the $\WW_0$ matrix, defined in \cref{eq:Wmatrix_def} and reduced here to a single number since there is only one inelastic channel. In more detail:

\paragraph{Renormalized unregulated annihilation cross section.}
From \cref{eq:AnnCS_unreg}, we obtain
\begin{align}
|\hat{M}_{0,\unreg} (k)|^2 =  
\dfrac12
\, \sqrt{1-\dfrac{\tau^2}{4}}
\, \qty(\dfrac{k}{\mu})
\, S_0 (k) .
\label{eq:yUnregRenorm}
\end{align}

\paragraph{Finite part of the projected Green's function.}
For the $s$-wave contact interaction being resummed here, it is convenient to compute $W_0^\fin$ using the position-space Green's function and express it in terms
of the irregular solution of the baseline Schr\"odinger problem
\cite{Blum:2016nrz,Flores:2025uoh}. We define the dimensionless function
$T(k;k_\diamond)$ by
\begin{align}
W_0^{\fin}(k;k_\diamond) 
\equiv 
-\dfrac
{4 k \, k^{\varrho}}
{\sym^{\varrho\varrho^\dagger} \, \mT^2} 
\, T(k;k_\diamond)     
\simeq -\dfrac{\vrel}{2}
\, \sqrt{1-\dfrac{\tau^2}{4}}
\, T(k;k_\diamond) ,   
\end{align}
where we recall that $W_0^\fin$ is defined via \cref{eq:Wmatrix_def,eq:W0fin}, and we took into account the final-state symmetry factor, $\sym^{\varrho\varrho^\dagger}=1$, and CM momentum, $k^\varrho \simeq \mX \sqrt{1-\tau^2/4}$ at leading nonrelativistic order. The relation of $T(k;k_\diamond)$ to the irregular solution can be found in \cite{Blum:2016nrz,Flores:2025uoh} and we do not repeat it here. For a Yukawa potential, $T(k;k_\diamond)$ depends on $\zeta$ and $\xi$, and the subtraction variable
$\zeta_\diamond\equiv\mu\alpha/k_\diamond$. Its numerical evaluation is discussed in detail in Ref.~\cite{Flores:2026num}.

Although we evaluate $T(k;k_\diamond)$ numerically for the Yukawa potential, we present its closed-form expressions for the Hulth\'en and Coulomb potentials here to provide analytic insight. Below, $\psi^{(0)}$ denotes the digamma function. For a Hulth\'en potential, 
$V^H(r) = -\alpha m_* \, e^{-m_* r}/(1-e^{-m_* r})$,
we let $\zeta=\mu\alpha/k$ and $\xiH\equiv 2\mu\alpha/m_*$, where we allow both $\zeta$ and $\xiH$ to carry the sign of $\alpha$. Then, 
\begin{align}
\label{eq:z0_Hulthen}
&T^H (k;k_\diamond) = 
\\
&-
2\zeta
\Re\qty[
 \psi^{(0)}\qty(\!
 \sqrt{\xiH-\qty(\dfrac{\xiH}{2\zeta})^2} 
-\dfrac{\im \, \xiH}{2\zeta}
)
\!+\!
\psi^{(0)}\qty(\!
-\sqrt{\xiH-\qty(\dfrac{\xiH}{2\zeta})^2} 
-\dfrac{\im \, \xiH}{2\zeta}
)
\!-\! 
\{k\to k_\diamond\}
],
\nonumber 
\end{align}
where the principal branch is used for the square roots. For $\alpha>0$, the bound-state thresholds occur at $\xiH_n=n^2$, $n\in\NN^*$; to relate the Yukawa and Hulth\'en potentials, we may use the piecewise-linear map
\begin{align}
\xiH =
\xiH_n
+
(\xiH_{n+1}-\xiH_n)
\dfrac{\xi-\xi_{n,0}}
{\xi_{n+1,0}-\xi_{n,0}},
\qquad
\xi_{n,0}\leqslant \xi < \xi_{n+1,0},
\qquad
\alpha >0,
\label{eq:YukawaHulthen_GlobalMapping}
\end{align}
with $\xi_{0,0}\equiv\xiH_{0}\equiv0$, such that below the first resonance, $\xiH=\xi/\xi_{1,0}$. By construction, \cref{eq:YukawaHulthen_GlobalMapping} maps every $s$-wave Yukawa resonance, $\xi=\xi_{n,0}$, exactly onto the corresponding Hulth\'en resonance, $\xiH=n^2$, is continuous across interval boundaries, and maps the detuning linearly within each interval. This alignment is important because $T(k;k_\diamond)$ is highly sensitive to the model parameters near resonance (cf.~\cref{fig:z0Diagnostics}).\footnote{
A common alternative prescription is $m_*=(\pi^2/6)m_\med$, corresponding to $|\xiH|=(12/\pi^2)\xi$. It is numerically close to \cref{eq:YukawaHulthen_GlobalMapping} for attractive potentials, particularly in the first resonance interval, but does not reproduce the Yukawa resonance positions exactly.}
Even after matching the resonances, however, the Hulth\'en potential does not approximate $T(k;k_\diamond)$ satisfactorily in the resonant region relevant here. We therefore use the numerical Yukawa result throughout~\cite{Flores:2026num}.
For a Coulomb potential, $V^C(r)=-\alpha/r$,
\begin{align}
\label{eq:z0_Coulomb}
T^{C}(k;k_\diamond)
=
-2\zeta
\left[
\Re\psi^{(0)}(-\im\zeta)
-\ln|\zeta|
-\{k\to k_\diamond\}
\right]
.
\end{align}
This also follows from \cref{eq:z0_Hulthen} in the limit $|\xiH|\to\infty$, using $\psi^{(0)}(z)=\ln z+{\cal O}(z^{-1})$. For $|\zeta|\gg1$, \cref{eq:z0_Coulomb} becomes
$T^{C}(k;k_\diamond)\simeq
2\zeta\qty[
\Re\psi^{(0)}(-\im\zeta_\diamond)
-\ln|\zeta_\diamond|
]$.

\paragraph{Dispersive coefficient.} The renormalized $s$-wave contact interaction is specified by two independent real short-distance parameters, $[\eta_{A,0}^{\ren}]^2$ and $\eta_{H,0}^{\ren}$, which must be fixed by two independent renormalization conditions. An elastic and an inelastic observable provide a natural choice. In general, however, each observable depends on both coefficients, so the two renormalization conditions must be imposed jointly.

\paragraph{Running of the couplings.} 
The physical cross sections must be independent of the choice of renormalization momentum. This implies
\begin{align}
\label{eq:RGeqs_sol}
\dfrac{1}
{\eta^\ren_{H,0} (k_\diamond)
+\im [\eta^\ren_{A,0} (k_\diamond)]^2} 
= 
\dfrac{1}
{\eta^\ren_{H,0} (\tilde{k}_\diamond)
+\im [\eta^\ren_{A,0} (\tilde{k}_\diamond)]^2} 
+ W_{0}^{\fin} (k;\tilde{k}_{\diamond}) 
- W_{0}^{\fin} (k;k_\diamond) ,
\end{align}
where we note that the difference on the right-hand side is independent of $k$. From this, we deduce the running of the dispersive coefficient and the absorptive coupling, 
\begin{subequations}
\label{eq:Running}
\label[pluralequation]{eqs:Running}
\begin{gather}
\beta_0(k_\diamond) =
\beta_0(\tilde{k}_\diamond)
+
[\eta_{A,0}^{\ren}(\tilde{k}_\diamond)]^2
\qty[1+\beta_0^2(\tilde{k}_\diamond)]
\qty[
 W_0^{\fin}(k;\tilde{k}_\diamond)
-W_0^{\fin}(k;k_\diamond)],
\label{eq:Running_Beta0}
\\[0.5ex]
[\eta_{A,0}^{\ren}(k_\diamond)]^2 
\qty[1+\beta_0^2(k_\diamond)]
=
[\eta_{A,0}^{\ren}(\tilde{k}_\diamond)]^2 
\qty[1+\beta_0^2(\tilde{k}_\diamond)]
.
\label{eq:Running_AbsorptiveCoupling}
\end{gather}
\end{subequations}
\Cref{eq:Running_Beta0} shows that even if the dispersive coefficient vanishes at one subtraction scale, it is generally nonzero at another.

\subsection{Benchmark prescription}
\label{Sec:UnitarityRestoration_Benchmarks}

We now apply the renormalization prescription of \cref{Sec:UnitarityRestoration_Renorm} to the $s$-wave Sommerfeld resonances and antiresonances. The baseline phase shifts, Sommerfeld factors and $T$ functions are computed numerically for the Yukawa potential~\cite{Flores:2026num}.  

\paragraph{Subtraction scale.} 
We choose the renormalization scale
\begin{align}
k_\diamond = \mu\arho,
\qquad
\zetaRenorm \equiv \dfrac{\mu\arho}{k_\diamond} = 1.
\label{eq:RenormScale}
\end{align}
This scale lies at the onset of the nonperturbative long-range regime and remains within the nonrelativistic domain provided that $\arho\ll1$, since the corresponding relative velocity is $\vrelren=\arho$.

\paragraph{Short-distance benchmark.}
For $\ell=0$, the absorptive coupling receives contributions from both the trilinear and biquadratic annihilation interactions, governed by $\alpha_\varrho$ and $\lambda_{X\varrho}$, respectively. The same annihilation interactions also generate dispersive contributions through off-shell
$\varrho\varrho^\dagger$ intermediate states, while the independent quartic coupling $\lambda_X$ provides an additional Hermitian contact interaction and the counterterm required to renormalize the dispersive coefficient. Rather than scanning over the two independent renormalized contact couplings, we adopt a simple benchmark at the subtraction scale.
For the absorptive coupling, we set the finite contribution proportional to $\lambda_{X\varrho}$ to zero and retain only the trilinear contribution,
\begin{subequations}
\label{eq:Benchmark}
\label[pluralequation]{eqs:Benchmark}
\begin{align}
\eta_{A,0}^{\ren}(k_\diamond)
\to
\dfrac{\alpha_\varrho}{2(1-\tau^2/8)}.
\label{eq:Benchmark_Absorptive}
\end{align}
For the dispersive interaction, we choose
\begin{align}
\beta_0(k_\diamond)=0.
\label{eq:Benchmark_Dispersive}
\end{align}
\end{subequations}
This condition does not imply a purely absorptive correction at all momenta, since $z_0(k;k_\diamond)$ generally generates a nonzero dispersive contribution away from $k=k_\diamond$. Equivalently, shifting the subtraction scale generates a nonzero $\beta_0$, as follows from \cref{eq:Running_Beta0}. We briefly discuss the regularization of the elastic and inelastic cross sections beyond this benchmark in \cref{Sec:AnnFeedback_BeyondBenchmark}.

\paragraph{Viscosity cross section.}
The viscosity cross section couples partial waves separated by two units.  In particular, its lowest-mode contribution involves the regulated $s$- and $d$-wave amplitudes, as seen from \cref{eqs:xregulated_Main}. However, in the low-velocity region where annihilation feedback near an isolated $s$-wave resonance is important, the centrifugal barrier suppresses the $d$-wave annihilation amplitude, which is also generally off resonance. We therefore neglect annihilation feedback in the $d$ wave while retaining its baseline phase shift. The regulated viscosity $\ell=0$ mode is obtained from \cref{eq:xregulated_Viscosity_Main}, using \cref{eqs:RegulRenorm_w0s} for the $s$ wave and $w_2=\tilde{w}_2=0$. This approximation requires the $d$ wave to be sufficiently far from resonance, and hence $y_{2,\unreg}\ll1$, as is satisfied in the parameter region considered below.

\paragraph{Total elastic and viscosity cross sections.}
Within this $s$-wave feedback approximation, the total regulated elastic and viscosity cross sections are
\begin{subequations}
\label{eqs:Ann_TotalCrossSections_SwaveFeedback}
\begin{align}
\sigma_{\reg}^{\elas}
&\simeq
\sigma_0^{\U} \, x_{0,\reg}^\elas
+
\sum_{\ell \geqslant 1}
\sigma_\ell^{\U} \, x_{\ell,\unreg}^\elas,
\label{eq:Ann_TotalElastic_SwaveFeedback}
\\
\sigma_{\reg}^{\V}
&\simeq
\sigma_0^{\U,\V}x_{0,\reg}^{\V}
+
\sum_{\ell\geqslant 1}
\sigma_\ell^{\U,\V}x_{\ell,\unreg}^{\V}.
\label{eq:TotalViscosity_SwaveFeedback}
\end{align}
\end{subequations}
Thus, all higher partial-wave phase shifts are retained at their baseline values, while annihilation feedback is included only in the resonant $s$-wave amplitude. The summation over $\ell$ is truncated when a $1\%$ precision is reached, as described in \cite{Flores:2026num}.

\paragraph{Regularization formulas for the short-distance benchmark.}
For easy reference, the $\ell=0$ regularization formulas for the benchmark $\beta_0(k_\diamond)=0$ of \cref{eqs:Benchmark}, adopted throughout, are
\begin{subequations}
\label{eq:Regulated_s-wave_Elas+Visc+Inel}
\label[pluralequation]{eqs:Regulated_s-wave_Elas+Visc+Inel}
\begin{align}
x_{0,\reg}^{\elas}
&=
\dfrac
{x_{0,\unreg}^{\elas} 
+
\qty(
z_0\sqrt{x_{0,\unreg}^{\elas}}
\pm
y_{0,\unreg}^{\ren}
\sqrt{1-x_{0,\unreg}^{\elas}}
)^2}
{(1+y_{0,\unreg}^{\ren})^2+z_0^2},
\label{eq:x0_Elas_Reg}
\\[1ex]
x_{0,\reg}^{\V}
&=
\dfrac{
x_{0,\unreg}^{\V}
+
\qty(
z_0\sqrt{x_{0,\unreg}^{\V}}
\pm
y_{0,\unreg}^{\ren}
\sqrt{1-x_{0,\unreg}^{\V}}
)^2}
{(1+y_{0,\unreg}^{\ren})^2+z_0^2},
\label{eq:x0_Visc_Reg}
\\[1ex]
y_{0,\reg}
&=
\dfrac
{y_{0,\unreg}^{\ren}}
{(1+y_{0,\unreg}^{\ren})^2+z_0^2} ,
\label{eq:y_Reg}
\end{align}
\end{subequations}
where we recall that the upper and lower signs in \cref{eq:x0_Elas_Reg} apply for $\sin(2\theta_0)>0$ and $\sin(2\theta_0)<0$, respectively, whereas those in \cref{eq:x0_Visc_Reg} apply for $\sin[2(\theta_0-\theta_2)]>0$ and $\sin[2(\theta_0-\theta_2)]<0$, respectively. We recall that the phase shifts, and hence these sign conditions, depend on the momentum.  Moreover, we define
\begin{align}
r_0^a
\equiv
\sqrt{
\dfrac
{x_{0,\unreg}^a}
{1-x_{0,\unreg}^a}
},
\qquad
a=\elas,{\rm V},
\label{eq:xreg_CrossoverRatio}
\end{align}
where $r_0^a \in [0,\infty)$ is a monotonically increasing function of $x_{0,\unreg}^a$. With this, the regulated elastic and viscosity cross sections of \cref{eq:x0_Elas_Reg,eq:x0_Visc_Reg} can equivalently be expressed as
\begin{align}
x_{0,\reg}^a
=
x_{0,\unreg}^a
\times
\dfrac{1+
\qty(z_0 \pm y_{0,\unreg}^{\ren}/r_0^a )^2
}
{(1+y_{0,\unreg}^{\ren})^2+z_0^2},
\qquad
a=\elas,{\rm V}.
\label{eq:x0Reg_r0}
\end{align}

\begin{figure}[t!]
\centering
\includegraphics[width=0.95\linewidth]{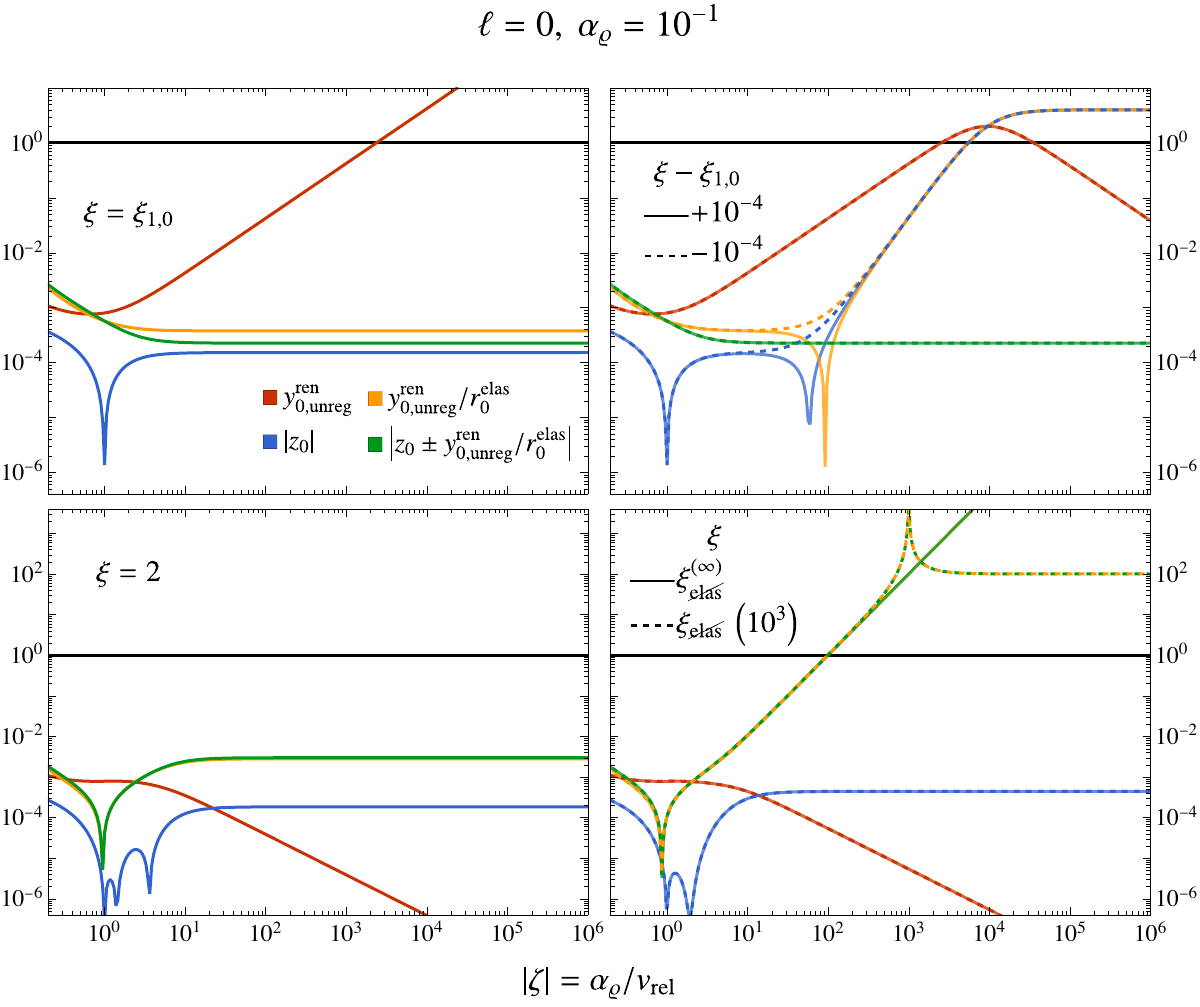}
\caption{
Comparison of the quantities entering \cref{eq:x0Reg_r0} for the $\ell=0$ regulated elastic cross section:
exactly on the $n=1$, $\ell=0$ resonance, $\xi = \xi_{1,0} \simeq 0.84$ (\emph{top left}); 
at very small detuning from this resonance, $\xi-\xi_{1,0} = \pm 10^{-4}$ (\emph{top right});
at a non-resonant point, $\xi = 2$ (\emph{bottom left});
and for two values of $\xi$ corresponding to elastic-scattering zeros: the zero-velocity antiresonance,
$\xi_{\cancel{\rm elas}~}^{(\infty)}\simeq 2.23$, and the zero at $\zeta=10^3$ (\emph{bottom right}).
}
\label{fig:z0Diagnostics}
\end{figure}

\paragraph{Interpretation of regularization formulas.} 
The terms $(1+y_{0,\unreg}^{\ren})^2$ and $z_0^2$ in the denominators of \cref{eq:Regulated_s-wave_Elas+Visc+Inel,eq:x0Reg_r0} encode the absorptive and dispersive effects, respectively; the former is associated with flux loss, while the latter modifies the elastic phase. For the benchmark $\beta_0(k_\diamond)=0$ adopted here, both are generated by the absorptive coupling. In the regulated elastic and viscosity cross sections, \cref{eq:x0_Elas_Reg,eq:x0_Visc_Reg,eq:x0Reg_r0}, the numerator contains the baseline contribution, $x_{0,\unreg}$, together with the combined dispersive and absorptive enhancement of the elastic probability, $\qty(z_0\sqrt{x_{0,\unreg}}\pm y_{0,\unreg}^{\ren}\sqrt{1-x_{0,\unreg}})^2$, whose cross term describes their interference. In \cref{fig:z0Diagnostics}, we compare the quantities entering \cref{eq:x0Reg_r0} in four representative cases: exactly on the $\{n=1,\ell=0\}$ resonance, at very small detuning from this resonance, at a generic non-resonant point, and at an elastic-scattering zero. This comparison will help identify which terms control the regularization in each regime and will guide the discussion of annihilation feedback in the following two sections.

\clearpage
\section{Annihilation feedback in dark-matter self-interactions}
\label{Sec:AnnFeedback}

\subsection{Near $s$-wave resonances}
\label{Sec:AnnFeedback_Resonances}

\begin{figure}[b!]
\centering
\includegraphics[width=0.95\linewidth]{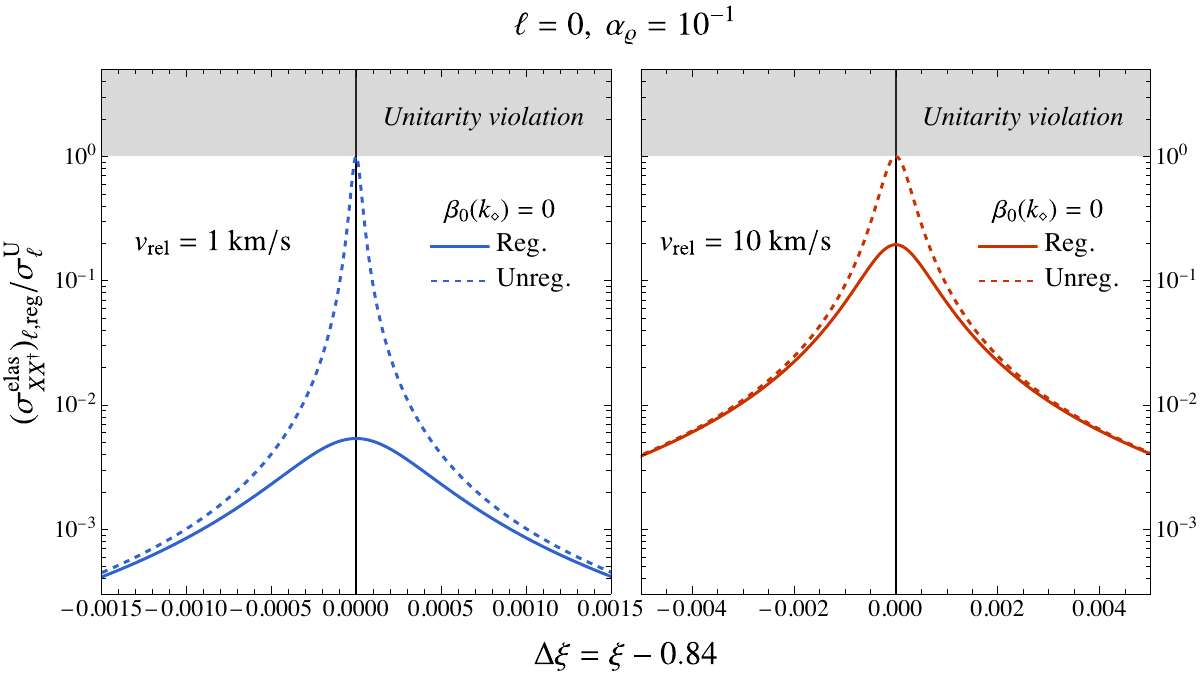}
\caption{
$s$-wave elastic cross section normalized to the unitarity cross section, as a function of the detuning from the first resonance, $\Delta\xi = \xi - \xi_{1,0}$, where $\xi=\mu\arho/m_{\varrho}$, for $\arho=10^{-1}$. We compare the regulated and unregulated results for $\vrel=1~{\rm km\,s^{-1}}$ (\emph{left}) and $10~{\rm km\,s^{-1}}$ (\emph{right}). At resonance, the regulated cross section can be suppressed by orders of magnitude, while rapidly approaching the unregulated result away from resonance. The suppression becomes more pronounced at lower velocities.
}
\label{fig:ElasticReg_FirstRes_vsXi}
\end{figure}

\Cref{fig:ElasticReg_FirstRes_vsXi} illustrates that annihilation feedback can strongly modify elastic scattering in the immediate vicinity of a Sommerfeld resonance, while rapidly becoming negligible away from it. The effect becomes increasingly pronounced towards lower velocities. We focus on the first $s$-wave resonance, $\xi=\xi_{1,0}$; higher resonances exhibit similar behavior. 

At a point sufficiently detuned from the resonance,  
\cref{fig:z0Diagnostics} shows that the dispersive and absorptive corrections do not significantly enhance the numerator of \cref{eq:x0Reg_r0}, $\abs{z_0\pm y_{0,\unreg}^\ren / r_0^\elas}\ll1$. Similarly, the corrections to the denominator are negligible, $y_{0,\unreg}^\ren, |z_0| \ll 1$. Consequently, $x_{0,\reg}^{\elas}\simeq x_{0,\unreg}^{\elas}$ and $y_{0,\reg}\simeq y_{0,\unreg}^{\ren}$.

\subsubsection*{On resonance}

\begin{figure}[t!]
\centering
\includegraphics[width=0.95\linewidth]{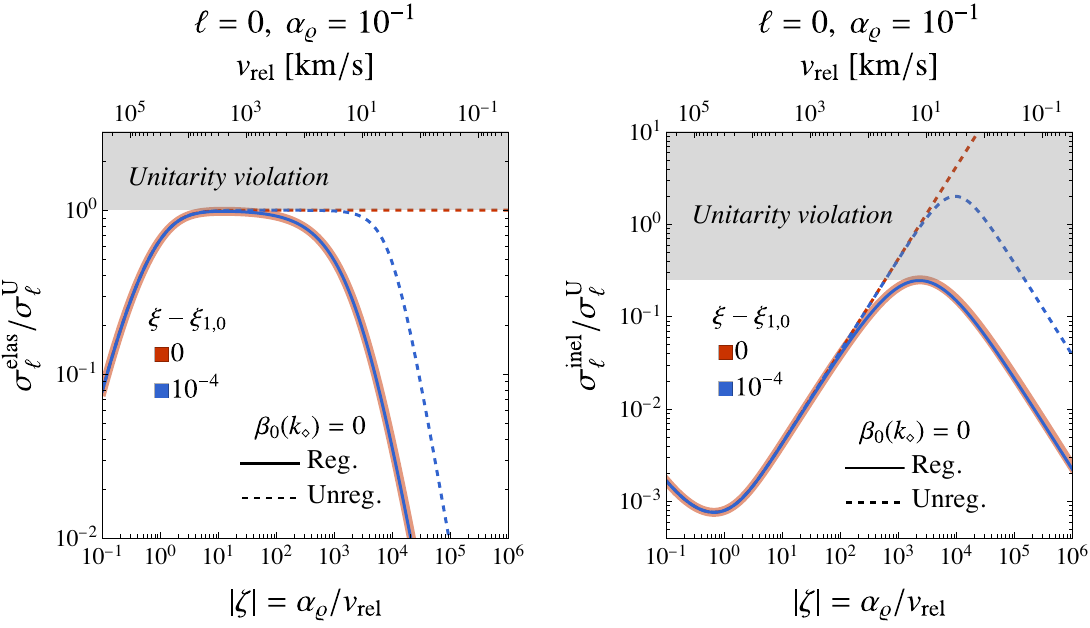}
\caption{
Unregulated and regulated $s$-wave cross sections, normalized to the unitarity cross section, as functions of $\abs{\zeta}=\arho/\vrel$, for $\arho=10^{-1}$ near the first $s$-wave resonance.
\emph{Left:} Elastic cross sections.
\emph{Right:} Annihilation cross sections.
On resonance, the unregulated annihilation cross section grows towards arbitrarily low velocities, increasingly suppressing elastic scattering through feedback. At finite detuning, the Sommerfeld enhancement eventually saturates and the ordinary contact-like scaling is recovered. Regularization strongly reduces the sensitivity to $\xi$, yielding very similar cross sections for nearby values around the resonance.
}
\label{fig:ElasticReg_FirstRes_vsZeta}
\end{figure}

Exactly on resonance, the baseline elastic cross section approaches its unitarity limit at low velocities, $x_{0,\unreg}^{\elas}\to1$, while the baseline annihilation cross section grows faster than the unitarity limit with decreasing velocity, $y_{0,\unreg}^{\ren}\propto k^{-1}\propto\zeta$.
As seen in \cref{fig:z0Diagnostics}, throughout this regime $|z_0|\ll1$ and $\left|z_0\pm y_{0,\unreg}^{\ren}/r_0^\elas\right|\ll1$; the dispersive correction and the modification of the numerator in \cref{eq:x0Reg_r0} are therefore subdominant, and the cross sections are regulated essentially only by the flux loss. 
As the velocity decreases, $y_{0,\unreg}^{\ren}$ grows and the feedback becomes important. When $y_{0,\unreg}^{\ren}\simeq1$, $y_{0,\reg}$ reaches its global unitarity maximum, $y_{0,\reg}\simeq1/4$, which also sets $x_{0,\reg}^{\elas}\simeq1/4$. At still lower velocities, $y_{0,\unreg}^{\ren}\gg1$, resulting in 
$x_{0,\reg}^{\elas} \propto \zeta^{-2}$ 
and
$y_{0,\reg} \propto \zeta^{-1}$.  
This is clearly visible in \cref{fig:ElasticReg_FirstRes_vsZeta}.
Thus, annihilation feedback cuts off the growth of the baseline rates at low velocities: since $\sigma_0^\U\propto k^{-2}$, both $\sigma_{0,\reg}^{\elas}$ and $\sigma_{0,\reg}^{\ann}\vrel$ approach constant values.

\subsubsection*{Very close to resonance}

At small but fixed detuning from the resonance, the system initially follows the resonant behavior described above. In contrast to the exact-resonance case, however, both $|z_0|$ and $y_{0,\unreg}^{\ren}/r_0^\elas$ grow as the velocity decreases and approach values much larger than unity. Nevertheless, the dispersive and absorptive contributions interfere destructively in the numerator of \cref{eq:x0Reg_r0}: as seen in \cref{fig:z0Diagnostics}, the combination $z_0\pm y_{0,\unreg}^{\ren}/r_0^\elas$ remains small despite the large magnitude of the individual terms. The potentially large correction to the numerator is therefore largely cancelled. By contrast, the large $z_0^2$ term dominates the denominator at sufficiently low velocities and regulates the cross section.

\medskip

\noindent
A notable consequence is that regularization greatly reduces the extreme sensitivity of the baseline cross sections to the resonance detuning. Very small variations of $\xi$ around a resonant value can change the unregulated low-velocity cross sections by orders of magnitude, because they determine whether the resonant growth persists or eventually saturates. Once annihilation feedback is included, however, these parametrically different baseline predictions are driven to remarkably similar regulated cross sections. Thus, the strong sensitivity of the baseline calculation near resonance is largely tamed by the self-consistent restoration of unitarity.

\subsubsection*{Velocity profile of the viscosity cross section}

\begin{figure}[t!]
\centering
\includegraphics[width=0.95\linewidth]{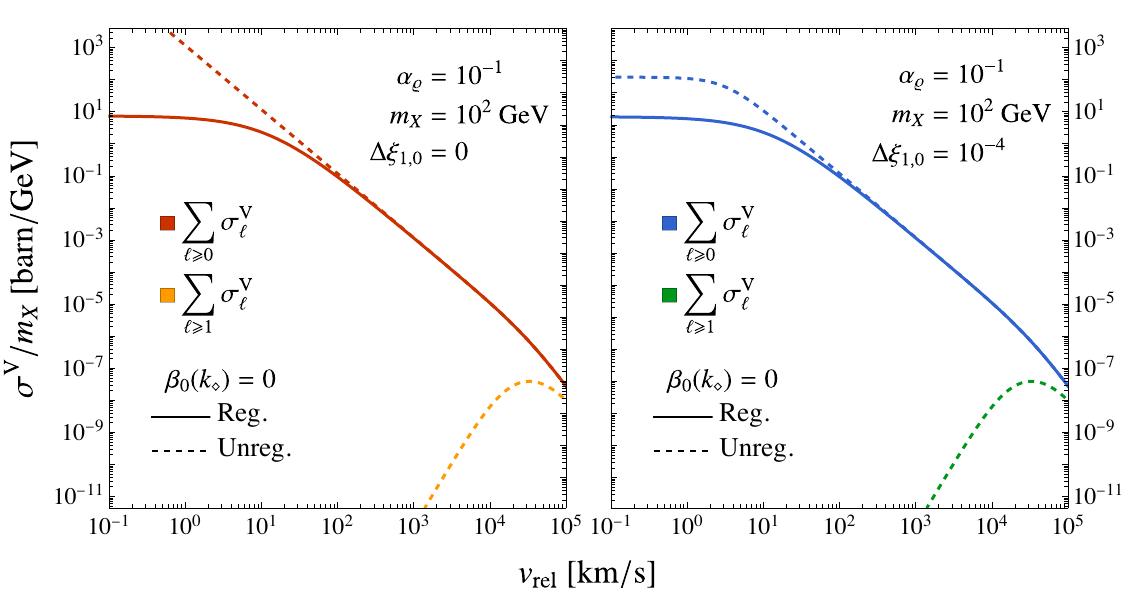}
\caption{
Total viscosity cross section per unit mass as a function of $\vrel$. We compare the regulated and unregulated results in the vicinity of the first $s$-wave resonance, $\xi_{1,0} \simeq 0.84$: exactly at the resonance (\emph{left}) and at very small fixed detuning, $\Delta\xi_{1,0} = 10^{-4}$ (\emph{right}). The difference between the regulated and unregulated results arises solely from the $\ell=0$ mode. We have set $\arho=0.1$, $\mX =100$~GeV and $\mrho \simeq 5.95$~GeV.
}
\label{fig:ElasticReg_FirstRes_VelocityProfile}
\end{figure}

In \cref{fig:ElasticReg_FirstRes_VelocityProfile}, we present the velocity profile for $\sigma^{\V}/m_{\DM}$, which is commonly employed to characterize the effect of DM self-interactions on heat transport inside halos. Over the velocity range where annihilation feedback is important, the cross section is dominated by its lowest mode, so that the behavior closely follows the $s$-wave discussion above. Exactly on resonance, the unregulated viscosity cross section continues to grow towards low velocities, whereas annihilation feedback cuts off this growth and produces a finite plateau. At small fixed detuning, the unregulated cross section already approaches a finite low-velocity plateau, but regularization suppresses its value substantially. Despite the very different baseline behavior of the on-resonance and detuned cases, the regulated cross sections approach remarkably similar low-velocity plateaus for the benchmark considered here. This illustrates how unitarization substantially reduces the extreme sensitivity of resonant SIDM cross sections to small variations of the model parameters.

\subsection{Near $s$-wave elastic-scattering zeros}
\label{Sec:AnnFeedback_Antiresonances}

\begin{figure}[t!]
\centering
\includegraphics[width=0.95\linewidth]{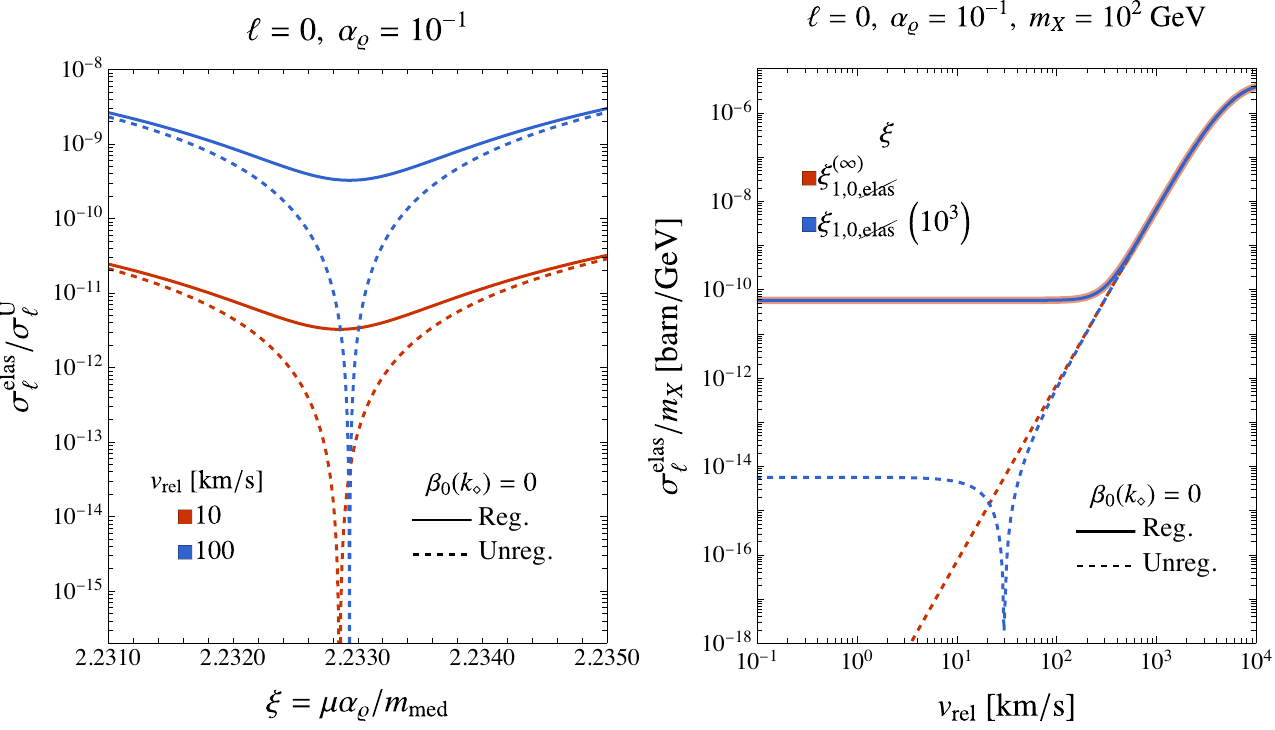}
\caption{
Annihilation feedback near the first $s$-wave elastic antiresonance, for $\arho=10^{-1}$.
\emph{Left:} Baseline and regulated $s$-wave elastic cross sections, normalized to the unitarity cross section, as functions of $\xi=\mu\arho/m_{\varrho}$ for $\vrel=10$ and $100~{\rm km\,s^{-1}}$. Annihilation feedback fills in the baseline zeros.
\emph{Right:} $s$-wave elastic cross section per unit DM mass as a function of $\vrel$, for $m_X=10^2~{\rm GeV}$, comparing the exact antiresonance with a nearby point whose baseline cross section vanishes at finite velocity.
}
\label{fig:ElasticReg_ElasticZeros}
\end{figure}

The baseline $s$-wave elastic cross section,
$x_{0,\unreg}^{\elas}=\sin^2\theta_0$, vanishes whenever
$\theta_0(\xi,\zeta)$ is an integer multiple of $\pi$.
Unlike Sommerfeld resonances, these zeros do not in general occur at fixed values of $\xi$, but follow trajectories
$\xi=\xi_{j,\ell,\cancel{\elas}}(\zeta)$, where $j$ enumerates the particular trajectory. We distinguish two behaviors.

\begin{description}

\item[Attractive potential.]
As $\zeta$ increases, the elastic-scattering zeros shift to smaller values of $\xi$ until they leave the Coulomb regime,
$\xi_{j,\ell,\cancel{\elas}}(\zeta)<\zeta$, and approach finite limits,
$\xi_{j,\ell,\cancel{\elas}}^{(\infty)} \equiv
\lim_{\zeta\to\infty} \xi_{j,\ell,\cancel{\elas}}(\zeta)$, 
which lie between successive $s$-wave resonances.
We refer to these zero-momentum limits as antiresonances; this suppression is also known as the Ramsauer--Townsend effect~\cite{sakurai2020modern}.

\item[Repulsive potential.]
As the velocity decreases, or equivalently as $|\zeta|$ increases, the values
$\xi_{j,\ell,\cancel{\elas}}(\zeta)$ initially decrease, reach a minimum, and then grow without bound while remaining in the quasi-Coulomb regime.
In contrast to the attractive case, the elastic-scattering zeros therefore do not approach finite antiresonances at low velocities.
We refer to Ref.~\cite{Flores:2026num} for further discussion.

\end{description}
The annihilation amplitude generally remains nonzero at an elastic-scattering zero.
Annihilation feedback can therefore regenerate an elastic contribution where the baseline one vanishes.
This is required by the coupled unitarity constraint
\eqref{eq:UnitarityConstraint_Combined_xy}, which, for
$y_{\ell,\reg}\ll1/4$, implies
\begin{align}
y_{\ell,\reg}^2
\lesssim
x_{\ell,\reg}
\lesssim
1-2y_{\ell,\reg}.
\end{align}
Indeed, setting $x_{0,\unreg}^{\elas}=0$ in
\cref{eq:x0_Elas_Reg} gives
\begin{align}
x_{0,\reg}^{\elas}
=
y_{0,\unreg}^{\ren}\,y_{0,\reg}
\geq
y_{0,\reg}^2 .
\label{eq:ElasticFloorZero}
\end{align}
Thus, annihilation feedback fills in an exact baseline elastic zero.
When $y_{0,\unreg}^{\ren},|z_0|\ll1$, the lower unitarity bound is saturated at leading order,
\begin{align}
x_{0,\reg}^{\elas}
\simeq
\left(y_{0,\unreg}^{\ren}\right)^2
\simeq
y_{0,\reg}^2 .
\label{eq:ElasticFloorWeak}
\end{align}
This behavior is illustrated in \cref{fig:ElasticReg_ElasticZeros} for an attractive Yukawa potential. Annihilation feedback fills in the sharp baseline elastic zeros, replacing them with nonzero minima, and substantially reduces the sensitivity to small variations of $\xi$: cross sections that differ strongly between the exact and slightly detuned antiresonance in the baseline calculation become very similar once annihilation feedback is included.

\subsubsection*{On the antiresonance}

Exactly on the antiresonance, the leading $s$-wave threshold contribution vanishes as
\begin{align}
x_{0,\unreg}^{\elas}
\propto
k^6
\propto
\zeta^{-6}.
\end{align}
By contrast, the Sommerfeld factor saturates at low velocities, so that
\begin{align}
y_{0,\unreg}^{\ren}
\propto
k
\propto
\zeta^{-1}.
\end{align}
Since $y_{0,\unreg}^{\ren},|z_0|\ll1$ in this regime, as seen in \cref{fig:z0Diagnostics}, we obtain
\begin{align}
x_{0,\reg}^{\elas}
\simeq
\left(y_{0,\unreg}^{\ren}\right)^2
\propto
k^2
\propto
\zeta^{-2}.
\label{eq:ElasticAntiresonanceScaling}
\end{align}
Thus, annihilation feedback changes the asymptotic normalized elastic scaling from
$x_{0,\unreg}^{\elas}\propto k^6$ to
$x_{0,\reg}^{\elas}\propto k^2$.
Since $\sigma_0^\U\propto k^{-2}$, the baseline physical elastic cross section vanishes as $k^4$, whereas the regulated one approaches a constant at low velocities, $\sigma_{0,\reg}^{\elas} \sim$~const; this is the plateau visible in the right panel of \cref{fig:ElasticReg_ElasticZeros}.

\subsubsection*{Very close to the antiresonance}

At small but fixed detuning from the antiresonance, the ordinary low-velocity $s$-wave behavior is eventually recovered, $\sigma_{0,\unreg}^{\elas} \sim$~const.~or equivalently $x_{0,\unreg}^{\elas} \propto k^2$.
The annihilation-induced contribution has the same asymptotic scaling, $\left(y_{0,\unreg}^{\ren}\right)^2\propto k^2$. Consequently, both the baseline and regulated physical cross sections approach finite low-velocity plateaus, but their normalizations can differ substantially. For the nearby benchmark shown in \cref{fig:ElasticReg_ElasticZeros}, annihilation feedback strongly enhances the plateau, bringing it remarkably close to that obtained exactly on the antiresonance.

\subsubsection*{Effect of higher partial waves}

At an $s$-wave antiresonance, the additional suppression of the baseline $s$ wave removes its usual low-velocity dominance over the higher partial waves. Indeed, $x_{0,\unreg}^{\elas}\propto k^6$, which is the same threshold scaling as a generic $p$ wave. The corresponding physical $s$- and $p$-wave cross sections therefore both scale as $k^4$, and higher partial waves may compete with the baseline $s$-wave contribution. Annihilation feedback restores the low-velocity dominance of the $s$ wave: since $x_{0,\reg}^{\elas}\propto k^2$, the regulated physical $s$-wave cross section approaches a constant, whereas the higher $\ell$ remain suppressed by positive powers of $k$.

An $s$-wave antiresonance does not imply an exact zero of the lowest viscosity mode, $x_{0,\unreg}^{\V}=\sin^2(\theta_0-\theta_2)$, at finite velocity. Nevertheless, in the zero-momentum limit $\theta_0\propto k^3$ while $\theta_2\propto k^5$, such that $x_{0,\unreg}^{\V} \simeq \theta_0^2 \propto k^6$. The annihilation-induced contribution therefore also gives $x_{0,\reg}^{\V}\propto k^2$ and generates a finite low-velocity viscosity cross section.

\subsubsection*{Velocity dependence and SIDM phenomenology}

The velocity dependence around an antiresonance is opposite to that typically sought in SIDM.
At sufficiently low velocities, annihilation feedback generates a finite elastic plateau; as the velocity increases, however, the antiresonant suppression becomes less effective and the cross section grows, as seen in the right panel of
\cref{fig:ElasticReg_ElasticZeros}.
Consequently, parameter choices for which
$\sigma_{\reg}^{\elas}/\mX\gtrsim1~{\rm barn/GeV}$
at dwarf-galaxy velocities yield still larger cross sections at cluster velocities, conflicting with the corresponding constraints.

\subsection{Short-distance couplings and higher resonances}
\label{Sec:AnnFeedback_BeyondBenchmark}

We now briefly discuss the dependence of the annihilation feedback on the short-distance benchmark and on the resonance considered.

\subsubsection*{Absorptive coupling}

In the present model, the absorptive coupling is not fixed by the long-range coupling $\arho$, since $\lambda_{X\varrho}$ provides an independent short-distance annihilation contribution. Our benchmark sets this contribution to zero, retaining only the trilinear one, and thus resembles models without an independent renormalizable contact inelastic interaction, such as fermionic DM coupled to a scalar mediator. Increasing the absorptive coupling enhances both $y_{0,\unreg}^{\ren}$ and $z_0$, which scale as $[\eta_{A,0}^{\ren}]^2$, and may therefore make annihilation feedback important at higher velocities and over a broader region around resonances and elastic-scattering zeros.

\subsubsection*{Dispersive coupling}

In the absence of an independent Hermitian contact interaction, the absorptive and dispersive short-distance interactions arise from the same dynamics, suggesting
$\eta^{\ren}_{H,0}\sim[\eta^{\ren}_{A,0}]^2$, or equivalently $\beta_0\sim{\cal O}(1)$, up to possible model-dependent cancellations in the principal-value contribution \cite[Eq.~(3.11)]{Flores:2025uoh}. At the subtraction point, where $z_0=0$, \cref{eqs:RegulRenorm_w0s} give $\tilde w_0=\beta_0 w_0$. Thus, for $|\beta_0|\sim{\cal O}(1)$, the absorptive and dispersive corrections are parametrically comparable and become important at the same parametric scale, although the latter can modify the interference with the baseline elastic amplitude, particularly near an elastic-scattering zero. Our choice $\beta_0(k_\diamond)=0$ is therefore only a benchmark: away from $k=k_\diamond$, $z_0(k;k_\diamond)$ generates a nonzero dispersive contribution, and changing the subtraction scale induces a nonzero $\beta_0$ according to \cref{eq:Running_Beta0}.

This power counting need not hold in the presence of an independent Hermitian short-distance interaction. For example, $\lambda_X$ contributes to $\eta^{\ren}_{H,0}$ but not to the absorptive coupling, so $\beta_0$ may be parametrically different from unity. At the subtraction point, 
$\tilde{w}_0 
= \beta_0\,y_{0,\unreg}^{\ren}
= \eta^{\ren}_{H,0} \,
|\hat{\cal M}_{0,\unreg}|^2$, 
so the absorptive coupling cancels and the correction is controlled directly by the Hermitian short-distance interaction. Depending on its sign and on the baseline phase shift, it may interfere constructively or destructively with the long-range elastic amplitude.

\subsubsection*{Higher $s$-wave resonances}

The behavior described for the first $s$-wave resonance extends to higher $s$-wave resonances, with the main difference being the velocity scale at which annihilation feedback becomes important. For the Hulth\'en potential, exactly on the $n$th resonance and at large $\zeta$, $S_0^{\rm H}\simeq4\zeta^2/n^2$, and hence $y_{0,\unreg}^{\ren}\propto\zeta/n^2$. A fixed feedback strength is therefore reached at $\zeta\propto n^2$, implying that higher resonances are affected at lower velocities.

\subsection{Self-interactions in dark-matter halos}
\label{Sec:AnnFeedback_SIDM_Halos}

\subsubsection*{SIDM benchmarks}

To identify regions relevant for SIDM phenomenology, we use the representative benchmarks at galaxy-cluster, dwarf-galaxy and subhalo scales,
\begin{subequations}
\label{eq:SIDM_Benchmarks}
\label[pluralequation]{eqs:SIDM_Benchmarks}
\begin{align}
\dfrac{\overline{\sigma}_{\SIDM}^{\V}}{\mX}
&\lesssim
1~{\rm barn\,GeV^{-1}},
&&
\text{at}~
v_c \sim 10^3~{\rm km/s},
&&
\text{galaxy-cluster scale},
\label{eq:SIDM_Benchmark_Cluster}
\\
\dfrac{\overline{\sigma}_{\SIDM}^{\V}}{\mX}
&\gtrsim
1~{\rm barn\,GeV^{-1}},
&&
\text{at}~
v_c \sim 10~{\rm km/s},
&&
\text{dwarf-galaxy scale},
\label{eq:SIDM_Benchmark_Dwarf}
\\
\dfrac{\overline{\sigma}_{\SIDM}^{\V}}{\mX}
&\sim
10^2~{\rm barn\,GeV^{-1}},
&&
\text{at}~
v_c \sim 1~{\rm km/s},
&&
\text{subhalo scale}.
\label{eq:SIDM_Benchmark_Subhalos}
\end{align}
\end{subequations}
Here, $v_c$ is the Maxwellian velocity scale defined in \cref{Sec:Unitarity_AveragedCrossSections}. At dwarf-galaxy velocities, self-interaction cross sections per unit mass of this order or larger are typically required to appreciably modify halo structure~\cite{Tulin:2017ara,Kaplinghat:2015aga}, whereas observations of galaxy clusters constrain the cross section per unit mass at the larger characteristic velocities of these systems~\cite{Markevitch:2003at,Randall:2008ppe}. The larger subhalo-scale benchmark is motivated by gravothermal core-collapse interpretations of recently observed compact substructures~\cite{Vegetti:2026mmx,Gilman:2026wul,Zeng:2023fnj}. These criteria should be regarded as phenomenological benchmarks rather than universal bounds, since the relation between the microscopic scattering cross section and halo evolution depends on the system, observable, and relevant transport process~\cite{Rosenlyst:2026koq}.

\subsubsection*{Annihilation feedback across the SIDM parameter space}

We consider symmetric DM, with equal number densities, $n_X=n_{X^\dagger}=n_{\rm DM}/2$. Assuming that the two species follow the same velocity distribution and taking into account that $\overline{\sigma}_{XX}^{\V}=\overline{\sigma}_{X^\dagger X^\dagger}^{\V}$, the halo-averaged viscosity cross section is
\begin{align}
\overline{\sigma}_{\rm SIDM}^{\V}
=
\frac{1}{2}
\left(
\overline{\sigma}_{XX^\dagger}^{\V}
+
\overline{\sigma}_{XX}^{\V}
\right).
\label{eq:SIDM_Symmetric_CrossSection}
\end{align}
The two contributions have different origins. The $XX^\dagger$ cross section is already present at baseline through the Yukawa interaction, whereas, under the assumptions of \cref{Sec:Model}, the baseline $XX$ elastic amplitude vanishes and $\overline{\sigma}_{XX}^{\V}$ is generated by BSF feedback. We neglect the latter contribution here and study it in detail in our companion paper~\cite{Flores:2026SIDMbsf}.

\begin{figure}[t!]
\centering
\includegraphics[width=0.93\linewidth]{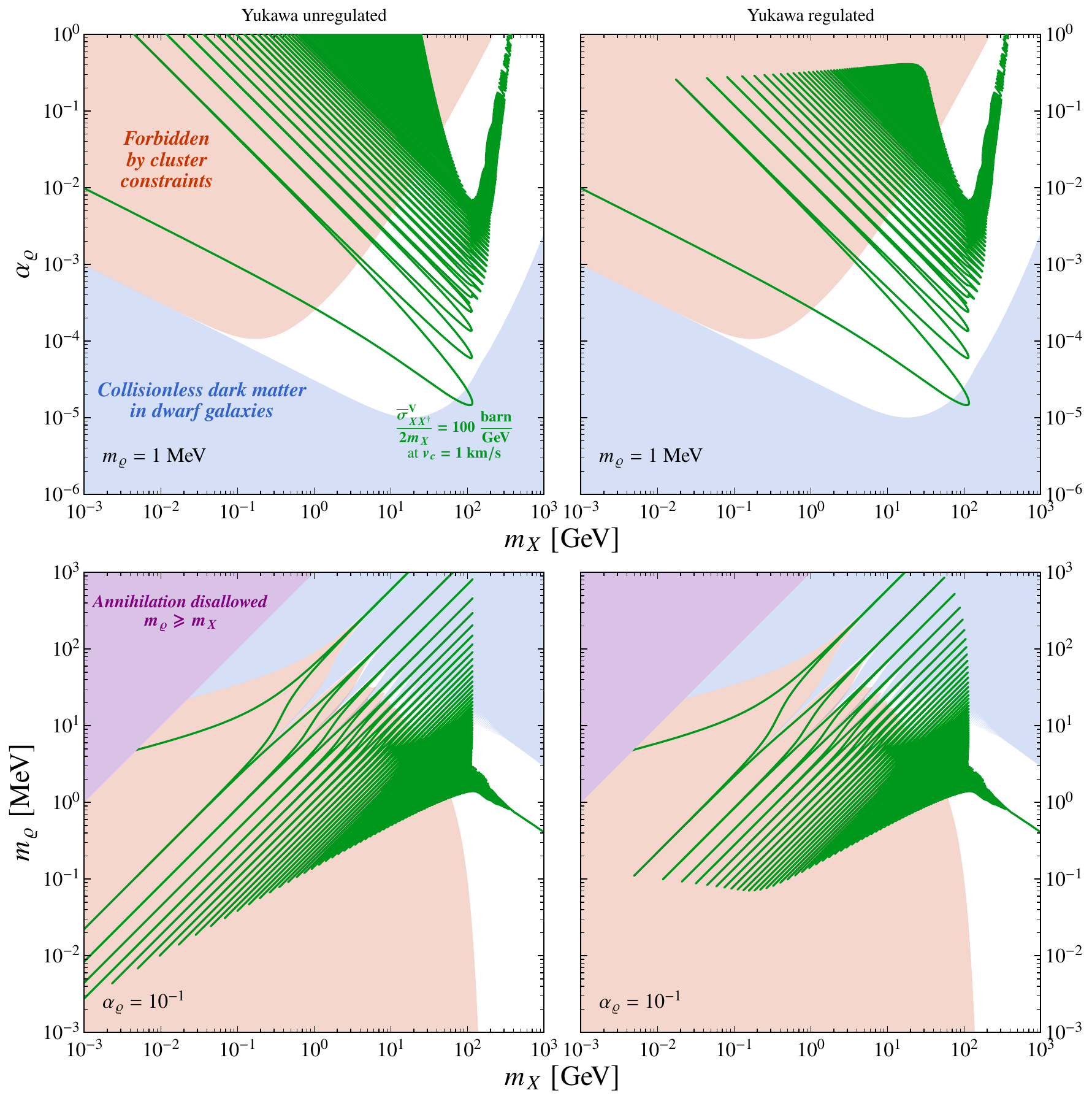}
\caption{
Halo-averaged $XX^\dagger$ Yukawa contribution to the viscosity cross section per unit DM mass, $\overline{\sigma}_{XX^\dagger}^{\V}/(2m_X)$, for symmetric DM. We show the DM mass--coupling plane at fixed mediator mass (\emph{upper row}) and the DM mass--mediator mass plane at fixed coupling (\emph{lower row}), comparing the baseline result (\emph{left}) with that including $s$-wave annihilation feedback using the benchmark prescription of \cref{Sec:UnitarityRestoration_Benchmarks} (\emph{right}); higher partial waves remain at their baseline values. Red regions exceed the cluster benchmark of \cref{eq:SIDM_Benchmark_Cluster}, blue regions fall below the dwarf benchmark of \cref{eq:SIDM_Benchmark_Dwarf}, and white regions satisfy both; green contours mark the subhalo benchmark of \cref{eq:SIDM_Benchmark_Subhalos}. Near antiresonances, feedback fills the $s$-wave suppression and truncates excursions toward larger coupling, until higher partial waves dominate and wash out the effect at large $\mX$. Near resonances, feedback curtails the baseline $s$-wave enhancement as it approaches its unitarity limit, but shifts the benchmark regions less strongly than near antiresonances. BSF-induced $XX$ scattering is neglected here and studied in \cite{Flores:2026SIDMbsf}. In the purple region, $m_\varrho\geq m_X$, annihilation into $\varrho\varrho^\dagger$ is kinematically forbidden.
}
\label{fig:SIDM_Yukawa_ParameterSpace}
\end{figure}

The $XX^\dagger$ Yukawa cross section is strongly velocity dependent. Away from exact resonances and antiresonances, it generally grows as the relative velocity decreases, until the finite interaction range cuts off this growth and the low-velocity cross section approaches an $s$-wave-dominated plateau. This velocity dependence determines whether a given choice of $(\mX,\mrho,\arho)$ can simultaneously yield sizable self-interactions in dwarf galaxies or subhalos while satisfying the cluster-scale benchmark. We illustrate this in \cref{fig:SIDM_Yukawa_ParameterSpace}, using the benchmarks of \cref{eq:SIDM_Benchmarks}, and compare the baseline Yukawa result with that including $s$-wave annihilation feedback. As anticipated from \cref{Sec:AnnFeedback_Resonances,Sec:AnnFeedback_Antiresonances}, annihilation feedback appreciably modifies the Yukawa cross section mainly near $s$-wave resonances and antiresonances.

Near antiresonances, feedback lifts the $s$-wave suppression and truncates the narrow excursions of the benchmark regions toward larger $\arho$ or smaller $\mrho$. At sufficiently large $\mX$, higher partial waves become important and wash out the $s$-wave antiresonance, reducing the impact of annihilation feedback. These modifications occur mainly in regions already disfavored by the cluster-scale benchmark.

The resonant bands relevant to the SIDM benchmarks occur at comparatively small $\arho$, where annihilation remains coupling-suppressed, and are therefore only weakly affected. For fixed $\arho$ (\emph{lower row}), however, the baseline $s$-wave contribution can approach its unitarity limit near a resonance, around $\mX\sim100~{\rm GeV}$ for the benchmark shown. Annihilation feedback then curtails the resonant enhancement, particularly toward larger mediator masses. At still larger $\mX$, higher partial waves again become important and reduce the relative impact of the \(s\)-wave unitarization.

\clearpage
\section{Conclusions}
\label{Sec:Conclusions}

Self-interacting DM models commonly invoke light force mediators to generate sizable and velocity-dependent scattering cross sections. The same mediators can also enhance inelastic processes, making their contribution to the two-particle dynamics important. Unitarity then implies that elastic and inelastic scattering cannot be computed independently: the interactions responsible for inelasticity must also be included in the self-energy kernel that determines the elastic amplitude. This feedback is conceptually distinct from the direct phenomenological consequences of inelastic reactions actually taking place in the early Universe or inside halos.

In this work, we investigated the feedback of DM annihilation on elastic scattering in the presence of a long-range Yukawa force. We identified two qualitatively distinct effects, illustrated in \cref{fig:ElasticReg_FirstRes_vsXi,fig:ElasticReg_FirstRes_vsZeta,fig:ElasticReg_FirstRes_VelocityProfile,fig:ElasticReg_ElasticZeros,fig:SIDM_Yukawa_ParameterSpace}. Near $s$-wave threshold resonances, where Sommerfeld enhancement can drive the baseline annihilation cross section towards or beyond its partial-wave unitarity bound, resumming the annihilation interaction curtails the low-velocity growth of both the annihilation and elastic cross sections. Conversely, near zeros of the baseline elastic amplitude, annihilation feedback generates a nonzero, albeit typically small, elastic contribution, as required by the coupled elastic--inelastic unitarity constraint. In both regimes, unitarization also substantially reduces the extreme sensitivity to the model parameters: baseline cross sections that differ by orders of magnitude for very small variations of $\xi$ around resonances or antiresonances become much more similar once annihilation feedback is included. Away from resonances and elastic-scattering zeros, annihilation feedback is generally negligible throughout the parameter range considered here.

The effect near threshold resonances is particularly relevant for SIDM phenomenology. Sommerfeld-like resonances have often been invoked to obtain strongly enhanced elastic scattering at low velocities, despite the finite mediator mass, which otherwise cuts off the long-range growth below a characteristic velocity. The reliability of this mechanism must therefore be reassessed whenever the same resonant wavefunction also enhances an inelastic channel. This question is especially timely in view of recent strong-lensing observations of exceptionally compact perturbers and galaxy-scale lenses. If such objects are manifestations of gravothermal core collapse in SIDM halos, their formation may require very large self-interaction cross sections at low characteristic velocities. Our results become relevant precisely in this regime: the feedback of inelasticity must be included in order to determine reliably whether resonant particle-physics models can generate the required elastic scattering.

The ingredients responsible for these effects are common to many SIDM models: models with a symmetric DM population generally contain annihilation channels, whose short-distance amplitudes are dressed by the long-range dynamics of the incoming state. We therefore expect the qualitative features identified here---the suppression of resonantly enhanced scattering, the regeneration of a nonzero elastic contribution near baseline zeros, and the reduced sensitivity to small parameter variations in both regions---to persist across a broad class of models, even though their quantitative impact depends on the underlying particle physics.

A companion paper considers the feedback generated by radiative BSF. Annihilation and BSF feedback differ both technically and phenomenologically. Annihilation is a hard, short-distance process whose nonrelativistic contact description requires renormalization. Radiative BSF is instead an ultrasoft process, whose baseline amplitudes possess a richer nonanalytic structure that affects unitarization. Correspondingly, annihilation feedback is important mainly around Sommerfeld resonances or elastic-scattering zeros, whereas BSF feedback can remain significant over a broader region of parameter space and can generate sizable elastic scattering even in channels for which no baseline long-range elastic interaction is present. However, the occurrence and properties of BSF are more model dependent.

The two studies isolate the feedback of these two qualitatively different types of inelasticity. More generally, however, several inelastic processes may contribute simultaneously to the self-energy, and hence to the regularization, of the same elastic scattering channel. In particular, hard processes such as annihilation and ultrasoft processes such as radiative BSF may both become important for the same incoming state. This occurs already in simple dark-QED-like models. In the present model, for example, including the gauge interaction, $\aV\neq0$, allows  $XX^\dagger$ pairs to both annihilate and form bound states radiatively through vector emission. Treating such contributions jointly, including their interplay in the unitarized elastic and inelastic amplitudes, constitutes a natural extension of the present studies.

\section*{Acknowledgments}
We thank Marcos M. Flores for discussions and collaboration on the numerical framework underlying the baseline Yukawa calculations~\cite{Flores:2026num}. This work was supported by the European Union’s Horizon 2020 research and innovation program under grant agreement No 101002846, ERC CoG CosmoChart.

\clearpage
\begin{appendices}
\addtocontents{toc}{\protect\setcounter{tocdepth}{1}}
\section{Regulated and unregulated partial-wave cross sections
}
\label{App:PWanalysis}

\subsection{Partial-wave expansions of unregulated cross sections}
\label{App:PWanalysis_Expansions}

The elastic, transfer and viscosity cross sections are defined as
\begin{subequations}
\label{eq:CrossSection_def}
\label[pluralequation]{eqs:CrossSection_def}
\begin{align}
    \sigma^\elas(k) &=
    \int \dd\Omega 
    \, \frac{\dd\sigma(k)}{\dd\Omega} ,
    \label{eq:CrossSection_Elastic_def}
    \\
    \sigma^{\T}(k) &= 
    \int \dd\Omega 
    (1-\cos \vartheta) 
    \, \frac{\dd\sigma(k)}{\dd\Omega} ,
    \label{eq:CrossSection_Transfer_def}
    \\
    \sigma^{\V}(k) &=
    \int \dd\Omega \sin^2\vartheta 
    \, \frac{\dd\sigma(k)}{\dd\Omega} ,
    \label{eq:CrossSection_Viscosity_def}
\end{align}
\end{subequations}
where ${\bf k}$ is the momentum of one of the particles in the CM frame, with $k=|{\bf k}|$, and $\vartheta$ and $\Omega$ are the polar and solid scattering angles, respectively. The differential elastic cross section is 
\begin{equation}
\frac{\dd\sigma(k)}{\dd\Omega} = 
\abs{f_{\vb{k}}(\vartheta)}^2 .
\label{eq:DiffCrossSection}
\end{equation}
We decompose the amplitude in partial waves as follows:
\begin{equation}
f_{\vb{k}}(\vartheta) =
\sum_{\ell=0}^\infty (2\ell+1) 
\, f_{k,\ell}
\, P_\ell(\cos \vartheta) ,
\label{eq:f_PWanalysis}
\end{equation}
where $P_\ell$ is the $\ell$-th Legendre polynomial. For identical particles, the wavefunctions and amplitudes are understood to be properly symmetrized or antisymmetrized, so that the above expressions apply to both distinguishable and identical particles. The partial-wave amplitudes are related by $f_{k,\ell} = \sqrt{\sym_\ell}\, f_{k,\ell}^{\rm DP}$, where DP denotes distinguishable particles and $\sym_\ell$ are the partial-wave symmetry factors defined in \cref{eq:SymmetryFactors}.

\subsubsection*{Elastic cross section.}
Inserting \eqref{eq:f_PWanalysis} into \eqref{eq:CrossSection_Elastic_def}, and using the orthogonality of Legendre polynomials, we find
\begin{equation}
\sigma^\elas(k) = 4\pi \sum_{\ell=0}^\infty 
(2\ell+1) \abs{f_{k,\ell}}^2 .
\label{eq:CrossSection_Elastic_PW}
\end{equation}
As expected, different partial waves do not interfere. This will not be true for the transfer and viscosity cross sections.

\subsubsection*{Transfer cross section.}
Inserting \eqref{eq:f_PWanalysis} into \eqref{eq:CrossSection_Transfer_def}, we find
\begin{align}
    \sigma^{\T}(k) &=
    \sum_{\ell,m=0}^\infty (2\ell+1)(2m+1) 
    \, f_{k,\ell}^* \, f_{k,m}^{} 
    \, 2\pi \int_{-1}^1 dx (1-x)P_{\ell}(x)P_m(x)\nonumber\\
    &=4\pi \sum_{\ell,m=0}^\infty(2\ell+1)(2m+1)
    \, f_{k,\ell}^* \, f_{k,m}^{} 
    \left[\ThreejSymbol{0}{0}{\ell}{0}{m}{0}^2-\ThreejSymbol{1}{0}{\ell}{0}{m}{0}^2\right],
    \label{sigmaT ell and m}
\end{align}
where in the second line we used the fact that $(1-x)=P_0(x)-P_1(x)$. The Wigner-$3j$ symbol $\ThreejSymbol{j}{0}{\ell}{0}{m}{0}$ is nonvanishing only when 
\begin{align}
\begin{cases}
j+\ell+m = \text{even}, 
\\
\text{and}
\\
j+\ell\geqslant m, \quad
\ell+m\geqslant j, \quad     
j+m \geqslant \ell \quad
\text{(triangular rule)}.
\end{cases}
\label{eq:3j wigner rule}
\end{align}
Using this rule, we get: 
\begin{align}
    \sigma^{\T}(k) &=
    4\pi \sum_{\ell=0}^\infty  \left[
    (2\ell+1)^2 \ThreejSymbol{0}{0}{\ell}{0}{\ell}{0}^2 \abs{f_{k,\ell}^{}}^2
    -(2\ell+3)(2\ell+1) \ThreejSymbol{1}{0}{\ell+1}{0}{\ell}{0}^2 f_{k,\ell+1}^* f_{k,\ell}^{} \right.
    \nonumber\\
    &\left.
    -(2\ell+1)(2\ell+3) \ThreejSymbol{1}{0}{\ell}{0}{\ell+1}{0}^2 f_{k,\ell}^* f_{k,\ell+1}^{} \right]
    \nonumber \\
    &=4\pi \sum_{\ell=0}^\infty \left[
    (2\ell+1)\abs{f_{k,\ell}}^2-2(\ell+1) 
    \Re(f_{k,\ell+1}^* f_{k,\ell}^{})\right]
    \nonumber \\
    &=4\pi\sum_{\ell=0}^\infty (\ell+1)
    \left[ \abs{f_{k,\ell}}^2
    +\abs{f_{k,\ell+1}}^2
    -2\Re(f_{k,\ell+1}^* f_{k,\ell}^{}) \right], 
\end{align}
where we isolated the term $\ell |f_{k,\ell}|^2$ and shifted the index of the corresponding sum. We arrive at
\begin{align}
\sigma^{\T} (k)
=4\pi\sum_{\ell=0}^\infty (\ell+1) 
\abs{f_{k,\ell+1}-f_{k,\ell}}^2 .
\label{eq:CrossSection_Transfer_PW}
\end{align}
We see that in the transfer cross section, even and odd partial waves interfere.

\subsubsection*{Viscosity cross section.}
Inserting \eqref{eq:f_PWanalysis} into \eqref{eq:CrossSection_Viscosity_def}, we find
\begin{align}
\sigma^{\V}(k) &= 2\pi \sum_{\ell,m=0}^\infty(2\ell+1)(2m+1)f_{k,\ell}^*f_{k,m}\int_{-1}^1 dx (1-x^2)P_{\ell}(x)P_{m}(x)
\nonumber\\
&=\frac{8\pi}{3} 
\sum_{\ell,m=0}^\infty(2\ell+1)(2m+1)
\, f_{k,\ell}^* \, f_{k,m}^{}
\left[
\ThreejSymbol{0}{0}{\ell}{0}{m}{0}^2
-\ThreejSymbol{2}{0}{\ell}{0}{m}{0}^2
\right] ,
\label{sigma V before integration}
\end{align}
where in the second line we used the fact that $1-x^2=\frac{2}{3}(P_0(x)-P_2(x))$. Using the selection rule \eqref{eq:3j wigner rule}, we obtain 
\begin{align}
\sigma^{\V}(k) &= 
\frac{8\pi}{3} \sum_{\ell=0}^\infty \Bigg\{
(2\ell+1)^2 
\left[\ThreejSymbol{0}{0}{\ell}{0}{\ell}{0}^2-\ThreejSymbol{2}{0}{\ell}{0}{\ell}{0}^2\right]
\abs{f_{k,\ell}}^2
\nonumber\\
&
-(2\ell+5)(2\ell+1)
\ThreejSymbol{2}{0}{\ell+2}{0}{\ell}{0}^2 
f_{k,\ell+2}^* f_{k,\ell}^{}
\nonumber\\
&
-(2\ell+1)(2\ell+5) 
\ThreejSymbol{2}{0}{\ell}{0}{\ell+2}{0}^2 
f_{k,\ell}^* f_{k,\ell+2}^{} 
\Bigg\}
\nonumber\\[1em]
&=8\pi\sum_{\ell=0}^\infty
\left\{(2\ell+1)
\left[\frac{\ell^2+\ell-1}{(2\ell-1)(2\ell+3)}\right] 
\abs{f_{k,\ell}^{}}^2
-\frac{(\ell+1)(\ell+2)}{(2\ell+3)} 
\Re(f_{k,\ell}^* \, f_{k,\ell+2}^{})
\right\} .
\end{align}
We remark that 
\begin{align}
    \frac{(2\ell+1)(\ell^2+\ell-1)}{(2\ell-1)(2\ell+3)}=\frac{1}{2}\left(\frac{(\ell+1)(\ell+2)}{(2\ell+3)}+\frac{\ell(\ell-1)}{(2\ell-1)}\right) .
\end{align}
By shifting the index for one term, we find that 
\begin{align}
    \sigma^{\V} (k)
    &=4\pi  \sum_{\ell=0}^\infty \frac{(\ell+1)(\ell+2)}{(2\ell+3)}\abs{f_{k,\ell+2}-f_{k,\ell}}^2.
\label{eq:CrossSection_Viscosity_PW}
\end{align}
Thus, for the viscosity cross section, partial waves of the same parity interfere; the even and odd sectors remain separate.

\subsection{Unregulated partial-wave  cross sections in terms of phase shifts}
\label{App:PWanalysis_UnregulatedCrossSections}

The unregulated scattering amplitudes, $f_{k,\ell}^{\unreg}$, can be rewritten in terms of the phase shifts, $\theta_\ell (k)\in \mathbb{R}$, as follows
\begin{equation}
e^{2\im \theta_\ell(k)} \equiv 
1+\dfrac{2\im k}{\sqrt{\sym_\ell}} f_{k,\ell}^{\unreg} .
\label{eq:PhaseShift-Amplitude}
\end{equation}
Taking into account that $\sym_{\ell+2}=\sym_\ell$, and that for identical particles adjacent partial waves do not simultaneously contribute, the partial-wave cross sections of \cref{eq:CrossSection_Elastic_PW,eq:CrossSection_Transfer_PW,eq:CrossSection_Viscosity_PW} become 
\begin{subequations}
\label{eq:CrossSections_PW_unreg}
\label[pluralequation]{eqs:CrossSections_PW_unreg}
\begin{align}
\sigma^\elas_{\ell,\unreg}(k) &=  
\sigma_{\ell}^{\U}(k) \sin^2\theta_\ell(k) ,
\label{eq:CrossSection_Elastic_PW_unreg}
\\
\sigma^{\T}_{\ell,\unreg}(k) &= 
\sigma_{\ell}^{\U,\T}(k) 
\sin^2[\theta_{\ell+1}(k)-\theta_\ell(k)], 
\qquad \text{for distinguishable particles,}
\label{eq:CrossSection_Transfer_PW_unreg}
\\
\sigma^{\V}_{\ell,\unreg}(k) &= 
\sigma_{\ell}^{\U,\V}(k)
\sin^2[\theta_{\ell+2}(k)-\theta_\ell(k)] ,
\label{eq:CrossSection_Viscosity_PW_unreg}
\end{align}
\end{subequations}
where we defined the unitarity cross sections 
\begin{subequations}
\label{eq:CrossSections_PW_Unitarity}
\label[pluralequation]{eqs:CrossSections_PW_Unitarity}
\begin{align}
\sigma^{\U}_\ell(k) &\equiv
\sym_\ell 
\frac{4\pi}{k^2}(2\ell+1) ,
\label{eq:CrossSection_Elastic_PW_Unitarity}
\\
\sigma^{\U,\T}_\ell(k) &\equiv
\phantom{\sym_\ell}
\frac{4\pi}{k^2}(\ell+1) , 
\qquad \text{for distinguishable particles,}
\label{eq:CrossSection_Transfer_PW_Unitarity}
\\
\sigma^{\U,\V}_\ell(k) &\equiv 
\sym_\ell
\frac{4\pi}{k^2}\frac{(\ell+1)(\ell+2)}{(2\ell+3)} .
\label{eq:CrossSection_Viscosity_PW_Unitarity}
\end{align}
\end{subequations}
Here, $\sigma_\ell^{\U}$ is the usual elastic partial-wave
unitarity limit, while $\sigma_\ell^{\U,\T}$ and
$\sigma_\ell^{\U,\V}$ provide term-by-term upper bounds on the
transfer and viscosity contributions, respectively. The transfer formulas hold for distinguishable particles only; for identical particles, it is easy to check that 
\begin{align}
\sigma^{\T}(k) = \sigma^{\elas}(k) .
\label{eq:IdenticalParticles_Elas=Transfer}
\end{align}

\subsection{Unitarization of elastic, transfer and viscosity cross sections}
\label{App:PWanalysis_RegulatedCrossSections}

When both elastic and inelastic processes are resummed, the elastic
partial-wave amplitude generalizes to
\begin{equation}
e^{2\im \Delta_\ell(k)} \equiv 
1 + \dfrac{2\im k}{\sqrt{\sym_\ell}} 
f_{k,\ell}^{\reg} ,
\label{eq:ScatteringAmplitude_Regulated}
\end{equation}
with
\begin{equation}
\Delta_\ell(k)=\theta_\ell(k)+\delta_\ell(k) ,
\label{eq:PhaseShift_Real+Complex}
\end{equation}
where $\theta_\ell\in\RR$ is generated by the purely elastic interactions, while $\delta_\ell\in\CC$ is generated by the inelastic processes. The latter is given by~\cite{Flores:2025uoh}
\begin{align}
e^{2\im\delta_\ell (k)}
&=
\dfrac
{1-[w_\ell(k)-\im\tilde w_\ell(k)]}
{1+[w_\ell(k)-\im\tilde w_\ell(k)]} ,
\label{eq:PhaseShift_Complex}
\end{align}
where $w_\ell \in \mathbb{R}$ and $\tilde{w}_\ell \in \mathbb{R}$ are related to the inelastic amplitudes, as described in \cref{Sec:Unitarity_Regularization}. 
For convenience, we define
\begin{equation}
x_{\ell,\UNreg}^{\elas} \equiv 
\dfrac
{\sigma_{\ell,\UNreg}^{\elas}}
{\sigma_\ell^{\U}},
\quad
x_{\ell,\UNreg}^{\T} \equiv 
\dfrac
{\sigma_{\ell,\UNreg}^{\T}}
{\sigma_\ell^{\U,\T}},
\quad
x_{\ell,\UNreg}^{\V} \equiv 
\dfrac
{\sigma_{\ell,\UNreg}^{\V}}
{\sigma_\ell^{\U,\V}},
\label{eq:ReducedCrossSections}
\end{equation}
and
\begin{subequations}
\label{eq:RegularizationFunctions}
\label[pluralequation]{eqs:RegularizationFunctions}
\begin{align}
{\cal U}
(X;~\{w,\tilde w\},\{w',\tilde w'\})
&\equiv
\dfrac{\left[
\sqrt{X}\,(1-ww'+\tilde w\,\tilde w')
\pm
\sqrt{1-X}\,(\tilde w-\tilde w')
\right]^2}
{\left[(1+w)^2+\tilde w^2\right]
\left[(1+w')^2+\tilde w'^2\right]}
\nonumber
\\
&+
\dfrac{\left[
\sqrt{X}\,(w\tilde w'+\tilde w\,w')
\pm
\sqrt{1-X}\,(w-w')
\right]^2}
{\left[(1+w)^2+\tilde w^2\right]
\left[(1+w')^2+\tilde w'^2\right]},
\label{eq:RegularizationFunction}
\\[1.5em]
\bar{\cal U} (X;~y,y') 
&\equiv 
{\cal U} (X;~\{y,0\},\{y',0\}) 
\nonumber \\
&= 
\dfrac
{X (1-yy')^2+ (1-X) (y-y')^2}
{(1+y)^2 (1+y')^2} .
\label{eq:RegularizationFunction_simple}
\end{align}
\end{subequations}
Then, combining \cref{eq:CrossSections_PW_unreg,eq:ScatteringAmplitude_Regulated,eq:PhaseShift_Real+Complex,eq:PhaseShift_Complex}, we find
\begin{subequations}
\label{eq:xRegularized}
\label[pluralequation]{eqs:xRegularized}
\begin{align}
\label{eq:xRegularized_Elastic}
x_{\ell,\reg}^{\elas} 
&=
{\cal U} (
x_{\ell,\unreg}^{\elas};~
\{w_\ell,\tilde{w}_\ell\},\{0,0\}
),
\\
\label{eq:xRegularized_Transfer}
x_{\ell,\reg}^{\T} 
&=
{\cal U} (
x_{\ell,\unreg}^{\T};~
\{w_\ell,\tilde{w}_\ell\},
\{w_{\ell+1},\tilde{w}_{\ell+1}\}
),
\quad \text{for distinguishable particles,}
\\
\label{eq:xRegularized_Viscosity}
x_{\ell,\reg}^{\V} 
&=
{\cal U} (
x_{\ell,\unreg}^{\V};~
\{w_\ell,\tilde{w}_\ell\},
\{w_{\ell+2},\tilde{w}_{\ell+2}\}
) .
\end{align}
\end{subequations}
In the limit $w_\ell\to y_{\ell,\unreg}$ and $\tilde w_\ell\to 0$, \cref{eqs:xRegularized} reduce to
\begin{subequations}
\label{eq:xRegularized_Anal+Converge}
\label[pluralequation]{eqs:xRegularized_Anal+Converge}
\begin{align}
\label{eq:xRegularized_Elastic_Anal+Converge}
x_{\ell,\reg}^{\elas} &=
\bar{\cal U} (x_{\ell,\unreg}^{\elas};~y_{\ell,\unreg}^{},0),
\\
\label{eq:xRegularized_Transfer_Anal+Converge}
x_{\ell,\reg}^{\T} &=
\bar{\cal U} (
x_{\ell,\unreg}^{\T};~
y_{\ell,\unreg}^{},
y_{\ell+1,\unreg}^{}),
\quad \text{for distinguishable particles,}
\\
\label{eq:xRegularized_Viscosity_Anal+Converge}
x_{\ell,\reg}^{\V} &=
\bar{\cal U} (
x_{\ell,\unreg}^{\V};~
y_{\ell,\unreg}^{},
y_{\ell+2,\unreg}^{}).
\end{align}
\end{subequations}
The signs in \cref{eq:RegularizationFunction}, when used in \cref{eqs:xRegularized}, are chosen according to the signs of
\begin{equation}
\sin(2\theta_\ell),
\qquad
\sin\left[2(\theta_\ell-\theta_{\ell+1})\right],
\qquad
\sin\left[2(\theta_\ell-\theta_{\ell+2})\right] ,
\end{equation}
respectively. If one of these quantities vanishes, the corresponding sign-dependent term vanishes. In \cref{eq:xRegularized,eq:xRegularized_Anal+Converge}, the formulas for the transfer cross sections apply to distinguishable particles; for identical particles, 
$\sigma_{\reg}^{\T}=\sigma_{\reg}^{\elas}$, according to \cref{eq:IdenticalParticles_Elas=Transfer}.

\clearpage
\section{Velocity averages inside a halo}
\label{App:VelocityAverage}

\subsection{Relative velocity distribution}

We assume that the DM particles follow an isotropic truncated Maxwellian distribution,
\begin{equation}
F(\mathbf{v},v_c,\vesc) \equiv 
{\cal N}(v_c,\vesc)
\, \Theta(\vesc-v) 
\, e^{-v^2/v_c^2},
\label{eq:Maxwellian distrib with cut off}
\end{equation}
where $v_c/\sqrt{2}$ is the one-dimensional velocity dispersion of the corresponding untruncated Maxwellian, denoted often $v_0$~\cite{Colquhoun:2020adl}, so that $v_c=\sqrt{2}\,v_0$. The Heaviside function $\Theta(\vesc-v)$ removes the energetic particles that escape the halo. The normalization factor,
\begin{equation}
{\cal N} (v_c,\vesc)
\equiv
\left(\sqrt{\pi}v_c\right)^{-3}
\left[
\erf\left(\dfrac{\vesc}{v_c}\right)
-\dfrac{2}{\sqrt{\pi}}
\left(\dfrac{\vesc}{v_c}\right)
\, e^{-\vesc^2/v_c^2}
\right]^{-1}, 
\end{equation}
ensures that
$\int \dd^3 {\bf v} \, F({\bf v},v_c,\vesc)=1$. 
For any function of the relative velocity,
\begin{equation}
\average{h(\mathbf{v}_{\rm rel})}=
\int \dd^3\mathbf{v}_1
\int \dd^3\mathbf{v}_2 
\, h(\mathbf{v}_1-\mathbf{v}_2)
\, F(\mathbf{v}_1,v_{c},\vesc)
\, F(\mathbf{v}_2,v_{c},\vesc) .
\label{eq:vrel_average_def}
\end{equation}
Setting
$\mathbf{v}_1 = \mathbf{V}+\mathbf{v}_{\rm rel}/2$ and 
$\mathbf{v}_2=\mathbf{V}-\mathbf{v}_{\rm rel}/2$, 
the cutoff $v_1,v_2 \leqslant \vesc$ implies 
\begin{align}
V\vrel \, 
\abs{\hat{\mathbf{V}}\cdot\hat{\mathbf{v}}_{\rm rel}} 
\leqslant \vesc^2 -V^2-\vrel^2/4. 
\end{align}
We can now separate the motion of the center of mass from the relative motion, 
\begin{align}
\average{h ({\bf v}_{\rm rel})}
=\int \dd^3\mathbf{v}_{\rm rel} 
\, h(\mathbf{v}_{\rm rel}) 
\, F_{\rm CM} ({\bf v}_{\rm rel},v_c,\vesc)
\, e^{-\vrel^2/(2v_c^2)},
\end{align}
with 
\begin{align*}
\dfrac{F_{\rm CM}
(\mathbf{v}_{\rm rel}, v_c,\vesc)}
{{\cal N}(v_{c},\vesc)^2}
&=\int \dd^3\mathbf{V}
\, e^{-2V^2/v_c^2}
\ \Theta\qty(
\vesc^2-V^2-\frac{\vrel^2}{4}
-V\vrel\abs{\hat{\mathbf V}\cdot\hat{\bf v}_{\rm rel}}
)
\\
&=
\Theta(2\vesc-\vrel)
\left(\frac{\pi v_c^2}{2}\right)^{\frac{3}{2}}
\Bigg[
\erf\left(
\frac{2\vesc-\vrel}{\sqrt{2}v_c}
\right)
\\
&-\sqrt{\frac{2}{\pi}}
\frac{v_c}{\vrel}
e^{-(2\vesc-\vrel)^2/(2v_c^2)}
\left(
1-e^{-\vrel(2\vesc-\vrel)/v_c^2} 
\right)
\Bigg].
\end{align*}
Setting
$x\equiv \vrel/(\sqrt{2}v_c)$ and
$x_f\equiv \sqrt{2}\vesc/v_c$, and defining
\begin{equation}
f(x,x_f)\equiv
\frac{4}{\sqrt{\pi}}
\, e^{-x^2}
\, \Theta(x_f-x)
\ \dfrac
{\erf(x_f-x)-\frac{e^{-(x_f-x)^2}}
{x\sqrt{\pi}} \qty[1-e^{-2x(x_f-x)}]}
{\qty[
\erf\qty(\frac{x_f}{\sqrt{2}}) 
-\sqrt{\frac{2}{\pi}}x_fe^{-x_f^2/2}]^2}
\end{equation}
we arrive at
\begin{align}
\average{h ({\bf v}_{\rm rel})} =
\int \frac{\dd\Omega_{{\bf v}_{\rm rel}}}{4\pi}
\int_0^\infty \frac{ \dd \vrel \, \vrel^2}{(\sqrt{2}v_c)^3}
\ f\left(\frac{\vrel}{\sqrt{2} v_c},\frac{\sqrt{2}\vesc}{v_c}\right)
\, h({\bf v}_{\rm rel}).
\end{align}
We note that 
$\int_0^\infty \dd x\,x^2 f(x,x_f) =1$, 
and in the limit $\vesc \to \infty$, 
\begin{equation}
f_\infty(x) \equiv
\lim_{x_f\to\infty} f(x,x_f)
=\frac{4}{\sqrt{\pi}}e^{-x^2} .
\end{equation}
\subsection{Useful averages}
Using the above, we now define and compute the following averages: 
\begin{subequations}
\label{eq:VelocityAverages_def}
\label[pluralequation]{eqs:VelocityAverages_def}
\begin{align}
\average{v_{\DM}} &=
\int \dd^3\mathbf{v}
\ F(\mathbf{v},v_c,\vesc)
\ v ,
\label{eq:vDM_Average_def}
\\
\average{v_{\DM}^2} &=
\int \dd^3\mathbf{v}
\ F(\mathbf{v},v_c,\vesc)
\ v^2 ,
\label{eq:vDMsq_Average_def}
\\
\average{\vrel} &=
\int \frac{\dd\Omega_{{\bf v}_{\rm rel}}}{4\pi}
\int_0^\infty \frac{\dd\vrel \, \vrel^2}{(\sqrt{2}v_c)^3}
\ f \left(\frac{\vrel}{\sqrt{2}v_c},\frac{\sqrt{2}\vesc}{v_c}\right) 
\ \vrel .
\label{eq:vrel_Average_def}
\end{align}
\end{subequations}
\Cref{eq:vDM_Average_def,eq:vDMsq_Average_def} are averages over the velocity distribution of a single particle, while \cref{eq:vrel_Average_def} is an average over the relative-velocity distribution defined above. We find
\begin{subequations}
\label{eq:VelocityAverages}
\label[pluralequation]{eqs:VelocityAverages}
\begin{align}
\average{v_{\DM}} &=
\dfrac{2v_c}{\sqrt{\pi}}
\ \dfrac{
1-\left(1+\dfrac{x_f^2}{2}\right)e^{-x_f^2/2}
}
{\erf\left(\dfrac{x_f}{\sqrt{2}}\right)-\sqrt{\dfrac{2}{\pi}} \, x_f \, e^{-x_f^2/2}},
\label{eq:vDM_Average}
\\[1ex]
\average{v_{\DM}^2} &=
\dfrac{3v_c^2}{2}
\left(1-
\dfrac
{\dfrac{\sqrt{2} x_f^3}{3\sqrt{\pi}} \, e^{-x_f^2/2}}
{\erf\left(\dfrac{x_f}{\sqrt{2}}\right)-\sqrt{\dfrac{2}{\pi}} \, x_f \, e^{-x_f^2/2}}
\right),
\label{eq:vDMsq_Average}
\\[1ex]
\average{\vrel} &= 
\dfrac{2\sqrt{2}v_c}{\sqrt{\pi}}
\ \dfrac{
\erf(x_f)
+ e^{-x_f^2} 
\, \dfrac{x_f(3+2x_f^2)}{3\sqrt{\pi}}
- e^{-x_f^2/2} 
\, \dfrac{(3+x_f^2)}{\sqrt{2}}
\, \erf\left(x_f/\sqrt{2}\right)}
{\left[
\erf\left(\dfrac{x_f}{\sqrt{2}}\right)-\sqrt{\dfrac{2}{\pi}} \, x_f \, e^{-x_f^2/2}
\right]^2}. 
\label{eq:vrel_Average}
\end{align}
\end{subequations}
We also note the exact relation
\begin{equation}
\average{\vrel^2}
=
2\average{v_{\DM}^2},
\label{eq:vrel2_vDM2_relation}
\end{equation}
which follows from isotropy and the statistical independence of the two single-particle velocities. As a consistency check, in the untruncated limit $\vesc\to\infty$,
\begin{equation}
\label{eq:UntruncatedMaxwellian_VelocityAverages}
\average{v_{\DM}} 
= \dfrac{2v_c}{\sqrt{\pi}},
\qquad
\langle v_{\DM}^2 \rangle 
= \dfrac{3v_c^2}{2},
\qquad
\langle\vrel \rangle
= \dfrac{2\sqrt{2}v_c}{\sqrt{\pi}}.
\end{equation}
\clearpage
\section{Short-distance annihilation and elastic amplitudes}
\label{App:ShortDistanceAmplitudes}

\subsection{Annihilation amplitude}
\label{App:ShortDistanceAmplitudes_Annihilation}

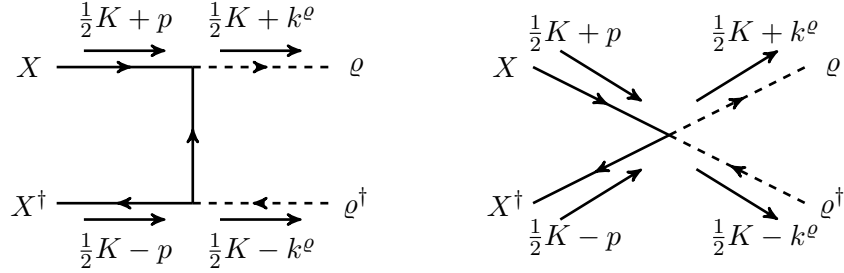
\begin{figure}[h!]
\centering
\begin{tikzpicture}[line width=1pt, scale=1.8]

\begin{scope}[shift={(-3,0.5)}]
\node at (-1.2,1){$X$};
\node at (-1.2,0){$X^{\dagger}$};
\draw[fermion] (-1,1) -- (0,1);
\draw[fermionbar] (-1,0) -- (0,0);
\draw[fermion] (0,0) -- (0,1);
\draw[scalar] (0,1) -- (1,1);
\draw[scalarbar] (0,0) -- (1,0);
\node at (1.2,1){$\varrho$};
\node at (1.2,0){$\varrho^{\dagger}$};
\draw[->] (-0.8,1.12) -- (-0.2,1.12);
\node at (-0.5,1.35){$\small \frac{1}{2}K+p$};
\draw[->] (-0.8,-0.12) -- (-0.2,-0.12);
\node at (-0.5,-0.35){$\small \frac{1}{2}K-p$};
\draw[->] (0.2,1.12) -- (0.8,1.12);
\node at (0.5,1.35){$\small \frac{1}{2}K+k^{\varrho}$};
\draw[->] (0.2,-0.12) -- (0.8,-0.12);
\node at (0.5,-0.35){$\small \frac{1}{2}K-k^{\varrho}$};
\end{scope}

\begin{scope}[shift={(0.5,0.5)}]
\node at (-1.2,1){$X$};
\node at (-1.2,0){$X^{\dagger}$};
\draw[fermion] (-1,1) -- (0,0.5);
\draw[fermionbar] (-1,0) -- (0,0.5);
\draw[scalar] (0,0.5) -- (1,1);
\draw[scalarbar] (0,0.5) -- (1,0);
\node at (1.2,1){$\varrho$};
\node at (1.2,0){$\varrho^{\dagger}$};
\draw[->] (-0.8,1.12) -- (-0.2,0.75);
\node at (-0.7,1.25){$\small \frac{1}{2}K+p$};
\draw[->] (-0.8,-0.12) -- (-0.2,0.25);
\node at (-0.7,-0.25){$\small \frac{1}{2}K-p$};
\draw[->] (0.2,0.75) -- (0.8,1.12);
\node at (0.7,1.25){$\small \frac{1}{2}K+k^{\varrho}$};
\draw[->] (0.2,0.25) -- (0.8,-0.12);
\node at (0.7,-0.25){$\small \frac{1}{2}K-k^{\varrho}$};
\end{scope}

\end{tikzpicture}
\caption{Leading-order short-distance contributions to DM annihilation, $XX^\dagger\to \varrho\varrho^\dagger$. The arrows on the field lines denote the charge flow.}
\label{fig:DMannihilation_Perturbative}
\end{figure}

\noindent
The leading-order contributions to the irreducible annihilation amplitude $XX^\dagger\to \varrho\varrho^\dagger$ are shown in \cref{fig:DMannihilation_Perturbative}. We compute the corresponding amplitude, allowing the incoming $XX^\dagger$ particles to be off-shell while requiring the outgoing $\varrho\varrho^\dagger$ to be on-shell.  In the CM frame, taking the total energy to be $\sqrt{s}$ and the energy of each incoming particle to be $\sqrt{s}/2$, the amplitude is
\begin{align}
{\cal A}(\vb{p},\vb{k}^\varrho)
&= 
\dfrac{16\pi\arho \, \mX^2}
{s/4+\mX^2-\mrho^2+p^2-2p k^\varrho
\cos \vartheta}
-\lambda_{X\varrho}
, \qquad
s \geqslant 4\mrho^2,
\label{eq:AnnihilationAmplitude_Tree}
\end{align}
where $\cos\vartheta = \hat{\vb{p}}\cdot \hat{\vb{k}}^\varrho$, $p=\abs{\vb{p}}$ is the momentum of the initial $X,X^\dagger$ particles, and $k^\varrho=\abs{\vb{k}^\varrho}=\sqrt{s/4-\mrho^2}$ is the momentum of the final $\varrho,\varrho^\dagger$ particles.

\subsubsection*{Partial-wave analysis}
We expand ${\cal A}(\vb{p},\vb{k}^\varrho)$ in partial waves as follows:
\begin{subequations}
\label{eq:AnnihilationAmplitude_Tree_PWanalysis}
\begin{align}
    {\cal A}(\vb{p},\vb{k}^\varrho) &=
    16\pi\sum_{\ell=0}^{\infty}
    (2\ell+1) 
    \, {\cal A}_\ell(p,k^\varrho) 
    \, P_\ell(\cos \vartheta),
    \\
    {\cal A}_\ell (p,k^\varrho) &=
    \dfrac{1}{64\pi^2} 
    \int \dd\Omega 
    \, P_\ell(\cos \vartheta) 
    \, {\cal A}(\vb{p},\vb{k}^\varrho).
\end{align}
\end{subequations}
From \cref{eq:AnnihilationAmplitude_Tree,eq:AnnihilationAmplitude_Tree_PWanalysis}, we find
\begin{align}
{\cal A}_\ell (p,k^\varrho) 
=\dfrac{\arho\mX^2}{2(s/4+\mX^2-\mrho^2 +p^2)} 
\, I_\ell \qty(
\dfrac
{2pk^\varrho}
{s/4+\mX^2-\mrho^2 + p^2}
) 
- \dfrac{\lambda_{X\varrho}}{16\pi} \, \delta_{\ell,0}
,
\label{eq:AnnihilationAmplitude_Tree_PW}
\end{align}
where the function $I_\ell$ is defined via
\begin{align}
I_\ell(z) 
&\equiv
\int_{-1}^{1} \dd x \,
\frac{P_\ell(x)}{1-zx}
=
\dfrac{2^{\ell+1}\ell!^2}{(2\ell+1)!}
\ z^\ell
\ {}_2F_1\qty(\dfrac{\ell+1}{2}, \dfrac{\ell+2}{2}; \ell+\frac{3}{2}; z^2).
\label{eq:Iell_def}
\end{align}
\Cref{eq:AnnihilationAmplitude_Tree_PW} can be expanded as
\begin{align}
{\cal A}_\ell (p,k^\varrho) =
\sum_{r=0}^\infty 
\eta_{A,\ell,r} (s)
\ (p/\mu)^{\ell+2r}, 
\label{eq:AnnihilationAmplitude_Tree_PW_exp}
\end{align}
where $\mu=\mX/2$ is the reduced mass of the $X,X^\dagger$ pair, with 
\begin{align}
\eta_{A,\ell,r}(s)
&=
\arho
\sum_{n=0}^{r}
\dfrac{
(-1)^{r-n}
\, 2^{2(\ell+n)}
\,(\ell+n)!
\,(\ell+r+n)!
}
{n! \,(r-n)! \,(2\ell+2n+1)!}
\ 
\dfrac
{\mX^2 \mu^{\ell+2r} \, \qty(s/4-\mrho^2)^{\ell/2+n}}
{\qty(s/4+\mX^2-\mrho^2)^{\ell+r+n+1}}
\nonumber \\
&-\dfrac{\lambda_{X\varrho}}{16\pi}
\,\delta_{\ell,0}
\,\delta_{r,0}
, \qquad s\geqslant 4\mrho^2.
\label{eq:Annihilation_AbsorptiveCouplings}
\end{align}

\subsubsection*{Nonrelativistic approximation}

In the nonrelativistic regime,
\begin{align}
\label{eq:EnergyMomentum_NRapprox}
s \simeq 4\mX^2
\qquad \text{and} \qquad
k^\varrho &\simeq \sqrt{\mX^2-\mrho^2}
= \mX \sqrt{1-\tau^2/4},
\end{align}
where $\tau\equiv 2\mrho/\mX$. Setting $r=0$ in \cref{eq:Annihilation_AbsorptiveCouplings} and evaluating the result in this limit, we obtain the leading absorptive coupling for each partial wave,
\begin{align}
\eta_{A,\ell} 
\equiv \eta_{A,\ell,0}
\simeq    
\dfrac{\arho}{2}\frac{\ell!^2}{(2\ell+1)!}
\qty(1-\frac{\tau^2}{4})^{\ell/2}
\qty(1-\frac{\tau^2}{8})^{-\ell-1}
- \dfrac{\lambda_{X\varrho}}{16\pi} \, \delta_{\ell,0} 
.
\label{eq:Annihilation_AbsorptiveCoupling_NR}
\end{align}

\subsection{Elastic scattering}
\label{App:ShortDistanceAmplitudes_Elastic}

The short-distance Hermitian kernel receives two distinct contributions. First, the quartic self-interaction generates an elastic contact term, ${\cal K}^{\twoPI} (\vb{p'},\vb{p}) \supset -\lambda_{X}$. Second, the annihilation channel induces a dispersive optical kernel. Retaining only the leading coefficient, $\eta_{A,\ell,0}(s)$, in the $p/\mu$ expansion of the partial-wave irreducible annihilation amplitudes, we find that the dispersive and absorptive partial-wave optical kernels have the same dependence on the incoming and outgoing momenta, while their energy-dependent coefficients are related by a Hilbert transform~\cite[Sec.~3.3]{Flores:2025uoh}. The $\lambda_X$  contribution also has the same $s$-wave structure and may therefore be incorporated into the $\ell=0$ dispersive coefficient, as in \cref{eq:Kernel2PI_AA_ComplexCoupling}, with $\hat{A}_\ell (p) = (p/\mu)^\ell$ and
\begin{align}
\eta_{H,\ell} (s) = 
\dfrac{1}{(1-4\mrho^2/s)^{1/2}}
\qty[
-\dfrac{\lambda_{X}^{\rm bare}}{16\pi}
\delta_{\ell,0}
+
\dfrac{\PV}{\pi}
\int_{2\mrho}^\Lambda
\dfrac{\dd \varepsilon}{\varepsilon-\sqrt{s}}
\, [\eta_{A,\ell,0} (\varepsilon^2)]^2
\, (1-4\mrho^2/\varepsilon^2)^{1/2}
].
\label{eq:DispersiveShortRangeCoupling}
\end{align}
Here, $\Lambda$ is a UV cutoff and $\lambda_X^{\rm bare}(\Lambda)$ is the corresponding bare quartic coupling. For the first term, we have used the partial-wave convention of \cref{eq:AnnihilationAmplitude_Tree_PWanalysis}. Using the full energy dependence of \cref{eq:Annihilation_AbsorptiveCouplings} for $r=0$, we find that the contributions to the dispersive integral proportional to $\alpha_\varrho^2$ and $\alpha_\varrho\lambda_{X\varrho}$ remain finite as $\Lambda\to\infty$, whereas the contribution proportional to $\lambda_{X\varrho}^2$ diverges logarithmically. This divergence is absorbed into the cutoff dependence of $\lambda_X^{\rm bare}(\Lambda)$. For $\ell\geqslant 1$, all terms involving $\lambda_{X\varrho}$ are absent, and the dispersive integral is UV convergent at this order.

\clearpage
\section{Unitarization of higher partial waves}
\label{App:UnitHigherWaves}

In \cref{Sec:AnnFeedback_Resonances}, we focused on $s$-wave resonances. Here we briefly explain why annihilation feedback from higher partial waves is expected to be less important in the present model.

For $\ell\geqslant1$, the centrifugal barrier suppresses the short-distance wavefunction and allows quasi-bound states. Consequently, finite-velocity resonance peaks can be displaced from the fixed bound-state thresholds $\xi_{n,\ell}$, approaching them in the zero-momentum limit~\cite{Beneke:2024iev}. Exactly at a bound-state threshold, the effective-range expansion gives $\theta_\ell\propto k^{2\ell-1}$ and $x_{\ell,\unreg}^{\elas}\propto k^{4\ell-2}$ for $\ell\geqslant 1$. Since the attractive partial waves in the present model are even, the first higher resonant mode is the $d$ wave, for which $x_{2,\unreg}^{\elas}\propto k^6$ on resonance, whereas at the same parameter point the $s$-wave is generically off resonance and scales as $x_{0,\unreg}^{\elas}\propto k^{2}$. The centrifugal suppression also limits annihilation feedback: since $y_{\ell,\unreg}\propto k^{2\ell+1}S_\ell$ and $S_\ell\propto k^{-4}$ for $\ell\geqslant1$ exactly at threshold resonance, $y_{\ell,\unreg}\propto k^{2\ell-3}$. Thus $y_{2,\unreg}\propto k$ decreases towards low velocities, in contrast to $y_{0,\unreg}\propto k^{-1}$ on an $s$-wave resonance.

\begin{figure}[b!]
\centering
\includegraphics[width=0.95\linewidth]{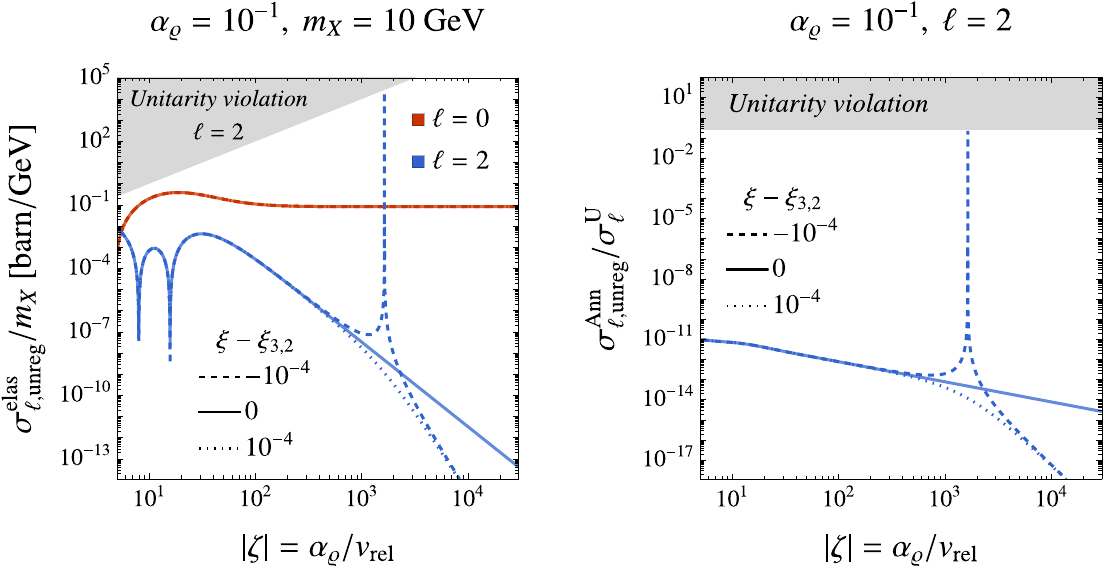}
\caption{
Unregulated partial-wave cross sections as functions of $|\zeta|=\arho/\vrel$, for $\arho=10^{-1}$, around the $(n,\ell)=(3,2)$ bound-state threshold.
\emph{Left:} Elastic $s$- and $d$-wave contributions, for $\mX=10~{\rm GeV}$. The $s$ wave dominates at low velocities, while a narrow finite-velocity enhancement appears for negative detuning.
\emph{Right:} Normalized $d$-wave annihilation cross section. It decreases towards low velocities even on resonance, but is sharply enhanced near the finite-velocity peak for negative detuning.
}
\label{fig:Elas_Ann_Res_Ell2_Enn3}
\end{figure}

A small negative detuning, $\Delta\xi\equiv\xi-\xi_{n,\ell}<0$, can nevertheless produce a narrow finite-velocity resonance associated with a quasi-bound state temporarily trapped by the centrifugal barrier, as seen in \cref{fig:Elas_Ann_Res_Ell2_Enn3}. As the velocity decreases, the value of $\xi$ at which this peak occurs approaches the threshold resonance $\xi_{n,\ell}$ from below, so that $|\Delta\xi_{\rm peak}|$ decreases towards zero. Around the peak, both the elastic and annihilation cross sections can become strongly enhanced over a narrow velocity interval.

For $\ell\geqslant1$, a direct unitarization is most naturally performed using the full irreducible partial-wave annihilation amplitude of \cref{eq:AnnihilationAmplitude_Tree_PW}, rather than its leading low-momentum expansion. The total CM energy must be kept fixed, $\sqrt{s}\simeq\mT$, and hence so must $k^\varrho(s)$, while allowing the incoming momentum $p$ to vary off shell. The resulting amplitude behaves as $\hat{\cal A}_\ell(p)\propto p^{-\ell-2}$ at large $p$. It can then be inserted into \cref{eq:InelasticAmplitude_M-Arelation} to obtain $\hat{\cal M}_{\ell,\unreg}(q)$ for the off-shell momenta entering the $\WW_\ell$ matrix of \cref{eq:Wmatrix_def}. The continuum contribution to $\WW_\ell$ may then be evaluated numerically along the real $q$ axis, together with the explicit bound-state contribution. The integral is UV convergent, avoiding the additional higher-derivative renormalization conditions that would be required if the leading behavior $\hat{\cal A}_\ell\propto p^\ell$ were extrapolated to arbitrarily large momenta.

We expect two main effects. First, unitarization should curtail the narrow finite-velocity peak. The annihilation cross section is itself strongly enhanced there, so its feedback regulates the inelastic amplitude, while the coupled unitarity constraint simultaneously requires the elastic peak to be reduced. Second, at velocities below the peak, unitarization is expected to reduce the strong sensitivity to the detuning. In the baseline calculation, the cross sections exactly at threshold resonance and at slight detuning can differ by many orders of magnitude. The self-energy correction shifts the finely tuned threshold pole, so by analogy with the $s$-wave case we expect the regulated on-resonance result to move closer to the nearby detuned ones. This may occur even when $y_{\ell,\unreg}\ll1$: a small on-shell absorptive cross section does not preclude a dispersive shift of a finely tuned threshold pole.

We do not pursue this calculation here because higher-partial-wave feedback appears unlikely to be important for SIDM phenomenology. Away from the narrow finite-velocity peaks, the centrifugal barrier renders the higher-$\ell$ contributions subdominant to the non-resonant $s$ wave even on their respective threshold resonances. The peaks themselves occupy a very narrow velocity interval and are strongly smoothed by averaging over the velocity distribution inside a halo, so their contribution to the halo-averaged total elastic and viscosity cross sections is expected to be small.
\clearpage
\end{appendices}

\newpage
\bibliography{biblio.bib}
\bibliographystyle{JHEP.bst}
\end{document}